\documentstyle[aps,prb,epsf,floats,eqsecnum]{revtex}
\begin{document}
\twocolumn[\hsize\textwidth\columnwidth\hsize\csname@twocolumnfalse%
\endcsname
\title{Ground states of quantum antiferromagnets in two dimensions}
\author{Subir Sachdev$^{1,2}$ and Kwon Park$^1$}
\address{$^1$Department of Physics, Yale University, P.O. Box 208120, New
Haven, CT 06520-8120\\
$^2$Department of Physics, Harvard University, Cambridge MA 02138}
\date{August 14, 2001}

\maketitle

\begin{abstract}
We explore the ground states and quantum phase transitions of
two-dimen\-sion\-al, spin $S=1/2$, antiferromagnets by
generalizing lattice models and duality transforms introduced by
Sachdev and Jalabert (Mod. Phys. Lett. B {\bf 4}, 1043 (1990)).
The `minimal' model for square lattice antiferromagnets is a
lattice discretization of the quantum non-linear sigma model,
along with Berry phases which impose quantization of spin. With
full SU(2) spin rotation invariance, we find a magnetically
ordered ground state with N\'{e}el order at weak coupling, and a
confining paramagnetic ground state with bond charge ({\em e.g.}
spin Peierls) order at strong coupling. We study the mechanisms
by which these two states are connected in intermediate coupling.
We extend the minimal model to study different routes to
fractionalization and deconfinement in the ground state, and also
generalize it to cases with a uniaxial anisotropy (the spin
symmetry group is then U(1)). For the latter systems,
fractionalization can appear by the pairing of vortices in the
staggered spin order in the easy-plane; however, we argue that
this route does not survive the restoration of SU(2) spin
symmetry. For SU(2) invariant systems we study a separate route
to fractionalization associated with the Higgs phase of a complex
boson measuring {\em non-collinear}, spiral spin correlations: we
present phase diagrams displaying competition between magnetic
order, bond charge order, and fractionalization, and discuss the
nature of the quantum transitions between the various states. A
strong check on our methods is provided by their application to
$S=1/2$ frustrated antiferromagnets in one dimension: here, our
results are in complete accord with those obtained by
bosonization and by the solution of integrable models.
\end{abstract}]

\tableofcontents

\section{Introduction}
\label{intro}

A central problem \cite{pwa} in the study of correlated electron
systems is the classification of the quantum phases and critical
points of two-dimensional quantum antiferromagnets at zero
temperature ($T$). Of particular interest, because of its
association with the physics of the cuprate superconductors, is
the spin $S=1/2$ antiferromagnet on the square lattice:
\begin{equation}
H = J \sum_{\langle i j \rangle} \hat{\bf S}_i \cdot \hat{\bf S}_j
+ \cdots \label{H}
\end{equation}
Here $\hat{\bf S}_j$ are $S=1/2$ quantum spin operators on the
sites, $j$, of a square lattice, the sum $\langle ij \rangle$ is
over nearest neighbor links, and $J>0$ is the antiferromagnetic
exchange energy. The ellipsis represents smaller further neighbor
or ring exchange terms that can be added to deform the purely
nearest neighbor model: we allow a wide class of such terms.
Without these additional terms, the ground state of $H$ is well
known: it is the N\'{e}el state with a spontaneous staggered
magnetic moment
\begin{equation}
\langle \hat{\bf S}_j \rangle = \eta_j N_0 {\bf e}~~~~,~~~~T=0,
\label{n0}
\end{equation}
where $\eta_j = \pm 1$ identifies the square sublattice of site
$j$, ${\bf e}$ is a unit vector with an arbitrary, fixed,
orientation in spin space, and $N_0>0$ is the magnitude of the
moment. We will refer to this state as the Heisenberg-N\'{e}el
(HN) state, the qualifier denoting the SU(2) (Heisenberg) symmetry
of $H$ (this allows us to distinguish from cases in which the
SU(2) spin symmetry has been reduced to U(1), which we consider
later in the paper). The low energy, elementary excitations of
the HN state are a doublet of gapless spin wave modes, associated
with slow modulations in the orientation ${\bf e}$.

This paper will address the following question and related
issues\cite{science}: what are the possible ground states of $H$
which are ``near'' the HN state~? More precisely, imagine slowly
reducing the value of $N_0$ (while (\ref{n0}) maintains its
spatial structure) by modifying some of the unspecified
non-nearest-neighbor terms in $H$; at some critical point in
parameter space, $N_0$ will vanish and spin rotation invariance
will be restored; how does the ground state of $H$ evolve towards
and beyond this critical point~? What is the field theoretic
description of the quantum critical point at which $N_0$
vanishes, and of any additional quantum critical points that may
be encountered~? How do these results get modified when exchange
anisotropies in $H$ (like an additional $J \zeta \sum_{\langle i j
\rangle} \hat{S}_{iz} \hat{S}_{jz}$ term) reduce the global spin
symmetry of $H$ from SU(2) to U(1)~?

Along some lines in parameter space, the spatial form of
(\ref{n0}) may change before spin rotation invariance is restored
{\em e.g.} the spontaneous magnetization can become incommensurate
and non-collinear. This is a situation of fundamental importance,
but in the interests of simplicity we will postpone discussion of
phases associated with non-collinear spin correlations to
Section~\ref{planar}, where we present in (\ref{f4}) our `global'
theory for $S=1/2$ quantum antiferromagnets in two dimensions.
The phases and phase transitions in Section~\ref{planar} are
generalizable to other lattices with non-collinear spin
correlations {\em e.g.} the triangular lattice. Also, although we
will only explicitly consider the square lattice antiferromagnet
in this paper, all our results can also be applied to other
two-dimensional, bipartite lattices, like the honeycomb lattice.

\begin{figure}
\epsfxsize=3in \centerline{\epsffile{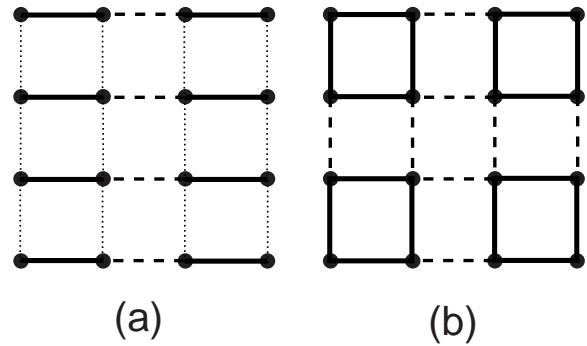}} \vspace{0.1in}
\caption{Sketch of the two simplest possible states with
bond-centered charge (BC) order: (a) the columnar spin-Peierls
states, and (b) plaquette state. The different values of the
$Q_{ij}$ on the links are encoded by the different line styles;
the state in (a) has 3 distinct values of $Q_{ij}$, while that in
(b) has 2. Both states are 4-fold degenerate; an 8-fold
degenerate state, with superposition of the above orders, also
appears as a possible ground state of the generalized action for
BC order in Appendix~\protect\ref{c}} \label{bcorder}
\end{figure}

Apart from the HN state, states which will play a central role in
our considerations are those with bond-centered charge order (BC
order); these states have been argued\cite{rsprl1,rodolfo,rsprb}
to be generically contiguous to the HN state, and we will also
find this here. See Refs.~\onlinecite{sushkov1,leiden}, and
references therein, for recent numerical evidence for BC order in
systems like (\ref{H}). The BC states are characterized by
spontaneous spatial modulations in the expectation values of the
bond fields
\begin{equation}
Q_{ij} = \hat{\bf S}_i \cdot \hat{\bf S}_j , \label{defQ}
\end{equation}
where $i,j$ are on nearest neighbor sites. Modulations in $Q_{ij}$
will lead to corresponding modulations in other spin-singlet,
charge-conserving observables, including the charge density
associated with the orbitals that reside on the link $(i,j)$ (in
the cuprates, this is the location of the O ions). In the simplest
cases, the modulation doubles the unit cell, with the largest
values of $\langle Q_{ij} \rangle$ residing either in parallel
columns (also referred to as a spin-Peierls state) or in disjoint
plaquettes (see Fig~\ref{bcorder}).

This paper will address the nature of the quantum critical points
the system must encounter in evolving from the HN to a BC state.
On the basis of numerical analysis of the square lattice
antiferromagnet with first and second neighbor exchange
interactions, Sushkov, Oitmaa, and Zheng \cite{sushkov1} have
proposed that this evolution occurs via an intermediate phase
with co-existence of the two order parameters {\em i.e.\/} there
is first a dimerization transition from the HN state to a state
with co-existing N\'eel and bond-charge orders (the HN+BC state;
see Table~\ref{t1} for our convention for identifying ground
states), and then a second transition to the BC state where spin
rotation invariance is restored. (Such a scenario was also
mentioned in the early work of Ref.~\onlinecite{rodolfo}.) As is
discussed at length below, we shall find a number of situations
in which our field theories display precisely this sequence of
transitions. The other possibility, which general field theoretic
considerations can never rule out, is a direct first-order
transition from the HN to the BC state. We will not find any
clear-cut scenario for a generic ({\em i.e.} not multicritical)
second order transition between the HN and BC states, but cannot
definitively rule out the possibility that such a transition
exists (further discussion of this point is at the end of
Section~\ref{sec:subo}).
\begin{table}
\begin{tabular}{c|c}
 Label & Property \cr \hline HN & \begin{minipage}{2.8in}
\vspace{0.07in} The Hamiltonian is invariant under SU(2) spin
rotations, and spin rotation symmetry is broken by
(\protect\ref{n0}) with $N_0 > 0$ and ${\bf e}$ an arbitrary unit
vector in spin space. There are two gapless spin-wave excitations.
\vspace{0.07in}
\end{minipage} \\
\hline XY & \begin{minipage}{2.8in} \vspace{0.05in} The
Hamiltonian has only a U(1) spin rotation symmetry about the $z$
axis (due to the presence of exchange anisotropies), and
(\protect\ref{n0}) is obeyed with $N_0
> 0$ and ${\bf e}$ a unit vector in the $x$-$y$ plane. There is
one gapless spin-wave excitation. \vspace{0.05in}
\end{minipage} \\
\hline I & \begin{minipage}{2.8in} \vspace{0.05in} The Hamiltonian
has only a U(1) spin rotation symmetry about the $z$ axis (due to
the presence of exchange anisotropies), and (\protect\ref{n0}) is
obeyed with $N_0 > 0$ an Ising order parameter, and ${\bf e}$ a
unit vector along the $\pm z$ directions. There are no gapless
spin-wave excitations. \vspace{0.05in}
\end{minipage} \\
\hline BC & \begin{minipage}{2.8in} \vspace{0.05in} Lattice space
group symmetry is broken by a spontaneous modulation in the values
of the bond-centered charge order parameter $Q_{ij}$ defined in
(\protect\ref{defQ}). The simplest patterns, in
Fig~\protect\ref{bcorder}, double the unit cell to two sites.
\vspace{0.05in}
\end{minipage} \\
\hline F & \begin{minipage}{2.8in} \vspace{0.05in} The ground
state has topological order (leading to a four-fold degeneracy on
a large torus) and fractionalization. There are one or more gapped
excitations with spin $S_z=\pm 1/2$, and also a spinless $Z_2$
vortex excitation. \vspace{0.05in}
\end{minipage} \\
\hline P & \begin{minipage}{2.8in} \vspace{0.05in} Spin rotation
symmetry is broken by a non-collinear, polarization of the spins.
There are three gapless spin-wave excitations if the Hamiltonian
has full SU(2) symmetry, and only one if the symmetry is U(1).
\vspace{0.05in}
\end{minipage}
\end{tabular}
\vspace{0.2in} \caption{We identify ground states of
two-dimensional antiferromagnets by a label which specifies all of
the properties from the above list which are satisfied by the
state. It is assumed that if a property not contained in a state's
label, that state does not satisfy its requirements; so {\em e.g.}
while both the HN and HN+F states obey (\protect\ref{n0}), the HN
state does not obey property F, while the HN+F state does. Along
the same lines, the F state is invariant under spin rotations, as
it does not obey properties HN, XY, I, or P.} \label{t1}
\end{table}

Another class of our main results concerns the issue of spin
fractionalization. These will be reviewed below in
Section~\ref{intro:frac}. We will find that paramagnetic
fractionalized states (F in the notation of Table~\ref{t1}) are
not connected by a second-order phase boundary to the HN state in
the SU(2) symmetric $H$. In contrast, in spin systems with a
reduced, easy-plane U(1) symmetry, there can be a second-order
transition between the XY and F states
\cite{senthil,lannert,five}.

We introduce the computations of this paper by recalling some
essential facts about quantum and thermal fluctuations near the
HN state. At long wavelengths, it is accepted that these are well
described by the O(3) non-linear sigma model in 2+1 dimensions
\cite{chn}, with the action
\begin{equation}
S_{\sigma} = \frac{1}{2 c \widetilde{g}} \int d^2 x d \tau \left[
(\partial_{\tau} {\bf n} )^2 + c^2 (\nabla_x {\bf n})^2 \right],
\label{sigma}
\end{equation}
where $x$ is the continuum spatial co-ordinate, $\tau$ is
imaginary time, $c$ is the spin-wave velocity, and $\widetilde{g}$
is a (dimensionful) coupling constant which controls the strength
of quantum fluctuations. The field ${\bf n} (x, \tau)$ satisfies
${\bf n}^2 = 1$ everywhere in spacetime, and denotes the local
orientation, ${\bf e}$, of the short-range N\'{e}el order; so
\begin{equation}
\hat{\bf S}_j \sim \eta_j {\bf n} (x_j, \tau).
\end{equation}
However, as the value of $N_0$ decreases with increasing
$\widetilde{g}$, the action $S_{\sigma}$ ceases to be a complete
description of the underlying antiferromagnet $H$. The essential
missing ingredients are residual Berry phases associated with the
precession of the lattice spins\cite{book}. These appear in the
coherent state path integral representation of the quantum spin
Hamiltonian $H$ in (\ref{H}); they do not have an obvious
continuum limit, and to specify them, we have to return to a
lattice regularization of $S_{\sigma}$ and its degrees of freedom.

For the spatial co-ordinates, we return to the original square
lattice associated with $H$, while we also discretize the
imaginary time direction: consequently, spacetime is replaced by a
cubic lattice, on which reside the ${\bf n}_j$ degrees of freedom
(the label $j$ is now a three-dimensional cubic lattice index,
rather than a two-dimensional index in (\ref{H})---the
interpretation of $j$ should usually be clear from the context).
On this cubic lattice, the Berry phase term in the partition
function is $e^{-S_B}$ with
\begin{equation}
S_B = i \sum_{j} \eta_j {\cal A}_{j \tau} , \label{berry}
\end{equation}
where, as before, $\eta_j = 1$ on one of the square sublattices
and $\eta_j=-1$ on the other (note $\eta_j$ is independent of
$\tau$), and the leading $i = \sqrt{-1}$. The quantity ${\cal
A}_{j \mu}$ is defined to be {\em half} (for $S=1/2$) the signed
area of the spherical triangle formed between ${\bf n}_j$, ${\bf
n}_{j + \hat{\mu}}$ and ${\bf n}_0$, with ${\bf n}_0$ an arbitrary
reference unit vector (the index $\mu$ extends over the spacetime
directions $x$,$y$,$\tau$). We can choose ${\bf n}_0$ at our
convenience, and a convenient choice is usually ${\bf n}_0 =
(0,0,1)$. The explicit expression for half the area, ${\cal A}$,
of the spherical triangle bounded by ${\bf n}_{1,2,3}$
is\cite{berg}
\begin{equation}
e^{i {\cal A}} \equiv \frac{1 + {\bf n}_1 \cdot {\bf n}_2 + {\bf
n}_2 \cdot {\bf n}_3 + {\bf n}_3 \cdot {\bf n}_1 + i {\bf n}_1
\cdot ( {\bf n}_2 \times {\bf n}_3 )}{\left\{ 2 ( 1+ {\bf n}_1
\cdot {\bf n}_2)(1 + {\bf n}_2 \cdot {\bf n}_3)(1 + {\bf n}_3
\cdot {\bf n}_1) \right\}^{1/2}} \label{area}.
\end{equation}
It can be shown from this definition, or more geometrically from
the interpretation of ${\cal A}$ as a spherical area, that changes
in the choice of ${\bf n}_0$ amount to a gauge transformation of
the ${\cal A}_{j \mu}$, with
\begin{equation}
{\cal A}_{j \mu} \rightarrow {\cal A}_{j \mu} + \phi_{j +
\hat{\mu}} - \phi_{j}, \label{gauge}
\end{equation}
where $\phi_j$ is half the area of the spherical triangle formed
by ${\bf n}_j$ and the old and new choices for ${\bf n}_0$. It is
easily seen that (\ref{berry}) is invariant under (\ref{gauge})
provided we choose periodic boundary conditions in the $\tau$
direction---this we will always do. As a companion to $S_B$, we
also write down the continuum $S_{\sigma}$ in (\ref{sigma}) on the
same hypercubic lattice:
\begin{equation}
S_{\bf n} = - \frac{1}{g} \sum_{j, \hat{\mu}} {\bf n}_j \cdot {\bf
n}_{j + \hat{\mu}} \label{sn}
\end{equation}
where the $\hat{\mu}$ are vectors of length equal to the lattice
spacing, and extending over the $x$, $y$, and $\tau$ directions.
The dimensionless coupling  constant $g$ is proportional to
$\widetilde{g}$, while $c$ has been absorbed by the choice of the
temporal lattice spacing. For the above discretizations to be
meaningful representations of the underlying antiferromagnet, we
exclude values of $g$ so large that the ${\bf n}$ correlation
length is of order a lattice spacing or smaller.

In the analogous formulation of {\em one}-dimensional
antiferromagnets\cite{book}, a natural continuum representation of
the Berry phase term does exist. A careful examination of $S_B$
for smooth configurations of ${\bf n}_j$ shows that it can be
written as
\begin{equation}
S_{\theta} = \frac{i \theta}{4 \pi} \int dx d \tau {\bf n} \cdot
\left( \frac{\partial {\bf n}}{\partial \tau} \times
\frac{\partial {\bf n}}{\partial x} \right)
\end{equation}
with $\theta = \pi$. This is a topological term, and the continuum
theory $S_{\sigma} + S_{\theta}$ (the integral in (\ref{sigma})
now extends over one spatial dimension) was used by Haldane to
argue for the fundamental distinction between integer and
half-integer spin chains. While the continuum theory has been
valuable in establishing this distinction, very little
information on the physics at $\theta = \pi$ has been obtained
directly from the continuum functional integral of
$S_{\sigma}+S_{\theta}$. Rather, the most efficient approach has
been to refer back to the lattice S=1/2 spin chains, and to
analyze them by the methods of abelian bosonization. This paper
will perform a direct analysis of the lattice actions $S_{\bf n}
+ S_{B}$, in both spatial dimensions $d=1$ and $d=2$. We will use
parallel methods in the two cases, and the comparison with the
known bosonization results in $d=1$ will act as a significant
consistency check of our conclusions.

Although it is not essential, it is convenient to introduce
another term in the final form lattice action we shall
study---this has the advantage of yielding an independent
dimensionless coupling constant (in addition to $g$) which can be
used to explore a potentially richer phase diagram. We wish to
write down a term associated with the gauge potential ${\cal A}_{j
\mu}$. Any such term must clearly be invariant under the gauge
transformation (\ref{gauge}). Further, as only the values of ${\bf
n}_{j}$ are observable, (\ref{area}) indicates that the term
should be periodic under ${\cal A}_{j \mu} \rightarrow {\cal A}_{j
\mu} + 2 \pi$. These requirements strongly constrain the allowed
terms, and the simplest permissible one is
\begin{equation}
S_{{\cal A}}^{\prime} = -\frac{1}{e^2} \sum_{\Box} \cos
(\epsilon_{\mu\nu\lambda} \Delta_{\nu} {\cal A}_{j\lambda}).
\label{sap}
\end{equation}
This term is written in notation standard in lattice gauge theory:
the sum over $\Box$ extends over all plaquettes of the cubic
lattice, the indices $\mu$, $\nu$, $\lambda$ extend over the $x$,
$y$, $\tau$ directions. The symbol $\Delta_{\mu}$ represents a
discrete lattice derivative ($\Delta_{\mu} f_j \equiv f_{j +
\hat{\mu}} - f_{j}$), and so $S_{\cal A}$ depends on the ${\cal
A}$ flux threading each plaquette. In principle, we should allow
for different coupling constants associated with fluxes in the
spatial and temporal directions, but we have chosen them equal for
convenience. As we will discuss shortly, an important tool in our
analysis is a dual lattice representation of (\ref{sap}). It is
somewhat more convenient to perform this in a ``Villain
representation'' of (\ref{sap}), which replaces the exponential of
a cosine by a periodic sum over gaussians. So we shall replace
$S_{{\cal A}}^{\prime}$ by
\begin{equation}
S_{{\cal A}} = \frac{1}{2e^2} \sum_{\Box}
(\epsilon_{\mu\nu\lambda} \Delta_{\nu} {\cal A}_{j\lambda}-2 \pi
q_{\bar{\jmath}\mu} )^2 , \label{sa}
\end{equation}
where $q_{\bar{\jmath}\mu}$ is an integer-valued vector field on
the sites, $\bar{\jmath}$, of the dual cubic lattice; $q$ couples
to the ${\cal A}$ flux on the plaquette it pierces. It is also
clear that $S_{{\cal A}}$ becomes singular as $e^2 \rightarrow 0$,
and we shall therefore exclude this limit from our considerations;
we are primarily interested in moderately large values of $e^2$,
where the connection to Heisenberg antiferromagnets is evident.
However, the limit $e^2 \rightarrow \infty$ is not expected to be
singular, and the properties at $e^2=\infty$ should be smoothly
connected to those for large $e^2$.

\begin{figure}
\epsfxsize=3in \centerline{\epsffile{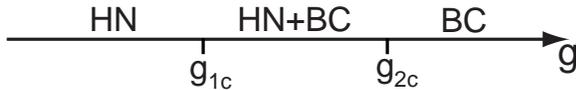}} \vspace{0.1in}
\caption{Proposed phase diagram of the model $Z$ in
(\protect\ref{z}) as a function of $g$ for a range of $e^2$. The
transition at $g=g_{1c}$ involves onset of BC order like that in
Fig~\protect\ref{bcorder}: this is in the universality class of
the 2+1 dimensional $Z_4$ clock model, which is in turn described
by the $\varphi^4$ field theory with O(2) symmetry. The HN order
vanishes at $g=g_{2c} > g_{1c}$ with a transition described by
the 2+1 dimensional $\varphi^4$ field theory with O(3) symmetry.
A direct first order transition between the HN and BC phases is
also possible. We have not find any clear-cut scenario for a
direct second order transition between the HN and BC states, but
cannot definitively rule out the possibility that such a
transition exists for some values of $e^2$ (further discussion of
this point is at the end of Section~\protect\ref{sec:subo}).}
\label{phaseZ}
\end{figure}

It is useful to collect all the terms introduced so far. One of
the central purposes of this paper is to understand the phase
diagram of the partition function
\begin{equation}
Z = \sum_{\{ q_{\bar{\jmath} \mu} \}} \int \prod_j d {\bf n}_j
\delta({\bf n}_j^2 - 1) \exp \left( - S_{\bf n} - S_{{\cal A}} -
S_B \right) \label{z}
\end{equation}
in the plane of the coupling constants $g$ and $e^2$. $Z$
constitutes a ``minimal model'' for a theory of $S=1/2$ quantum
antiferromagnets in two dimensions: it incorporates the essential
ingredients---quantum fluctuations of the short-range N\'{e}el
order represented by the non-linear sigma model plus the Berry
phases accounting for the quantization of spin. This model, and
the duality methods we shall employ, are closely related to those
introduced in Ref.~\onlinecite{rodolfo}; see also Appendix~\ref{a}
where we introduce an alternative formulation of $Z$ using fields
in $CP^1$---this formulation is more convenient for some purposes,
and is closer to the approach of Ref.~\onlinecite{rodolfo}. The
presence of the complex Berry phases associated with $S_B$ means
that a Monte Carlo evaluation of $Z$ will not be straightforward;
nevertheless, we hope that such evaluations will be attempted by
experts in the future. Our strategy will be to employ a variety
of duality and field-theoretic techniques to delineate the phase
diagram of $Z$. We will present strong arguments that only phases
with HN and BC order appear in this phase diagram, and that there
are no states with spin fractionalization in the SU(2) symmetric
$Z$. (see also Section~\ref{intro:frac} below). A likely scenario
for the sequence of phases as a function of increasing $g$ is HN,
HN+BC, BC, for all non-zero values of $e^2$, as shown in
Fig~\ref{phaseZ}.

As we have already indicated, the connection of our models with
the successful theory of one-dimensional antiferromagnets will be
an important guide to our analysis. For these systems, a complete
understanding of the physics was achieved by considering the phase
diagram in a wider space: the spin-symmetry was reduced to $U(1)$
by introducing an easy-plane or easy-axis anisotropy, and the
phase diagram was studied as a function of the strength of the
anisotropy. After mapping to continuum field theories by abelian
bosonization methods applicable for systems with only a U(1)
symmetry, it was found that one could nevertheless identify lines
in the parameter space of the field theories where the Heisenberg
SU(2) symmetry was restored. Thus one achieved an understanding of
SU(2) symmetric antiferromagnets by completely exploring the phase
diagram of less-symmetric, and simpler, systems. A similar
strategy will be an important ingredient in our arsenal here: we
will reduce the symmetry in the ${\bf n}$ space to U(1), by
introducing exchange anisotropy in $S_{\bf n}$ and so deforming
$Z$ to $Z_{\rm U(1)}$ (the explicit form is below in
(\ref{u11a})). We find then that duality transforms of $Z_{\rm
U(1)}$ can be carried through to completion. This yields models
which we will study by mappings to continuum field theories in
2+1 dimensions which are amenable to standard renormalization
group analyses. We will then search for lines in the parameter
space of these dual models at which SU(2) symmetry is present; the
properties of such lines will allow us to reach important
conclusions about the SU(2) symmetric partition function $Z$.

\subsection{Fractionalization}
\label{intro:frac}

Much attention has focused on ground states of two-dimensional
antiferromagnets with an ``order'' quite distinct from those
discussed so far. These are states with fractionalized spin
excitations (spin $S=1/2$) and associated topological,
spin-singlet, excitations. States with this property are denoted
by F, see Table~\ref{t1}). In a
theory\cite{rsprl2,sr,rodolfo2,triangle,japan} of paramagnetic F
states in systems with SU(2) symmetry, fractionalization was
linked with non-collinear spin correlations~\cite{note}. In
particular, restoring spin rotation invariance in a state with
long-range, non-collinear, magnetic order (a P state, see
Table~\ref{t1}) by a second-order quantum phase transition led
naturally to a paramagnetic F state. Such P states are not
expected to appear in the model $Z$ in (\ref{z}), and so the
theory of Ref.~\onlinecite{rsprl2} leads to the conclusion that
the SU(2) symmetric $Z$ does not contain the F state either. Our
results here will support this conclusion.

As already noted above, we will also study a deformation of $Z$ to
anisotropic spin models, $Z_{\rm U(1)}$ with only a U(1) spin
symmetry. For easy-plane anisotropy, this will deform the HN
state into the XY state (see Table~\ref{t1}). In such a
situation, the ${\bf n}_j$ field will fluctuate primarily in the
$x$-$y$ plane in spin space, and it is therefore useful to
parameterize in terms of an angular variable $\theta_j$ by
\begin{equation}
{\bf n}_j = (\cos \theta_j , \sin \theta_j, 0). \label{o3}
\end{equation}
We will show in Sections~\ref{sec:chain} and~\ref{sec:u1} that in
such an easy-plane limit, the model $Z$ in (\ref{z}) is
equivalent to the following XY model coupled to a $Z_2$ gauge
theory in 2+1 dimensions:
\begin{eqnarray}
&& Z_{\rm U(1)} = \sum_{\{ s_{j,j+\hat{\mu}} = \pm 1\}} \int
\prod_j d \theta_j \exp \Biggl( K \sum_{\Box} \prod_{\Box}
s_{j,j+\hat{\mu}} \nonumber
\\ &&+ \frac{4}{g} \sum_{j, \hat{\mu}} s_{j,j+\hat{\mu}} \cos \left(
\frac{\Delta_{\mu} \theta_j}{2} \right) - i \frac{\pi}{2} \sum_j
( 1 - s_{j,j+\hat{\tau}}) \Biggr), \label{u11a}
\end{eqnarray}
with the coupling $K$ is related to $e^2$ by (\ref{o5c});
actually the precise equivalence is to a model in which the
cosine in (\ref{u11a}) is replaced a periodic Gaussian-Villain
form, as in (\ref{o5b}). The $Z_2$ gauge field
$s_{j,j+\hat{\mu}}$ resides on the links of the direct lattice;
apart from the usual Maxwell term, it also carries the last Berry
phase term which is the remnant of $S_B$ in (\ref{berry}). Models
closely related to (\ref{u11a}) have recently been studied by
Senthil and Fisher and collaborators \cite{senthil,lannert,five}:
we have thus established a surprising connection between their
work and the easy-plane limit of model $Z$ of
Ref.~\onlinecite{rodolfo} and the present paper.

We make a brief aside to place the work of Senthil and Fisher in
the context of the models being studied here: their work has shed
new light on subject of fractionalization in spin systems with
U(1) symmetry.
 In their first paper\cite{senthil},
Senthil and Fisher considered a model with charge fluctuations
and SU(2) spin symmetry, but examined fractionalization of the
Cooper pairs associated with superfluid order and the U(1) charge
symmetry. Alternatively, we could replace the Cooper pair bosons
by the hard-core bosons associated with S=1/2 spin systems (with
the spin raising operators $S^{+} = b^{\dagger} \sim e^{i
\theta}$, the operator creating the hard-core boson (see
(\ref{o14}) later)), and transcribe their results to quantum spin
systems with only a U(1) symmetry. Precisely such a point of view
was taken in Section II of Ref.~\onlinecite{five}.

Our analysis of $Z_{\rm U(1)}$ will follow and generalize the
framework laid by these studies of $Z_2$ gauge theories. Duality
transformations of the model $Z_{\rm U(1)}$ show that, unlike
$Z$,  $Z_{\rm U(1)}$ does indeed contain a paramagnetic F state
in two dimensions; further, it appears possible to directly
connect the XY and F states by a second-order quantum phase
transition. The appearance of these F phases is intimately linked
to the half-angle variable appearing as the argument of the
cosine term in (\ref{u11a}); indeed $e^{i \theta/2}$ is the
`square root' of the $S_z = 1$ spin-flip boson $b^{\dagger}$, and
so the former is the creation operator a $S_z =1/2$ spinon. For
large $K$ and large $g$, the fluctuations of the $Z_2$ gauge field
$s_{j,j+\hat{\mu}}$ and the XY order will both be suppressed, and
this will introduce a paramagnetic F state in which such spinons
can propagate freely by the $1/g$ term in (\ref{u11a}).

Our main interest here is in spin systems with SU(2) symmetry,
and so in Sections~\ref{sec:kt} and~\ref{sec:su2} we will
consider a wide class of generalizations of $Z_{\rm U(1)}$ in the
hope of identifying special regimes of parameter space where the
full SU(2) symmetry may be restored. Our search for such spaces
for enhanced symmetry will be successful: one of our main findings
will be that such F states do not survive the restoration of SU(2)
symmetry. Along with other reasons to be discussed in
Sections~\ref{sec:bo} and~\ref{sec:hnbo}, this leads to our
conclusion that the SU(2) symmetric $Z$ in (\ref{z}) does not
contain any fractionalized states. Similar reasoning will also
allow us to conclude that the HN and F states cannot be separated
by a single, second-order quantum critical point, in contrast to
that found in $Z_{\rm U(1)}$ between the XY and F states.

To further clarify the issues of fractionalization in systems with
SU(2) symmetry, we will introduce a distinct lattice model, $Z_P$,
in Section~\ref{planar}. This model generalizes $Z$ to explicitly
allow for a non-collinear, magnetically ordered state (a P state).
We will argue that it also contains an F state, and discuss the
structure of the phase diagram of $Z_P$. This will allow us to
delineate the routes by which it is possible to connect states
with HN, BC, and/or F order in systems with full SU(2) spin
symmetry. We expect the conclusions in this section to be rather
general, and apply to a wide class of SU(2) symmetric models with
phases characterized as in Table~\ref{t1}.

\subsection{Outline}

This is a lengthy paper, and so it is helpful for the reader to
have an overview of the logical relationships between the sections
at the outset. We will begin in Section~\ref{sec:bo} by a direct
assault on the SU(2) invariant partition function $Z$ in
(\ref{z}). We will argue that BC order appears at large $g$ and
then proceed in Section~\ref{sec:hnbo} to a discussion of how the
system interpolates between the small $g$ HN phase and the large
$g$ BC phase. We will present a number of arguments based upon the
statistical mechanics of defects in the $\varphi^4$ field theory
in 3 spacetime dimensions that support the proposal of an
intermediate phase with co-existing HN and BC order. The next two
sections take the reader on a long detour. In Section~\ref{sec:1d}
we turn our attention to one dimensional antiferromagnets and
allow for the system to acquire a uniaxial anisotropy and a
reduced U(1) symmetry. For such systems, we are able to develop
rather precise duality methods which completely account for the
defects and their Berry phases; we are also able to restore SU(2)
symmetry along special lines in parameter space. The results
obtained in this manner are found to be in excellent accord with
those obtained earlier by bosonization methods. Emboldened by this
success, we proceed with the application of the same methods in
two dimensions in Section~\ref{sec:u1}. The main results again
support the co-existence of HN and BC order in the SU(2) invariant
systems, in agreement with Section~\ref{sec:hnbo}. Finally, in
Section~\ref{planar} we return to the main road, and focus direct
attention on SU(2) invariant antiferromagnets in two dimensions.
We introduce a generalization of $Z$, denoted $Z_P$, which
explicitly accounts for non-collinear spin correlations expected
in frustrated antiferromagnets. This model does contain
paramagnetic fractionalized phases, and we present phase diagrams
which describe competition between HN, BC, and F orders.

\section{Duality and Bond Order}
\label{sec:bo}

The model $Z$ in (\ref{z}) has a clear association with the HN
state: in the limit of small $g$, the coupling in $S_{\bf n}$
prefers that all the ${\bf n}_j$ align along a common direction
in spin space, and this leads to a ground state with HN order, as
in (\ref{n0}). This section will introduce a duality
transformation of $Z$ which demonstrates its intimate connection
with BC states. We will argue that BC order appears for large $g$
in the SU(2)-symmetric $Z$. We will also indicate why there is no
such requirement for $Z_{\rm U(1)}$.

This first step in the duality transformation, as usual, is to
rewrite (\ref{sa}) by the Poisson summation formula:
\begin{eqnarray}
&& \sum_{\{q_{\bar{\jmath}\mu}\}} e^{-S_{{\cal A}}} = \nonumber \\
&&~~~ \sum_{\{a_{\bar{\jmath}\mu}\}} \exp \left( - \frac{e^2}{2}
\sum_{\bar{\jmath}} a_{\bar{\jmath}\mu}^2 - i \sum_{\Box}
\epsilon_{\mu\nu\lambda} a_{\bar{\jmath}\mu} \Delta_{\nu} {\cal
A}_{j \lambda}\right), \label{d1}
\end{eqnarray}
where $a_{\bar{\jmath}\mu}$ (like $q_{\bar{\jmath}\mu}$) is an
integer-valued vector field on the links of the dual lattice.
Here, and henceforth, we drop overall normalization constants in
front of partition functions. The last term in (\ref{d1}) has the
structure of a `Chern-Simons' coupling, and connects
$a_{\bar{\jmath}\mu}$ to the ${\cal A}$ flux on the plaquette of
the direct lattice that it pierces.

\begin{figure}
\epsfxsize=2.5in \centerline{\epsffile{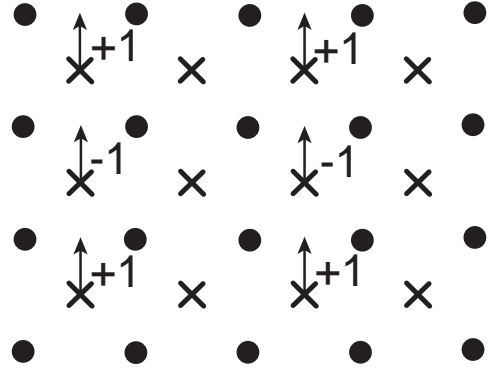}} \vspace{0.2in}
\caption{Specification of the non-zero values of the fixed field
$a_{\bar{\jmath}\mu}^0$. The circles are the sites of the direct
lattice, $j$, while the crosses are the sites of the dual
lattice, $\bar{\jmath}$; the latter are also offset by half a
lattice spacing in the direction out of the paper (the $\mu =
\tau$ direction). The $a_{\bar{\jmath}\mu}^0$ are all zero for
$\mu=\tau,x$, while the only non-zero values of
$a_{\bar{\jmath}y}^0$ are shown above. Notice that the $a^0$ flux
obeys (\protect\ref{d2}).}\label{figa0}
\end{figure}

Next, we write $S_B$ in a form more amenable to duality
transformations. Choose a `background' $a_{\bar{\jmath}
\mu}=a_{\bar{\jmath}}^0$ flux which satisfies
\begin{equation}
\epsilon_{\mu\nu\lambda} \Delta_{\nu} a_{\bar{\jmath}\lambda}^0 =
\eta_j \delta_{\mu \tau}, \label{d2}
\end{equation}
where $j$ is the direct lattice site in the center of the
plaquette defined by the curl on the left-hand-side. Any
integer-valued solution of (\ref{d2}) is an acceptable choice for
$a_{\bar{\jmath}\mu}^0$, and a convenient choice is shown in
Fig~\ref{figa0}.
With (\ref{d2}), the Berry phase term also takes the form of
Chern-Simons coupling:
\begin{equation}
S_B =  i \sum_{\Box} \epsilon_{\mu\nu\lambda}
a_{\bar{\jmath}\mu}^0 \Delta_{\nu} {\cal A}_{j \lambda} \label{d3}
\end{equation}
We can now combine (\ref{d1}) and (\ref{d3}) and, after a shift
of $a_{\bar{\jmath}\mu}$ by $a_{\bar{\jmath}\mu}^0$, obtain a new
exact representation of $Z$:
\begin{eqnarray}
&& Z = \sum_{\{ a_{\bar{\jmath} \mu} \}}  \int \prod_j d {\bf n}_j
\delta({\bf n}_j^2 - 1) \exp \Bigl( - S_{\bf n} - S_{a} \nonumber
\\&&~~~~~~~~~~~~~~~~~~~~~
- i \sum_{\Box} \epsilon_{\mu\nu\lambda} a_{\bar{\jmath}\mu}
\Delta_{\nu} {\cal A}_{j \lambda} \Bigr), \label{d4}
\end{eqnarray}
where
\begin{equation}
S_a = \frac{e^2}{2} \sum_{\bar{\jmath},\mu}
(a_{\bar{\jmath}\mu}-a_{\bar{\jmath}\mu}^0)^2 . \label{d5}
\end{equation}
The structure of this representation of $Z$ also allows us to give
a simple physical interpretation of $a_{\bar{\jmath}\mu}$. The
quantity $(a_{\bar{\jmath}\mu}-a_{\bar{\jmath}\mu}^0)^2$ for
$\mu=x,y$ is a term in action associated with a spatial link on
the dual lattice, and is therefore surely associated with a
measure of the energy of the bond on the direct lattice which
intersects it. We may therefore identify the bond variable in
(\ref{defQ}) as
\begin{equation}
Q_{j,j+\hat{x}} \sim (a_{\bar{\jmath}y}-a_{\bar{\jmath}y}^0)^2 ,
\label{d6}
\end{equation}
where the direct lattice link on the left-hand-side intersects the
dual lattice link on the right-hand-side. A similar relation holds
for $x \leftrightarrow y$.

We have now assembled the ingredients necessary to address the
physics of the large $g$ regime. Here the ${\bf n}$ field will
fluctuate strongly, and so it appears natural to integrate it
out. To do this, we need to evaluate the average
\begin{equation}
W(a_{\mu}) \equiv \left\langle \exp \left(- i \sum_{\Box}
\epsilon_{\mu\nu\lambda} a_{\bar{\jmath}\mu} \Delta_{\nu} {\cal
A}_{j \lambda} \right) \right\rangle_{S_{\bf n}}; \label{d7}
\end{equation}
In terms of $W(a_{\mu})$, the partition function is
\begin{equation}
Z = \sum_{\{ a_{\bar{\jmath} \mu} \}} W ( a_{\mu} ) \exp \left( -
S_a \right). \label{d7a}
\end{equation}
 Strong
fluctuations in ${\bf n}$ will lead to concomitant fluctuations
in the flux $\epsilon_{\mu\nu\lambda} \Delta_{\nu} {\cal
A}_{j\lambda}$ over all real values between 0 and $2 \pi$. This
means that the average in (\ref{d7}) will be strongly dominated by
configurations in $a_{\bar{\jmath} \mu}$ which obey
\begin{equation}
\epsilon_{\mu\nu\lambda} \Delta_{\nu} a_{\bar{\jmath} \lambda} =
0. \label{d8}
\end{equation}
In other words, the large $g$ region is in a ``$a$-Meissner''
phase which expels the flux of $a_{\bar{\jmath} \mu}$. We will
restrict attention to $a_{\bar{\jmath} \mu}$ configurations which
satisfy (\ref{d8}) in most of the remainder of this section. A
more careful discussion of the origin of (\ref{d8}) is given in
Appendix~\ref{b}; there we show that the configurations which
violate (\ref{d8}) lead mainly to a renormalization of the
coupling constant $e$ in $S_a$. The discussion in
Appendix~\ref{b} also shows that the analysis of this section
also applies to models which a bare value $e = \infty$; the
renormalized $e$ appearing in $S_a$ is finite even in that case.

This is also a convenient point to make a brief aside on systems
with an easy-plane anisotropy, and a reduced U(1) symmetry. Here
${\bf n}$ lies in the $x$-$y$ plane as indicated in (\ref{o3}),
and the flux $\epsilon_{\mu\nu\lambda} \Delta_{\nu} {\cal
A}_{j\lambda}$ on any plaquette is typically zero. Only if
plaquette happens to contain a {\em vortex} in the angle $\theta$
is the flux non-zero, and then it takes the {\em discrete} values
$\pm \pi$ (half the area of a hemisphere). More generally,
allowing for multiple vortices, the allowed values of the flux
$\epsilon_{\mu\nu\lambda} \Delta_{\nu} {\cal A}_{j\lambda}$ are $m
\pi$ with $m$ integer. In the large $g$ region, we can assume that
there were be strong fluctuations in the value of $m$ in the
plaquettes, and so the average in (\ref{d7}) will be dominated by
configurations in $a_{\bar{\jmath} \mu}$ which satisfy
\begin{equation}
\epsilon_{\mu\nu\lambda} \Delta_{\nu} a_{\bar{\jmath} \lambda} = 0
\mbox{ (mod 2)}, \label{d9}
\end{equation}
and the flux of $a_{\bar{\jmath} \mu}$ is only expelled modulo 2.
The difference between (\ref{d8}) and (\ref{d9}) is the key reason
from the strong distinction between the properties of SU(2) and
U(1) symmetric spin systems. We will show in Section~\ref{sec:u11}
that the partition function (\ref{d7a}), when evaluated under the
constraint (\ref{d9}), reduces exactly to an Ising model on the
cubic lattice in $D=3$ dimensions, with the signs of the couplings
chosen so that every spatial plaquette is frustrated; this Ising
model is dual to the Ising gauge theory plus Berry phases
obtained in the large $g$ limit of (\ref{u11a}). Precisely the
same Ising model was considered in
Refs.~\onlinecite{rodolfo2,japan} in a context somewhat different
from the present, but closely related to the models of frustrated
antiferromagnets to be considered in Section~\ref{planar}. We can
also already notice here that double vortices (with $m = \pm 2$)
don't couple at all to fluctuations in $a_{\bar{\jmath} \mu}$,
and so can selectively proliferate for large $g$: this is the
effect responsible for appearance of F states in $Z_{\rm U(1)}$.

Returning to SU(2)-invariant systems, and the $a$-Meissner phase
obeying (\ref{d8}), we can solve the constraint by writing
\begin{equation}
a_{\bar{\jmath} \mu} = \Delta_{\mu} h_{\bar{\jmath}} \label{d10}
\end{equation}
where $h_{\bar{\jmath}}$ is a ``height'' on the sites of the dual
lattice. The partition function $Z$ is now equivalent to the
three-dimensional height-model partition function
\begin{equation}
Z_h = \sum_{\{h_{\bar{\jmath}}\}} \exp \left ( - \frac{e^2}{2}
\sum_{\bar{\jmath}} \left( \Delta_{\mu} h_{\bar{\jmath}} -
a_{\bar{\jmath}\mu}^0 \right)^2 \right). \label{d11}
\end{equation}
The model $Z_h$ has been studied previously on a number of
occasions\cite{rsprb,zheng,fradkiv,srprl} and we now state the
main results. The most important property of $Z_h$ is that the
heights are in a {\em smooth} phase for all values of $e$; in
other words, any state of $Z_h$ has a definite value for its
average height $\langle h_{\bar{\jmath}} \rangle$, where the
average is over both quantum fluctuations and over sites of the
dual lattice---this is a generic property of height models in
three dimensions. Furthermore, the combination of the background
$a_{\bar{\jmath}\mu}^{0}$ terms in (\ref{d11}) and the existence
of an average height implies that the translational symmetry will
be broken by a modulation in the $Q_{ij}$ in (\ref{d6}). In other
words, the ground state of $Z_h$ has BC order.

A review of the prior analyses of $Z_h$ which led to the above
results is provided in Appendix~\ref{c}. There, we also establish
the close connection between $Z_h$ and a number of other models
for quantum paramagnets that have appeared in the literature. In
particular, there is a simple and direct connection between $Z_h$
and the ``quantum dimer'' model\cite{qd,sondhi1,sondhi2}.
Furthermore, $Z_h$ can also be written as a Coulomb gas of point
instanton charges, each of which carries a Berry phase: these
instantons are nothing but `hedgehog' defects in the ${\bf n}$
field, and the Berry phases implied by $Z_h$ are precisely equal
to the hedgehog Berry phases computed by Haldane\cite{berryh}.
Finally, $Z_h$ can also be written as a frustrated sine-Gordon
model. This last form is the most useful in determining the
ground states of $Z_h$. This simplest state which emerges is the
4-fold degenerate columnar BC state shown in Fig~\ref{bcorder}a.
However, within the generalized parameter space of the
sine-Gordon model we also find\cite{srprl} the 4-fold degenerate
plaquette ground state of Fig~\ref{bcorder}b, and an 8-fold
degenerate states involving co-existence of the orders in
Fig~\ref{bcorder}a and b. The symbol BC refers collectively to
any one of these states; an antiferromagnet could also have
quantum phase transitions between different BC states--these
transitions can also be discussed easily within the framework of
the frustrated sine-Gordon model\cite{srprl}, but we will not
dwell on them here.

\section{Connecting the N\'{e}el and Bond order states}
\label{sec:hnbo}

We have so far examined the phases of the SU(2) symmetric model
$Z$ (Eqn (\ref{z})) in the limits of small and large $g$. For
small $g$ we have the HN state with broken spin rotation symmetry,
while for large $g$ we have SU(2) symmetric ground states with BC
order. This section will comment on the evolution of the ground
state as $g$ moves between these two limits. Our analysis will
focus of the behavior of the function $W(a_{\mu})$ in (\ref{d7})
as a function of $g$: this will permit us to study the influence
of the HN order parameter ${\bf n}$ on the fluctuations of the
$a_{\bar{\jmath}\mu}$ which control the BC order. We will begin
in Section~\ref{sec:sw} with limit of small $g$, where ${\bf n}$
fluctuations can be computed in the spin-wave approximation.
Next, in Section~\ref{sec:crit}, we will consider the point where
${\bf n}$ fluctuations are critical {\em i.e.} at the point where
$SU(2)$ symmetry is broken, and describe the associated structure
of $W(a_{\mu})$. The critical point is a strongly coupled theory
in three spacetime dimensions, and so our arguments here do have
ad hoc steps which are based mainly on  physical arguments.

\subsection{Spin wave expansion}
\label{sec:sw}

First, we describe the small $g$ spin-wave expansion. Here, we
assume $\langle {\bf n} \rangle = N_0 (0,0,1)$. We perform the
familiar spin-wave expansion of the non-linear sigma model by
parameterizing ${\bf n} = (\pi_x, \pi_y , \sqrt{1 - \pi_x^2 -
\pi_y^2})$ and expanding the action, including the expressions
for ${\cal A}_{j \mu}$ in powers of $\pi_x , \pi_y$. Integrating
out the $\pi_x$, $\pi_y$ fluctuations to lowest order in $g$ we
obtain
\begin{eqnarray}
W(a_{\mu}) &=& \exp \left( - S_{sw} (a_{\mu})
\right)\nonumber \\
S_{sw} (a_{\mu}) &=& \frac{g}{8} \sum_{i,j,\mu,\nu}
(\epsilon_{\mu\rho\sigma} \Delta_{\rho} a_{\sigma})_i
(\epsilon_{\nu\lambda\gamma} \Delta_{\lambda} a_{\gamma})_j \nonumber \\
\times \bigl[ G(i-&j&) G(i-j+\hat\mu -\hat\nu) - G(i-j+\hat\mu)
G(i-j-\hat\nu)\bigr] \nonumber
\\
G(j) &=& \int_{-\pi}^{\pi} \frac{d^3 k}{8 \pi^3} \frac{ge^{i k
\cdot j}/2}{3 - \cos k_x - \cos k_y - \cos k_z} \label{m1}
\end{eqnarray}
We now need to study the $a_{\bar{\jmath}\mu}$ fluctuations, as
controlled by the action $S_a + S_{sw}$, to determine the fate of
the bond order in this framework. A closely related action was
numerically simulated earlier in Ref.~\onlinecite{rodolfo}, with
an action in which the double summation in $S_{sw}$ over $i$ and
$j$ was replaced with a single summation over the on-site term
with $i=j$, and $\mu=\nu$. Numerical evaluation of the expression
for $G$ in (\ref{m1}) shows that this on-site term is over 10
times larger than the off-site terms, and so the truncation
involved in mapping to the model of Ref.~\onlinecite{rodolfo} is
reasonable. The primary result of Ref.~\onlinecite{rodolfo},
which we expect to apply to the full action in (\ref{m1}), was
that there is a critical value of $g$ above which there is an
onset of BC order. However, the present small $g$ expansion does
not allow us to determine if this onset occurs in a regime where
the small $g$ spin-wave expansion about the HN state is valid. If
the transition in $S_a+S_{sw}$ occurs in its regime of
applicability, then the ground state has HN+BC order beyond the
critical point. So the structure of the spin-wave expansion in
powers of $g$ is compatible with both the HN and HN+BC states,
but does not allow us to definitively conclude that the HN+BC
state must exist.

\subsection{Critical $\varphi^4$ field theory}
\label{sec:crit}

To study the competition between the HN and BC states further, we
imagine turning up the value of $g$ to the critical point where
SU(2) symmetry is first restored, and $N_0$ vanishes. This
section will make the assumption that the fluctuations of ${\bf
n}$ at this critical point are described by the field theory
associated with the critical point in $S_{\bf n}$ itself {\em
i.e.} the 3-component $\varphi^4$ field theory in 2+1 dimensions
with O(3) symmetry. We will argue, by an evaluation of $W(a_{
\mu})$ at the critical point of the $\varphi^4$ field theory, that
this assumption is justified only if HN order vanishes at a
transition between the HN+BC and BC states. In other words, we
will find that a critical point described by the 3-component
$\varphi^4$ field theory has finite, non-critical BC order. The
alert reader will notice that these arguments do not logically
exclude the possibility that there is some other unknown critical
field theory describing the vanishing of HN order, and which
describes a second-order phase transition between the HN and BC
states; however, we will not find any likely candidates for such a
critical theory here.

To evaluate $W(a_{ \mu})$ we need some understanding of the
fluctuations of the ${\cal A}$ flux. As we saw in the large $g$
theory in Section~\ref{sec:bo}, these are controlled by the
fluctuations of the hedgehog point-defects in the ${\bf n}$
fluctuations. So as a first step, we should determine the
correlations of the hedgehogs at the critical point of the
$\varphi^4$ field theory. We will be able to make a number of
statements about these correlations in an expansion in $\epsilon
= 4- D$ (where $D$ is the dimensionality of spacetime) by an
extension of methods developed by Halperin\cite{bert} some time
ago, which are described in Appendix~\ref{d}. We begin by
outlining the physical ideas behind our analysis, and will then
apply them to the computation of $W(a_{ \mu})$.

It is useful to first contrast the critical behavior of the
hedgehogs with a well-known and familiar example: vortices in the
$D=2$ XY model which exhibit the Kosterlitz-Thouless (KT)
transition. Imagine we are examining a model with a short-distance
momentum cutoff, $\Lambda$. Let the mean density of vortices be
$\bar\rho_v$, and a dimensionless measure of this is $\bar\rho_v
\Lambda^{-2}$. Now we perform an RG transformation, integrating
out tightly-bound dipoles of vortices, and gradually reduce the
value of $\Lambda$. It is known that at the KT critical point the
vortex fugacity ultimately flows to zero; in other words, as we
scale $\Lambda$ to smaller values, the dimensionless vortex
density, $\bar\rho_v \Lambda^{-2}$, ultimately scales to zero.
This is the precise form of the statement that there are no free
vortices at the KT critical point. The behavior of the density of
hedgehogs, $\bar\rho_h$, at the critical point of the $D=3$ O(3)
model is known\cite{bert,yates} to be dramatically different:
$\bar\rho_h \Lambda^{-3}$ remains a finite number of order unity
even after RG scaling to the longest scales. So, loosely speaking,
there is finite density of ``free'' hedgehogs at the critical
point. We will compute correlations of this critical ensemble of
defects of below. Our main physical observation will be that
these critical decay correlations decay sufficiently rapidly (see
(\ref{m4}) and (\ref{m6}) below) that there is little material
difference from the corresponding correlations in the paramagnetic
phase. The consequences of the latter correlations were explored
earlier in the large $g$ theory: we saw in Section~\ref{sec:bo}
that they led to BC order in the paramagnetic phase. So it is
reasonable to conclude that finite BC order is also present at
the critical point of the HN order.

Let us now describe the correlations of the hedgehogs at the
critical point of the O(3) $\varphi^4$ field theory in 2+1
dimensions. Let $\rho_h (r)$ be the topological charge density
operator of hedgehogs at the spacetime point $r$ (so $\bar\rho_h
= \langle |\rho_h (r)| \rangle$). We consider first the two-point
correlator
\begin{equation}
C_h (r) = \langle \rho_h (r) \rho_h (0) \rangle \label{m2}
\end{equation}
The topological character of the $\rho_h (r)$, and its
association with a continuum field $\varphi (r)$ implies that
this correlator obeys an overall charge neutrality condition
\begin{equation}
\int d^3 r C_h (r) = 0. \label{m3}
\end{equation}
We explicitly evaluate $C_h (r)$ in an expansion in $\epsilon=4-D$
in Appendix~\ref{d}. The topological character is special to
$D=3$, and so the constraints of the conservation law (\ref{m3})
are not reflected in this computation. We find that the
long-distance decay of $C_h (r)$ is controlled by the scaling
dimension
\begin{equation}
\mbox{dim}[\rho_h(r)] \equiv (9 + \eta_h)/2 \label{m3a}
\end{equation}
where $\eta_h$ has the $\epsilon$ expansion
\begin{eqnarray}
\eta_h &=& 3 \eta + {\cal O}(\epsilon^3) \nonumber\\
&=& \frac{15}{242} \epsilon^2 + {\cal O}(\epsilon^3), \label{m5}
\end{eqnarray}
where $\eta$ is the anomalous dimension of the field $\varphi$.
This scaling dimension implies that $C_h$ has the long distance
behavior
\begin{equation}
C_h ( r \rightarrow \infty ) \sim \frac{1}{|r|^{9 + \eta_h}},
\label{m4}
\end{equation}
One might question the applicability of such an expansion to
$D=3$, given the neglect of the conservation law (\ref{m3}) to all
orders in the $\epsilon$ expansion. Indeed, if (\ref{m3}) was
obeyed by the scaling limit of the field theory, then we would
expect $\mbox{dim}[\rho_h (r)] = 3$, or $\eta_h = -3$. We think
this is extremely unlikely, and the more plausible scenario is
that the conservation (\ref{m3}) is not obeyed by the scaling
limit field theory (this is analogous to the the appearance of an
``anomaly'' in a field theory): to properly define the field
theory, some consistent short-distance regularization has to be
introduced, and this regularized theory does obey (\ref{m3})
after including short distance contributions which are beyond the
scaling limit. The reason for the presence of the ``anomaly''
becomes clear after implementing the consequences of (\ref{m3})
and (\ref{m4}) in momentum space: $C_h (k)$, the Fourier
transform of $C_h (r)$, has the following small $k$ expansion
\begin{equation}
C_h (k \rightarrow 0) \sim c_1 k^2 + \ldots + c_2 k^{6 + \eta_h} +
\ldots \label{m6}
\end{equation} where the first set
of ellipsis in (\ref{m6}) refer to various analytic terms in
integer powers $k^2$, and $c_{1,2}$ are constants. Even for
$\eta_h=-3$, note that the second singular term remains
subdominant to the analytic term proportional to $c_1$. The latter
is controlled by non-universal short-distance effects, and it
appears that these are the primary degrees of freedom controlling
the conservation of total topological charge. For the scaling
limit theory to be tightly constrained by the conservation law, we
would require $c_1=0$, something that does not seem likely in
general, given its dependence on non-universal features. So we
conclude that the critical theory can ignore the conservation law,
and the $\epsilon$ expansion for $\eta_h$ remains applicable in
$D=3$. The singular piece of $C_h$ decays very rapidly with $r$,
and the small momentum behavior of $C_h$ is dominated by a
non-universal, analytic term with coefficient $c_1 \neq 0$: these
are the primary conclusions on the properties of $C_h (r)$ that we
will need in our analysis below.

The $\epsilon$ expansion also allows computation of higher order
correlations of $\rho_h$. To leading order in $\epsilon$ and at
small $k$, we can neglect higher order cumulants in the
correlations of $\rho_h$. So we will assume that the fluctuations
of $\rho_h$ are controlled by the distribution controlled only by
the second cumulant, which is
\begin{equation}
\exp \left( - \frac{1}{2} \sum_k \frac{|\rho_h (k)|^2}{C_h (k)}
\right). \label{m7}
\end{equation}

It now remains to evaluate the distribution of the ${\cal A}$ flux
associated with the above distribution of $\rho_h$, and to thence
obtain $W (a_{ \mu})$. Let us define the `electromagnetic' field
\begin{equation}
{\cal E}_{\mu} = \epsilon_{\mu\nu\lambda} \partial_{\nu} {\cal
A}_{\lambda}. \label{m8}
\end{equation}
As we are focusing attention on long-wavelengths here, we have
replaced the lattice derivative by a continuum partial derivative.
In the presence of a fixed hedgehog density $\rho_h$, the ${\cal
E}$ field obeys the `Maxwell' equations of motion
\begin{equation}
\partial_{\mu} {\cal E}_{\mu} = 2 \pi
\rho_h~~~~~~\epsilon_{\mu\nu\lambda} \partial_{\nu} {\cal
E}_{\lambda} = 0 \label{m9}
\end{equation}
From these equations, and from (\ref{m7}), we can deduce that the
distribution of ${\cal E}$ is controlled by
\begin{equation}
\exp \left( - \frac{1}{2} \sum_k \frac{k^2 |{\cal E}_{\mu}
(k)|^2}{4 \pi^2 C_h (k)} \right) \label{m10}.
\end{equation}
At this point, it is tempting to compute $W ( a_{\mu} ) = \langle
e^{i a_{\mu} {\cal E}_{\mu}} \rangle$ from the distribution
(\ref{m10}) and to conclude that
\begin{equation}
W(a_{\mu}) \sim \exp\left( - \frac{1}{2} \sum_k \frac{4 \pi^2 C_h
(k)}{k^2} | a_{\mu} (k) |^2 \right) \label{m11}.
\end{equation}
However, this expression is not invariant under the integer-valued
gauge transformations $a_{\bar{\jmath}\mu} \rightarrow
a_{\bar{\jmath}\mu} - \Delta_{\mu} \ell_{\bar{\jmath}}$ which are
respected by the original definition in (\ref{d7}). To remedy
this, we will have to put (\ref{m10}) back on the lattice, and
compute the average over ${\cal E}_{\mu}$ while respecting the
periodicity ${\cal A}_{j \mu} \rightarrow {\cal A}_{j \mu} + 2
\pi$. This we do in the following paragraph. Readers not
interested in this technical step may skip ahead to the main
result (\ref{mmain}) which replaces (\ref{m11}).

First we decouple the distribution in (\ref{m10}) by a
Hubbard-Stratonovich field ${\cal S}_{\mu}$
\begin{equation}
\int {\cal D} {\cal S}_{\mu} \exp \left( - \frac{1}{2} \sum_k
\frac{4 \pi^2 C_h (k)}{k^2} | {\cal S}_{\mu} (k) |^2 + i \sum_k
{\cal E}_{\mu} (k) {\cal S}_{\mu} (-k) \right). \label{m12}
\end{equation}
Then we restore the proper flux-periodicity in the last term by
replacing (\ref{m12}) by
\begin{eqnarray}
\sum_{\{q_{\bar{\jmath}\mu}\}}  \int {\cal D} {\cal S}_{\mu} &&
\exp \left( - \frac{1}{2} \sum_k \frac{4 \pi^2 C_h (k)}{k^2} |
{\cal S}_{\mu} (k) |^2 \right.
\nonumber \\
&& \left.+ i \sum_{\bar{\jmath}} {\cal S}_{\bar{\jmath}\mu}
(\epsilon_{\mu\nu\lambda} \Delta_{\nu} {\cal A}_{j\lambda}-2 \pi
q_{\bar{\jmath}\mu}) \right). \label{m13}
\end{eqnarray}
We can now evaluate $W (a_{\mu})$ with the distribution of ${\cal
A}$ controlled by (\ref{m13}): the sum over $q_{\bar{\jmath}\mu}$
implies that ${\cal S}_{\bar{\jmath}\mu}$ must be an integer,
while the integral over ${\cal A}$ implies that ${\cal
S}_{\bar{\jmath}\mu} = a_{\bar{\jmath}\mu} - \Delta_{\mu}
h_{\bar{\jmath} \mu}$ where $h$ is integer-valued field on the
sites of the dual lattice. From this we obtain our main result,
\begin{equation}
W(a_{\mu}) \sim \sum_{ \{ h_{\bar{\jmath} \mu} \} } \exp\left( -
\sum_k \frac{2 \pi^2 C_h (k)}{k^2} | a_{\mu} (k) - \Delta_{\mu} h
(k) |^2 \right) \label{mmain}
\end{equation}
for $W(a_{\mu})$ at the critical point of the O(3) $\varphi^4$
field theory. The expression (\ref{mmain}) modifies (\ref{m11})
to allow invariance under integer-valued gauge transformations of
the $a_{\mu}$.

The status of the BC order at this critical point is now
controlled by the $a_{\mu}$ fluctuations associated with the
distribution $e^{-S_a} W(a_{\mu})$. It is surely acceptable to
replace $C_h$ in (\ref{mmain}) by its leading small $k$ expansion
in (\ref{m6}): $C_h (k) = c_1 k^2$, with $c_1 \neq 0$. Then, the
distribution (\ref{mmain}) for $a_{\mu}$ becomes precisely that in
(\ref{g4}) in Appendix~\ref{b} for the case of the large $g$
limit. Our conclusions here are therefore the same as those in the
discussion following (\ref{g4}): the systems expels $a$ flux, and
is in the $a$-Meissner phase. There is, therefore, {\em
well-developed} BC {\em order at the critical point of the} O(3)
$\varphi^4$ {\em model}. The BC order fluctuations are also not
critical and this justifies, a posteriori, that the fluctuations
of the HN order at the critical point at which $N_0$ vanishes are
described by the O(3) $\varphi^4$ model alone.

This section has therefore dilineated an explicit route for the
destruction of the HN order by second-order quantum transitions,
which was sketched in Fig~\ref{phaseZ}. With increasing $g$, and
{\em before} the HN order vanishes, there is an onset of BC order
at $g=g_{1c}$; for BC order of the two types shown in
Fig~\ref{bcorder}, this transition is described by the $Z_4$ clock
model, which is in turn in the universality class of the O(2)
$\varphi^4$ field theory. For $g> g_{1c}$, the system is in the
HN+BC state. At a larger value of $g=g_{2c}$ HN order disappears
by a continuous transition described by the O(3) $\varphi^4$
model; the BC order is finite on both sides of this transition,
as well as the critical point, and does not play an essential
role in its critical theory at $g_{2c}$.

\subsubsection{U(1) symmetry}
\label{subsec:u1}

We mention in this subsection the application of the analysis
above to systems with only a U(1) spin symmetry. We will consider
such models in much detail in Section~\ref{sec:u1} using rather
more precise methods. So we can test our present approach and see
if consistent conclusions are obtained. For $Z_{\rm U(1)}$, as we
already noted in Section~\ref{sec:bo}, the `flux' ${\cal E}_{\mu}
= \epsilon_{\mu\nu\lambda} \Delta_{\nu} {\cal A}_{\lambda}$
through any plaquette is simply $\pi$ times the vortex number
piercing that plaquette. In other words, we can write ${\cal
E}_{\mu} = \pi {\cal J}_{\mu}$, where ${\cal J}_{\mu}$ is the
vortex current of a 2+1 dimensional XY model. Assuming the latter
is described by the critical point of the O(2) $\varphi^4$ field
theory in 2+1 dimensions, we can compute the correlations of
${\cal J}_{\mu}$ by Halperin's analysis~\cite{bert}, as
generalized in Appendix~\ref{d} to the critical point. A
computation very similar to that leading to (\ref{m6}) now yields
\begin{eqnarray}
C_{v, \mu\nu} (k)  &\equiv& \left\langle {\cal J}_{\mu} (k)  {\cal
J}_{\nu} (-k) \right\rangle \nonumber \\
&=& (k^2 \delta_{\mu} - k_{\mu} k_{\nu})(\widetilde{c}_1 +
\widetilde{c}_2 k^{1 + \eta_v} + \ldots ), \label{m14}
\end{eqnarray}
where now we choose to define (see Appendix~\ref{d})
$\mbox{dim}[{\cal J}_{\mu} (r)] \equiv (6 + \eta_v)/2$.  The
transverse nature of the right-hand-side is a consequence of the
conservation of the vortex current, which is now expressed by the
local law
\begin{equation}
\partial_{\mu} C_{v, \mu \nu} = 0.
\label{m14a}
\end{equation}
If (\ref{m14a}) was not violated by an anomaly, then we would
expect that $\mbox{dim}[{\cal J}_{\mu} (r)]=2$ or $\eta_v=-2$.
Inserting the latter value in (\ref{m14}) we see that the
singular contribution from the scaling theory {\em dominates} the
non-singular piece proportional to $\widetilde{c}_1$. This is the
opposite of the situation in the SU(2) symmetric case considered
in the body of this section. In the present case, we do {\em not}
expect an anomaly in the conservation of vortex current as the
conservation law is dominated by the scaling theory contributions,
and so the result $\eta_v = -2$ should be exact in $D=3$.
Integrating out the ${\cal J}_{\mu}$ fluctuations using
(\ref{m14}) and this value of $\eta_v$, we conclude that
\begin{equation}
W(a_{\mu}) \sim \exp \left( - \frac{\widetilde{c}_2}{2} \sum_{k}
\frac{1}{k} a_{\mu} (-k) (k^2 \delta_{\mu} - k_{\mu} k_{\nu})
a_{\nu} (k) \right) \label{m15}
\end{equation}
for the case of $Z_{\rm U(1)}$. The strong contrast between
(\ref{m15}) and (\ref{mmain}) should be noted. The present result
(\ref{m15}) is midway between the $a$-Meissner result
(\ref{mmain}) which expels $a_{\mu}$ flux, and the ``$a$-Maxwell''
result (\ref{m1}) which allows penetration of $a_{\mu}$ flux. A
more careful analysis is required to understand the consequences
of (\ref{m15}) and this will be taken up in Section~\ref{sec:u1}.
We will find there that the BC order can either present, absent,
or critical at the point at which U(1) spin-rotation invariance is
restored; the situation depends upon the details of the model
being studied. For our purposes here, it is satisfying to note
that, unlike the SU(2) case, the present methods do not place the
U(1) critical point in the $a$-Meissner phase, which would have
required BC order.

\subsubsection{Remarks}
\label{sec:remarks}

We conclude this section by a few remarks on SU(2)-invariant
systems, motivated by the previous subsection on U(1) symmetry. It
is interesting to observe that if the SU(2) symmetric spin system
had $C_h(k) \sim k^3$ (the value suggested naively by (\ref{m3})),
then its $W(a_{\mu})$ would have had a power of $k$ in the
exponent which was the same as that in (\ref{m15}) for systems
with U(1) symmetry; in such a situation one could envisage a
direct second order phase transition between the HN and BC phases,
as had been conjectured in earlier work by one of
us~\cite{ganpathy}. We have argued above that this behavior of
$W(a_{\mu})$ is not present at the critical point described by the
O(3) symmetric $\varphi^4$ field theory. This paper does not find
a clear-cut scenario under which this behavior could arise
generically, as we will discuss further at the end of
Section~\ref{sec:subo}.

\section{Antiferromagnets in one dimension}
\label{sec:1d}

There is now a rather complete understanding of $S=1/2$ quantum
antiferromagnets, like those in (\ref{H}), in one spatial
dimension. This is largely due to the powerful bosonization
technique, which not only allows classification of various
gapless critical states, but also describes the renormalization
group flow between gapped states with numerous possible broken
discrete symmetries. We will use these well-established results as
a convenient laboratory for testing the methods developed here for
two-dimensional antiferromagnets. As we show below, when adapted
to one dimension, our methods reproduce all the important
universal results obtained by the bosonization method. This
agreement significantly boosts our confidence that our lattice
models and duality methods are also successfully capturing the
physics of antiferromagnets in two dimensions.

Our discussion is divided into two subsections. In
Section~\ref{sec:haldane}, we recall the phase diagram of a
$S=1/2$ quantum antiferromagnet in one dimension obtained by the
abelian bosonization method. In Section~\ref{sec:chain} we show
that an essentially identical phase diagram is obtained by an
analysis of the partition function $Z$ in (\ref{z}) when applied
to one dimension.

\subsection{Results from bosonization}
\label{sec:haldane}

A convenient frustrated $S=1/2$ antiferromagnet in one dimension
is that studied by Haldane \cite{xxz1d} with first and second
neighbor exchange interactions:
\begin{eqnarray}
H^{(1)} &=&  \sum_{ j } \Biggl[ J_1 \left( \hat{S}_{ix}
\hat{S}_{j+1,x} + \hat{S}_{iy} \hat{S}_{j+1,y} + \zeta
\hat{S}_{iz} \hat{S}_{j+1,z} \right) \nonumber \\
&~&~~~~~+  J_2 \hat{\bf S}_j \cdot \hat{\bf S}_{j+2} \Biggr].
\label{o1}
\end{eqnarray}
We have also introduced a uniaxial anisotropy parameter $\zeta$
so that the spin symmetry is SU(2) only at $\zeta=1$, and is U(1)
otherwise. Both exchange constants are positive, $J_{1,2} > 0$,
so the second neighbor coupling is frustrating.

\begin{figure}
\epsfxsize=3in \centerline{\epsffile{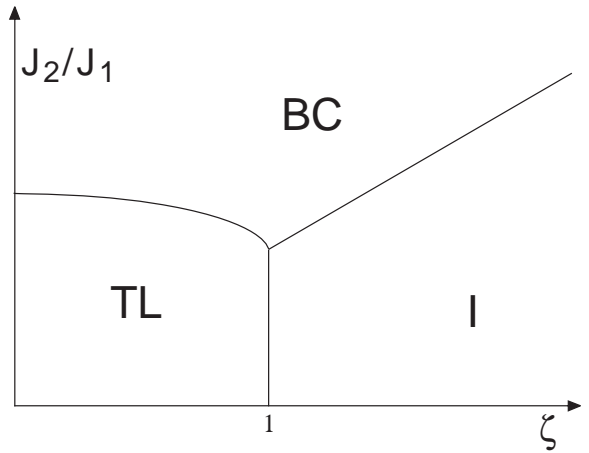}} \vspace{0.2in}
\caption{Phase diagram \protect\cite{xxz1d} of the $S=1/2$,
one-dimensional antiferromagnet $H^{(1)}$ in (\protect\ref{o1}).
The BC and I phases are as in Table~\protect\ref{t1}, and have a
gap to all excitations; these phases have gapped excitations with
$S_z=\pm 1/2$ but are not fractionalized in the two-dimensional
sense---the $S_z=\pm 1/2$ excitations confine immediately in pairs
when such one-dimensional chains are infinitesimally coupled
together to form a two-dimensional lattice. The Tomonaga-Luttinger
(TL) phase is special to one dimension: it is gapless and has
power-law spin correlations. There is SU(2) spin symmetry along
the line $\zeta=1$. }\label{figh}
\end{figure}

A review of the abelian bosonization analysis of (\ref{o1}) may be
found in Ref.~\onlinecite{bosobook}. The resulting phase diagram
is shown in Fig~\ref{figh}. We exclude the region of large $J_2
/J_1$ where additional phases with a time-reversal symmetry
breaking, spin chirality order parameter coexisting with BC or TL
order are found~\cite{leche}.
There are two phases with a gap to all excitations, the BC and I,
and these are as described in Table~\ref{t1}. The I phase occurs
for large $\zeta$, when the spins clearly prefer to be polarized
along the $\pm z$ directions. The BC phase appears for moderate
values of the frustration, $J_2 / J_1$, and the spatial pattern
of the bond order, $Q_{j,j+1}$, is sketched in Fig~\ref{figh}:
this phase contains the special Majumdar-Ghosh point\cite{majgh}
$\zeta=1$, $J_2/J_1=1/2$ where the fully dimerized wavefunction,
consisting simply of products of singlet pairs of nearest
neighbor sites, is the exact ground state.

The Tomonaga-Luttinger (TL) phase is gapless, and special to one
dimension. It is the analog of the XY phase in two dimensions,
but instead of exhibiting true long-range order, spin-correlations
in the $x$-$y$ plane have a slow power-law decay,
\begin{equation}
\langle \hat{S}_{ix} \hat{S}_{j+n,x} + \hat{S}_{iy}
\hat{S}_{j+n,y} \rangle \sim \frac{(-1)^n}{|n|^{\eta_{XY}}}
\label{o2}
\end{equation}
for large $|n|$, where the exponent $\eta_{XY} \leq 1$ varies
continuously within the TL phase ($\eta_{XY}=1/2$ at $\zeta=0$,
$J_2/J_1=0$). Similarly, the correlations of the Ising order
$(-1)^j \hat{S}_{j,z}$ decay with an exponent $\eta_I =
1/\eta_{XY}$, while those of the BC order decay with an exponent
also equal to $\eta_I$.

At boundary between the TL and BC phases, and also at the
boundary between the TL and I phases, we have
$\eta_{XY}=\eta_I=1$; actually, there is also a logarithmic
correction to (\ref{o2}) at these boundaries--we do not display
this here but note that our methods in Section~\ref{sec:chain}
also reproduce this correction. Within the TL phase we have
$\eta_{XY} <1$. The boundary between the BC and I phases is also
critical, and has $\eta_{XY} > 1$.

The manner in which the BC and I orders vanish at the boundaries
of their respective phases is also reviewed in
Ref~\onlinecite{bosobook}: both order parameters vanish with an
essential singularity at their boundaries to the TL phase, while
they vanish with the same continuously varying power-law on
opposite sides of the BC-I boundary.

\subsection{Dual lattice models}
\label{sec:chain}

We will now describe the properties of the lattice model $Z$ in
(\ref{z}) adapted to one dimension: this latter model will be
denoted $Z^{(1)}$. We will see that the phase diagram of a
generalized class of such models is identical in structure to
that of Fig~\ref{figh}.

First, let us explicitly write down the form of $Z^{(1)}$; we have
\begin{equation}
Z^{(1)} = \sum_{\{ q_{\bar{\jmath}} \}} \int \prod_j d {\bf n}_j
\delta({\bf n}_j^2 - 1) \exp \left( - S_{\bf n} - S_{{\cal
A}}^{(1)} - S_B \right). \label{z1}
\end{equation}
In present situation in two spacetime dimensions, the site indices
$j$ and $\bar{\jmath}$ extend over the square and dual-square
lattices respectively, and the Greek indices $\mu$, $\nu$,
$\lambda \ldots$ extend over $x$, $\tau$, and the parameter
$\eta_j$ in the Berry phase term is now $(-1)^{j_x}$.
Furthermore, the ${\cal A}$ flux can point in only one direction,
and so $S_{{\cal A}}$ in (\ref{sap}) is replaced by
\begin{equation}
S_{{\cal A}}^{(1)} = \frac{1}{2e^2} \sum_{\Box} (\epsilon_{\mu\nu}
\Delta_{\mu} {\cal A}_{j\nu}-2 \pi q_{\bar{\jmath}} )^2 ,
\label{sa1}
\end{equation}
so that the integer $q_{\bar\jmath}$ is simply a scalar and
carries no spacetime vector index. We wish to map the phase
diagram of $Z^{(1)}$ in a parameter space where the coupling in
$S_{{\bf n}}$ in (\ref{sn}) is allowed to acquire a uniaxial
anisotropy and reduce the spin symmetry to U(1).

We proceed with the duality transformations as in
Section~\ref{sec:bo}. Then the expression for the partition
function in (\ref{d4}) is replaced by
\begin{eqnarray}
Z^{(1)} &=& \sum_{\{ a_{\bar{\jmath}} \}}  \int \prod_j d {\bf
n}_j \delta({\bf n}_j^2 - 1) \exp \Biggl( - S_{\bf n} -
S_{a}^{(1)} \nonumber \\
&~&~~~~~~~~- i \sum_{\Box} \epsilon_{\mu\nu} a_{\bar{\jmath}}
\Delta_{\mu} {\cal A}_{j \nu} \Biggr), \label{d41}
\end{eqnarray}
where now the integers $a_{\bar{\jmath}}$ carry no spacetime
vector index, and the action $S_a$ is replaced by
\begin{equation}
S_a^{(1)} = \frac{e^2}{2} \sum_{\bar{\jmath}}
(a_{\bar{\jmath}}-a_{\bar{\jmath}}^0)^2 , \label{d51}
\end{equation}
with $a_{\bar{\jmath}}^0$ now satisfying $\epsilon_{\mu\nu}
a_{\bar{\jmath}}^0 = \eta_j \delta_{\mu\tau}$ instead of
(\ref{d2}); a convenient choice has $a_{\bar{\jmath}}^0$ taking
the values 1 and 0 on alternating spatial columns {\em i.e.\/}
$a_{\bar{\jmath}}^0 = (1 + (-1)^{\bar{\jmath}_x})/2$.

We now need to understand the consequences of the integration
over the ${\bf n}_j$ in (\ref{d41}). Because our purpose is to
compare with the properties of the model (\ref{o1}) which has a
uniaxial anisotropy and only a U(1) symmetry, we choose to
introduce a similar anisotropy in $S_{\bf n}$. Actually it is
convenient to take the easy-plane limit, and to assume that the
${\bf n}$ fields are completely localized in the $x$-$y$ plane in
spin space. In such a limit the integral over the ${\bf n}$
fields in (\ref{d41}) can be evaluated completely, and a precise
mapping to a dual lattice model is obtained. We will then examine
at the general structure of the resulting dual lattice model, and
discuss how its parameter space allows us to relax the easy-plane
restriction, and even find regimes in which the spins are
localized in the $\pm z$ directions, and also special parameter
values for which full SU(2) spin symmetry is restored. So, in the
easy-plane limit, we parameterize ${\bf n}_j$ in terms of an
angular field $\theta_j$ as noted in (\ref{o3}). To aid the
integration over the $\theta_j$, we perform the usual innocuous
change of writing the action $S_{\bf n}$ in the periodic Gaussian
(Villain) form
\begin{equation}
S_{{\bf n},{\rm U(1)}} = \frac{1}{2 g} \sum_{j, \hat{\mu}} \left(
\Delta_{\mu} \theta_j - 2 \pi m_{j \mu} \right)^2 ,\label{o4}
\end{equation}
where the $m_{j \mu}$ are the integers on the links of the direct
lattice which ensure the periodicity of the action in the angular
$\theta_j$. These integers also have the utility of allowing us to
measure the vorticity within each plaquette: this is simply
$\epsilon_{\mu\nu} \Delta_{\mu} m_{j \nu}$. Now recall the
discussion in Section~\ref{sec:hnbo}, in the paragraph containing
(\ref{m14}), where we noted that in the easy-plane limit the
${\cal A}$ flux through a plaquette is simply $\pi$ times the
vorticity. Using this result, and the modified from of the action
in (\ref{o4}), the partition function $Z^{(1)}$ transforms to the
following form in the easy-plane limit
\begin{eqnarray}
Z^{(1)}_{\rm U(1)} = && \sum_{\{ a_{\bar{\jmath}}, m_{j \mu} \}}
\int \prod_j d \theta_j \exp \Bigl( - S_{{\bf n},{\rm U(1)}}
\nonumber \\ &&~~~~~~~~~~~~~~~- S_{a}^{(1)} - i \pi \sum_{\Box}
\epsilon_{\mu\nu} a_{\bar{\jmath}} \Delta_{\mu} m_{j \nu} \Bigr).
\label{o5}
\end{eqnarray}
Before proceeding to a duality mapping of (\ref{o5}), we present
an alternative representation of (\ref{o5}) which exposes its
close relationship to the $Z_2$ gauge theories of
Refs.~\onlinecite{senthil,sedgewick}. We notice that the last term
in (\ref{o5}) is only sensitive to whether the integer $m_{j \mu}$
is even or odd. So we write
\begin{equation}
m_{j \mu} \equiv 2 f_{j \mu} + \frac{1 - s_{j,j+\hat{\mu}}}{2}
\label{o5a}
\end{equation}
where $f_{ j \mu}$ is an integer vector field on the direct
lattice, and $s_{j,j+\hat{\mu}} = \pm 1$ is an Ising variable on
the links of the direct lattice; notice that $s_{j,j+\hat{\mu}}$
is {\em unoriented} (unlike $m_{j \mu}$ and $f_{j \mu}$) and this
is reflected in the choice of notation. At this stage, it is now
easy to perform the summation over the $a_{\bar{\jmath}}$
independently on every dual lattice site, and to obtain the
alternative representation of $Z^{(1)}_{\rm U(1)}$:
\begin{eqnarray}
&& Z^{(1)}_{\rm U(1)} = \sum_{\{ s_{j,j+\hat{\mu}} = \pm 1\}}
\sum_{\{f_{j \mu} \}} \int \prod_j d \theta_j \exp \Biggl( K
\sum_{\Box} \prod_{\Box} s_{j,j+\hat{\mu}} \nonumber
\\ &&~~~~~~~~~~- \frac{2}{g} \sum_{j, \hat{\mu}} \left(
\frac{\Delta_{\mu} \theta_j}{2} - 2 \pi f_{j \mu} - \frac{\pi}{2}
( 1 - s_{j,j+\hat{\mu}}) \right)^2 \nonumber \\
&&~~~~~~~~~~- i \frac{\pi}{2} \sum_j ( 1 - s_{j,j+\hat{\tau}})
\Biggr). \label{o5b}
\end{eqnarray}
The first term in the action of this form of $Z^{(1)}_{\rm U(1)}$
is the standard Maxwell term of a $Z_2$ gauge theory, with
$s_{j,j+\hat{\mu}}$ the $Z_2$ gauge field. The coupling $K$ is
related to $e^2$ by
\begin{equation}
e^{2 K} \equiv \frac{\displaystyle \sum_{n=-\infty}^{\infty} e^{-
e^2 n^2/2}}{\displaystyle \sum_{n=-\infty}^{\infty} (-1)^n e^{-
e^2 n^2 /2}}; \label{o5c}
\end{equation}
$K$ is a monotonically decreasing function of $e^2$, with $K =
\pi^2/(4 e^2) - (\ln 2)/2$ as $e^2 \rightarrow 0$ and $K = 2
e^{-e^2 /2}$ as $e^2 \rightarrow \infty$. The last term in
(\ref{o5b}) is the Berry phase of the $Z_2$ gauge theory: this is
identical in form to that in Ref.~\onlinecite{senthil}, and was
obtained after using the identity $\epsilon_{\mu \nu} \Delta_{\nu}
a_{\bar{\jmath}}^0 = \eta_j \delta_{\mu \tau}$. The connection to
the models of Ref.~\onlinecite{senthil} becomes clearer if we
write (\ref{o5b}) using the simple cosine interaction of the angle
$\theta_j$, rather than the periodic Gaussian, Villain form:
\begin{eqnarray}
&& Z^{(1)}_{\rm U(1)} \approx \sum_{\{ s_{j,j+\hat{\mu}} = \pm
1\}} \int \prod_j d \theta_j \exp \Biggl( K \sum_{\Box}
\prod_{\Box} s_{j,j+\hat{\mu}} \nonumber
\\ &&~+ \frac{4}{g} \sum_{j, \hat{\mu}} s_{j,j+\hat{\mu}} \cos \left(
\frac{\Delta_{\mu} \theta_j}{2} \right) - i \frac{\pi}{2} \sum_j
( 1 - s_{j,j+\hat{\tau}})\Biggr) . \label{o5d}
\end{eqnarray}
This is the action of an XY model coupled to a $Z_2$ gauge field,
with an additional Berry phase term associated with the $S=1/2$
nature of the underlying antiferromagnet. Note that the angular
variable in the XY coupling is $\theta /2$, which is {\em half}
the angle determining the orientation of ${\bf n}$ in (\ref{o3}):
this half-angle variable is of course the reason for the
appearance of the $Z_2$ gauge degrees of freedom. Thus we have
established the rather surprising result that simply by
introducing an easy-plane anisotropy in the model $Z$ in
(\ref{z}) of Sachdev and Jalabert \cite{rodolfo} we obtain the
XY-$Z_2$ gauge model of Senthil and Fisher \cite{senthil}; this
result holds also in two dimensions, as we will see in
Section~\ref{sec:u1}. The model (\ref{o5d}) generalized to two
dimensions, but without the Berry phase, was studied recently by
Sedgewick {\em et al.} \cite{sedgewick}.

We now return to the original expression for $Z^{(1)}_{\rm U(1)}$
in (\ref{o5}) and proceed with the duality mapping: this will
allow evaluation of the $\theta_j$ integrals and will also write
it in a form with only positive weights. First, we perform the
summation over the $m_{j \mu}$ by using the following generalized
Poisson summation identity on every link of the square lattice
\begin{eqnarray}
&&\sqrt{2 \pi C} \sum_{m=-\infty}^{\infty} \exp \left[ -
\frac{C}{2} (x - 2 \pi
m)^2 + i m y \right] = \nonumber \\
&&~\sum_{J=-\infty}^{\infty} \exp \left[ - \frac{1}{2 C}\left( J+
\frac{y}{2 \pi} \right)^2 + i x \left( J + \frac{y}{2 \pi}
\right) \right], \label{o6}
\end{eqnarray}
where $m$, $J$ are integers, and $C>0$, $x$, $y$ are arbitrary
real numbers; this replaces the sum over the $m_{j \mu}$ by a
summation over a different set of integer ``currents'', $J_{j
\mu}$, also residing on the links of the square lattice. The
advantage of this form is that the integrals over the $\theta_{j}$
can be performed independently, and yield only the constraint that
the $J_{j \mu}$ currents are divergenceless. In this manner,
$Z^{(1)}_{\rm U(1)}$ in (\ref{o5}) is shown to be exactly
equivalent to
\begin{eqnarray}
&& Z^{(1)}_{\rm U(1)} = \sum_{\{ a_{\bar{\jmath}}, J_{j \mu}
\}}^{\prime} \exp \Biggl[ - \frac{e^2}{2} \sum_{\bar{\jmath}}
(a_{\bar{\jmath}}-a_{\bar{\jmath}}^0)^2 -\nonumber \\
&&~\frac{g}{2} \sum_{j,j'} K_{jj'}^{\mu\nu}\left(J_{j \mu} +
\frac{\epsilon_{\mu\lambda} \Delta_{\lambda} a_{\bar{\jmath}}}{2}
\right) \left(J_{j' \nu} + \frac{ \epsilon_{\nu\rho}
\Delta_{\rho} a_{\bar{\jmath}'}}{2} \right)\Biggr], \label{o7}
\end{eqnarray}
where the prime on the summation indicates the constraint
\begin{equation}
\Delta_{\mu} J_{j \mu} = 0. \label{o8}
\end{equation}
The mapping from (\ref{o5}) to (\ref{o7}) yields only a on-site
coupling between the currents with
\begin{equation}
K_{jj'}^{\mu\nu} = \delta_{jj'}\delta_{\mu\nu} \label{o9}
\end{equation}
(this $K_{jj'}^{\mu\nu}$ is not to be confused with the Ising
gauge coupling $K$ in (\ref{o5b},\ref{o5d}); the latter does not
carry any indices). However, to reproduce the full phase diagram
of the antiferromagnet $H^{(1)}$ in (\ref{o1}) it is necessary to
allow for additional near-neighbor couplings in
$K_{jj'}^{\mu\nu}$. As will become clear from our discussion in
Section~\ref{sec:obs} below, these new terms allow us to include
the couplings between the $z$ components of the ${\bf n}_j$ which
were dropped when we specialized the SU(2) invariant $S_{\bf n}$
in (\ref{sn}) to the easy-plane limit $S_{{\bf n}, {\rm U(1)}}$
in (\ref{o4}). With the off-site terms in $K_{jj'}^{\mu\nu}$ we
will be able to restore the SU(2) symmetry along special
parameter lines of $Z^{(1)}_{\rm U(1)}$, and even obtain regimes
where the spin anisotropy is easy-axis, and the spin-ordering is
Ising-like.

The expression (\ref{o7}) is one of the final dual forms of
$Z^{(1)}_{\rm U(1)}$: many physical properties will be most
transparent in this formulation. However, for future technical
purposes, it is useful to perform a few more manipulations on it.
We can solve the constraint (\ref{o8}) by writing
\begin{equation}
J_{j \mu} = \epsilon_{\mu\nu} \Delta_{\nu} p_{\bar{\jmath}}
\label{o10}
\end{equation}
where $p_{\bar{\jmath}}$ is an integer on the sites of the dual
lattice. Then, after defining $\ell_{\bar{\jmath}} =
2p_{\bar{\jmath}} + a_{\bar{\jmath}}$, it is easy to see that the
summation over the $a_{\bar{\jmath}}$ can be performed
explicitly. In this manner we find that (\ref{o7}) reduces
exactly to
\begin{eqnarray}
&& Z^{(1)}_{\rm U(1)} = \sum_{\{ \ell_{\bar{\jmath}} \}} \exp
\Biggl[ K_d \sum_{\bar{\jmath}}\varepsilon_{\bar{\jmath}}
\sigma_{\bar{\jmath}} \nonumber \\
&&~~~~~~~~~- \frac{g}{8} \sum_{j,j'}
K_{jj'}^{\mu\nu}\left(\epsilon_{\mu\lambda} \Delta_{\lambda}
\ell_{\bar{\jmath}} \right) \left( \epsilon_{\nu\rho}
\Delta_{\rho} \ell_{\bar{\jmath}'} \right)\Biggr], \label{o11}
\end{eqnarray}
where $\sigma_{\bar{\jmath}}$ is a fluctuating Ising variable
which denotes whether $\ell_{\bar{\jmath}}$ is even or odd,
\begin{equation}
\sigma_{\bar{\jmath}} \equiv 1 - 2 ( \ell_{\bar{\jmath}} \mbox{
mod 2}), \label{o12}
\end{equation}
while $\varepsilon_{\bar{\jmath}} = (-1)^{\bar{\jmath}_x}$ is a
fixed oscillating field which arises from the background
$a_{\bar{\jmath}}^0$ field, and is ultimately related to the spin
Berry phases. The coupling $K_d$ is the ``dual'' of the coupling
$K$
\begin{equation}
\tanh K_d \equiv e^{-2K} , \label{o13}
\end{equation}
where $K$ was defined as a function of $e^2$ in (\ref{o5c}); note
that $K_d$ is a monotonically increasing function of $e^2$, with
$K_d = 2 e^{-\pi^2/(2 e^2)}$ as $e^2 \rightarrow 0$, and $K_d =
e^2/4 - (\ln 2)/2$ for $e^2 \rightarrow \infty$.

\subsubsection{Observables and limiting cases}
\label{sec:obs}

We pause in our chain of duality mappings to discuss the physical
interpretation of the new degrees of freedom we have introduced.
The physical picture becomes clearest after an understanding of
the interpretation of the currents $J_{j \mu}$ in (\ref{o7}).
Analysis of the duality mapping from (\ref{o2}) to (\ref{o7})
along the lines of the dualities of quantum XY and boson models in
Refs.~\onlinecite{sorensen,otterlo} shows immediately that these
currents are precisely those of a boson representing the spin-flip
operator. In other words, if in the original spin model (\ref{o1})
we make the identification
\begin{equation}
\hat{S}_{j,x} + i \hat{S}_{j,y} = b^{\dagger}_j ,\label{o14}
\end{equation}
and perform the usual boson duality transformations (as discussed
by S{\o}rensen {\em et al.\/}\cite{sorensen}) to a model of
interacting integer-valued current loops, the $J_{j \mu}$ will be
the spatial and temporal ({\em i.e.} the density) components of
the current of the boson in (\ref{o14}). More precisely, we can
couple an external chemical potential to the boson charge and
thence deduce that
\begin{equation}
\hat{S}_{j,z} \sim \epsilon_{\tau \nu} \Delta_{\nu}
\ell_{\bar{\jmath}} = \Delta_x \ell_{\bar{\jmath}} \label{o15}
\end{equation}
So the spatial gradients of the $\ell_{\bar{\jmath}}$ are related
to the value of $\hat{S}_{j, z}$, and a phase with I order will
have a staggered, zigzag pattern of the $\ell_{\bar{\jmath}}$.

The above interpretation can be reinforced by examining the
limiting case $e^2 \rightarrow \infty$ (or equivalently, $K_d
\rightarrow \infty$). Then in (\ref{o7}), the $a_{\bar{\jmath}}$
fluctuations are quenched to $a_{\bar{\jmath}}=a_{\bar{\jmath}}^0$
and so $\epsilon_{\mu\nu}\Delta_{\nu} a_{\bar{\jmath}} =
(-1)^{\bar{\jmath}_x} \delta_{\mu\tau}$. With this value, the
partition function in (\ref{o7}) is seen to be precisely the boson
current loop model examined by Otterlo {\em et al.\/}
\cite{otterlo} (but in two spatial dimensions) at a mean boson
density of 1/2 per site. This last density is, of course,
precisely that needed to have a state with $\langle \sum_j
\hat{S}_{jz} \rangle = 0$ in the spin model. With the purely
on-site coupling in (\ref{o9}) this boson-loop model is expected
to always be superfluid: in one spatial dimension the superfluid
(or XY) correlations are only power-law, and so the spin system
is in the TL phase.

With the interpretation of the spin operators in
(\ref{o14},\ref{o15}), we have just seen that at $K_d = \infty$
each site has an average density of 1/2 spin-flip bosons. This is
just as expected from the usual microscopic interpretation of the
original spin antiferromagnet. However, what about the case when
$K_d < \infty$ ? In this picture, there now appear to be
fluctuations in the mean boson site density,
$\epsilon_{\tau\nu}\Delta_{\nu} a_{\bar{\jmath}}/2$. Where do the
missing bosons go ? The answer is that they reside on the links of
the lattice; indeed if the spins on sites $j$ and $j+\hat{x}$ form
a spin-singlet, this is expressed in terms of the bosons in
(\ref{o14}) as the single boson state
\begin{equation}
\frac{1}{\sqrt{2}} \left( b^{\dagger}_j - b^{\dagger}_{j+\hat{x}}
\right) |0 \rangle \label{o15a}
\end{equation}
where $|0 \rangle$ is the reference state with both spins down. So
a boson on a link corresponds to a spin-singlet bond, and each
site on the ends of the link will lose half a boson. So our
present model for spin systems goes beyond that of Otterlo {\em et
al.\/} \cite{otterlo} by allowing easily for coherent occupation
of bond orbitals. This sets the stage for states with BC order,
and the present interpretation is seen to be compatible with
identification in (\ref{d6}) of the ``bond-charge'' density with
the energy associated with the $a_{\bar{\jmath}}$ fluctuations on
the dual lattice site in the center of a bond.
 Using the variables
appearing in (\ref{o11}), it is evident that the bond modulation
is now measured by
\begin{equation}
Q_{j, j+\hat{x}} \sim \varepsilon_{\bar{\jmath}}
\sigma_{\bar{\jmath}}, \label{o16}
\end{equation}
where $\bar{\jmath}$ resides on the middle of the plaquette
between $j$ and $j+\hat{x}$. With this identification, we can see
that BC order is present in $Z^{(1)}_{\rm U(1)}$ in the limit $g
\rightarrow \infty$. From (\ref{o11}), we see that
$\ell_{\bar{\jmath}}$ is independent of $\bar{\jmath}$, and
therefore so is $\sigma_{\bar{\jmath}}$. From (\ref{o16}), a
constant $\sigma_{\bar{\jmath}}$ leads to an oscillating $Q_{ij}$
with BC order, with the values $\sigma_{\bar{\jmath}} = \pm 1$
identifying the two dimerized states.

\begin{figure}
\epsfxsize=2.7in \centerline{\epsffile{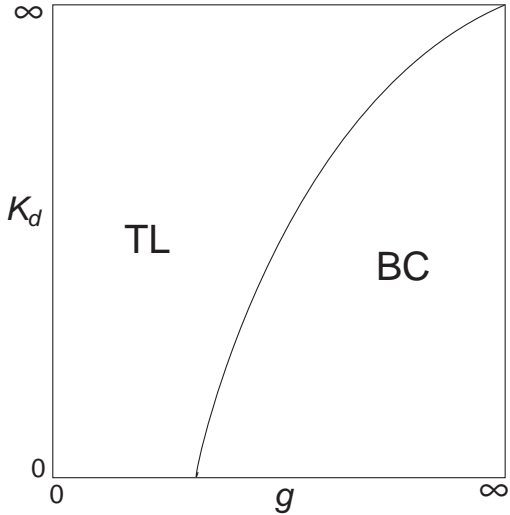}} \vspace{0.2in}
\caption{Phase diagram of the one spatial-dimensional model
$Z^{(1)}_{\rm U(1)}$ for the case of the on-site coupling
(\protect\ref{o9}). The phase boundary is in the
Kosterlitz-Thouless universality class, as discussed in
Section~\protect\ref{sec:kt}.}\label{figz1}
\end{figure}

We now have enough information to construct the phase diagram of
$Z^{(1)}_{\rm U(1)}$ for the case of the purely on-site coupling
(\ref{o9}). From (\ref{o11}) it is clear that there can be no
phase transition at $g=0$. Moreover, at $K_d=0$, (\ref{o11})
reduces to a height model in 1+1 dimensions, and this must have a
roughening transition at a critical value of $g$; this will be
discussed further in the following subsection. These results then
allow us to sketch the phase diagram in Fig~\ref{figz1}.
Note there is no I phase; this will appear below when we perform
generalized analysis which allows for further neighbor couplings
in the $K_{jj'}^{\mu\nu}$.

\subsubsection{Field theory and phase diagram}
\label{sec:kt}

We now perform a field theoretic analysis of $Z^{(1)}_{\rm U(1)}$
allowing for a general short-range coupling $K_{jj'}^{\mu\nu}$.
The main idea, as mentioned above, is to view the partition
function as a model of interfacial ``heights'',
$\ell_{\bar{\jmath}}$, in 1+1 dimensions. Then the
$K_{jj'}^{\mu\nu}$ control the roughness of the interface, while
the first term in (\ref{o11}) indicates that the heights are
preferentially even (odd) integers on every even (odd) spatial
column. This height model can be analyzed by entirely standard
technology, which is reviewed in Appendix~\ref{e}. The final
result is that $Z^{(1)}_{\rm U(1)}$ is related to the continuum
sine-Gordon model in 1+1 dimensions with action
\begin{equation}
S_{sG}^{(1)} = \int d^2 r \left( \frac{g}{8} (\partial_{\mu}
\phi_0)^2 - Y \cos (2 \pi \phi_0 ) \right), \label{o17}
\end{equation}
where $r$ is a spacetime co-ordinate. In terms of the continuum
field $\phi_0$, the I order is measured by $\sin(\pi \phi_0)$,
while the BC order is measured by $\cos(\pi \phi_0)$. The
coupling $Y$ is related to the parameters of $Z^{(1)}_{\rm U(1)}$
by (\ref{appe5}).

The action (\ref{o17}) is of the standard sine-Gordon type, and
its properties can be read off from existing results. For small
$g$, the coupling $Y$ scales to 0 and its only effect is a
renormalization of $g$ to $\widetilde{g}$. This yields the TL
phase, and it is stable provided $\widetilde{g} \leq 2 \pi$. It is
gapless and has power-law decay of correlations, controlled by the
fluctuations of the free field $\phi_0$. With the identification
of the BC and I order parameters below (\ref{o17}), we can now
compute that they both decay with the exponent $\eta_I = 2
\pi/\widetilde{g}$. To compute the correlations in the $x$-$y$
plane in spin space we note from (\ref{o4}) that fluctuations of
the angular variable $\theta$ are controlled by the continuum
action
\begin{equation}
S_{XY}^{(1)} = \frac{1}{2\widetilde{g}} \int d^2 r
(\partial_{\mu} \theta)^2 ;\label{o18}
\end{equation}
Now computing the correlator of $e^{i \theta}$ we deduce that
$x$-$y$ spin correlations decay as (\ref{o2}) with $\eta_{XY} =
\widetilde{g}/(2\pi) = 1/\eta_{I}$. Note that all of these
exponents are identical to those obtained by abelian
bosonization, as reviewed in Section~\ref{sec:haldane}.

\begin{figure}
\epsfxsize=3in \centerline{\epsffile{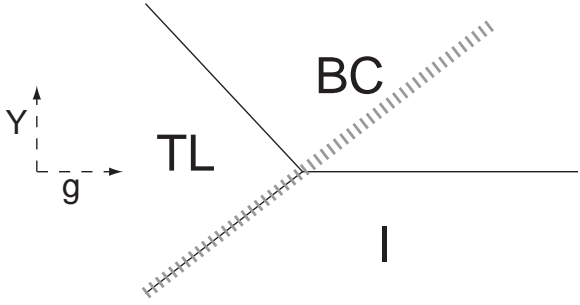}} \vspace{0.2in}
\caption{Phase diagram of the one spatial-dimensional model
$S^{(1)}_{sG}$ in (\protect\ref{o17}). The same phase diagram
also applies to $Z^{(1)}_{\rm U(1)}$ in (\protect\ref{o7}) with
general off-site couplings beyond the on-site coupling in
(\protect\ref{o9}). The topology of this diagram is identical to
that of Fig~\protect\ref{figh}, as are all the universal
long-distance properties of the TL phase, and of the phase
boundaries. There is SU(2) spin symmetry along the grey hatched
line. This line includes the boundary between the TL and I
phases, but does not represent a phase boundary within the BC
phase. }\label{figsg}
\end{figure}

The TL phase is unstable for $\widetilde{g} > 2\pi$, and the
ground state acquires a gap induced by the flow of $|Y|$ to large
values. For $Y>0$, the field $\phi_0$ is preferentially pinned at
the values $\phi_0=0,1$, and these are easily seen to correspond
to the BC phase; the case with on-site couplings as in (\ref{o9})
is therefore seen to correspond to $Y>0$, as (\ref{o17})
describes the phases in Fig~\ref{figz1}. Similarly, for $Y<0$, we
have $\phi_0 = 1/2, 3/2$, the values corresponding to the two I
states. These results imply the phase diagram shown in
Fig~\ref{figsg}.
There is complete consistency between all the universal
properties of Fig~\ref{figsg}, and those obtained by abelian
bosonization for the model of Fig~\ref{figh}. Establishing this
was the primary purpose of this section.

A notable feature of Figs~\ref{figh} and~\ref{figsg} is that
SU(2) spin symmetry is restored at the boundary between the TL
and I phases. Moving along this line we see that the ground state
initially has SU(2) symmetric spin correlations decaying with an
exponent $\eta_{XY} = \eta_I = 1$. However, there is eventually a
continuous phase transition to the phase with BC order. So we
have succeeded in understanding the phases of the SU(2) model via
a detour into models with only a U(1) symmetry.

We conclude this section by noting that closely related results
are obtained in an analysis based on the $CP^1$ formulation
presented in Appendix~\ref{a}. This extension is discussed in
Appendix~\ref{f}. This is in keeping with our contention that the
${\bf n}$ field formulation of Section~\ref{intro} and the $CP^1$
field formulation in Appendix~\ref{a} have similar phase diagrams.

\section{U(1) symmetry in two dimensions}
\label{sec:u1}

Bolstered by the success of our approach in one dimension, we
forge ahead to the application of the same methods in two
dimensions. We are interested in models of two-dimensional
antiferromagnets like (\ref{H}), but we now allow the spin
exchange terms to acquire a uniaxial anisotropy like that in the
first term in (\ref{o1}). We will carry out duality mappings on
such spin systems with U(1) symmetry, and then search for lines in
generalized parameter space where SU(2) symmetry can be restored.
This should considerable light on the properties of
SU(2)-symmetric quantum antiferromagnets.

Our initial analysis follows the same steps as in
Section~\ref{sec:chain}; the only change is that some of the
fields acquire additional vector indices, associated with the
change of spacetime dimension to 3. We introduce spin anisotropy
into the partition function $Z$ in (\ref{z}) by parameterizing the
${\bf n}$ field as in (\ref{o3}) and replacing $S_{\bf n}$ in
(\ref{sn}) by $S_{{\bf n},{\rm U(1)}}$ in (\ref{o4}). Then
proceeding with the duality mapping at the beginning of
Section~\ref{sec:bo}, we obtain the following U(1) symmetric
version of the partition function $Z$ in (\ref{d4}); this
generalizes (\ref{o5}) to two dimensions
\begin{eqnarray}
&& Z_{\rm U(1)} =\sum_{\{ a_{\bar{\jmath}\mu}, m_{j \mu} \}} \int
\prod_j d \theta_j \exp \Bigl( - S_{{\bf n},{\rm U(1)}} \nonumber
\\ &&~~~~~~~~- S_{a} - i \pi \sum_{\Box}
\epsilon_{\mu\nu\lambda} a_{\bar{\jmath}\mu} \Delta_{\nu} m_{j
\lambda} \Bigr). \label{u11}
\end{eqnarray}
As in Section~\ref{sec:chain}, we can obtain alternative
representations of (\ref{u11}) which expose its close connection
to a $Z_2$ gauge theory. Indeed, these representations for
$Z_{\rm U(1)}$ are identical to those for $Z^{(1)}_{\rm U(1)}$ in
(\ref{o5b}) and (\ref{o5d}) with the only change being that the
summation over $\mu$ extends over the three spacetime directions:
this alternative formulation of $Z_{\rm U(1)}$ was presented
earlier in (\ref{u11a}).

Next, we proceed with the two-dimensional generalization of the
duality mappings discussed below (\ref{o5d}); we generalize
(\ref{o7}) to
\begin{eqnarray}
&& Z_{\rm U(1)} = \sum_{\{ a_{\bar{\jmath}\mu}, J_{j \mu}
\}}^{\prime} \exp \Biggl[ - \frac{e^2}{2} \sum_{\bar{\jmath}}
(a_{\bar{\jmath}\mu}-a_{\bar{\jmath}\mu}^0)^2 -\nonumber \\
&&\frac{g}{2} \sum_{j,j'} K_{jj'}^{\mu\nu} \left(J_{j \mu} +
\frac{\epsilon_{\mu\lambda\kappa} \Delta_{\lambda}
a_{\bar{\jmath}\kappa}}{2} \right) \left(J_{j' \nu} + \frac{
\epsilon_{\nu\rho\gamma} \Delta_{\rho} a_{\bar{\jmath}'\gamma}}{2}
\right)\Biggr],\nonumber \\ \label{u12}
\end{eqnarray}
where, as before, the prime on the summation indicates the
constraint (\ref{o8}). The particular model (\ref{u11})
corresponds to the purely on-site values of the coupling
$K_{jj'}^{\mu\nu}$ as in (\ref{o9}). However, as discussed in
Section~\ref{sec:chain}, we find it useful to consider models
with near neighbor $K_{jj'}^{\mu\nu}$ to allow for phases with
easy-axis anisotropy and parameter values with enhanced SU(2)
symmetry. Another useful form of (\ref{u12}) is obtained by
solving the constraint (\ref{o8}) by
\begin{equation}
J_{j \mu} = \epsilon_{\mu\nu\lambda} \Delta_{\nu}
p_{\bar{\jmath}\lambda} \label{u13}
\end{equation}
with $p_{\bar{\jmath}\lambda}$ integer; this generalizes
(\ref{o10}). With this parameterization, the two-dimensional form
of (\ref{o11}) is
\begin{eqnarray}
Z_{\rm U(1)} &=& \sum_{\{ \ell_{\bar{\jmath}\mu} \}} \exp \Biggl[
K_d
\sum_{\bar{\jmath}}\varepsilon_{\bar{\jmath},\bar{\jmath}+\hat{\mu}}
\sigma_{\bar{\jmath},\bar{\jmath}+\hat{\mu}}  \nonumber \\
&-& \frac{g}{8} \sum_{j,j'}
K_{jj'}^{\mu\nu}\left(\epsilon_{\mu\lambda\kappa}
\Delta_{\lambda} \ell_{\bar{\jmath}\kappa} \right) \left(
\epsilon_{\nu\rho\gamma} \Delta_{\rho} \ell_{\bar{\jmath}'\gamma}
\right)\Biggr], \label{u14}
\end{eqnarray}
where $\ell_{\bar{\jmath}\mu}= 2 p_{\bar{\jmath}\mu}+
a_{\bar{\jmath}\mu}$ is an integer valued vector field on the
links of the dual cubic lattice, and
$\sigma_{\bar{\jmath},\bar{\jmath}+\hat{\mu}}$ is an Ising degree
of freedom which is pinned to the $\ell_{\bar{\jmath}\mu}$ by
\begin{equation}
\sigma_{\bar{\jmath},\bar{\jmath}+\hat{\mu}} \equiv 1 - 2 (
\ell_{\bar{\jmath}\mu} \mbox{ mod 2}), \label{u15}
\end{equation}
generalizing (\ref{o12}). Notice that, unlike
$\ell_{\bar{\jmath}\mu}$,
$\sigma_{\bar{\jmath},\bar{\jmath}+\hat{\mu}}$ is {\em
unoriented}, and has only a single value $\pm 1$ on each link;
this is the reason for the different choice of notation for these
two fields. The fixed field
$\varepsilon_{\bar{\jmath},\bar{\jmath}+\hat{\mu}}$ is likewise
unoriented and takes values $\pm 1$. It arises, as in
(\ref{o11}), from the Berry phase terms in $Z_{\rm U(1)}$, and
its values must satisfy
\begin{equation}
\prod_{\mbox{spatial~} \Box}
\varepsilon_{\bar{\jmath},\bar{\jmath}+\hat{\mu}} = -1 \label{u16}
\end{equation}
about every spatial plaquette. All other plaquettes have
$\prod_{\Box} \varepsilon_{\bar{\jmath},\bar{\jmath}+\hat{\mu}} =
1$. The coupling $K_d$ is a monotonic function of $e^2$ given, as
before, by (\ref{o13}) and (\ref{o5c}).

\subsection{Observables and limiting cases}
\label{sec:u11}

Continuing our parallel to the discussion of one-dimensional
antiferromagnets in Section~\ref{sec:chain}, we generalize the
discussion of Section~\ref{sec:obs} to two dimensions. The
generalization of the observables is immediate. The spin-flip site
boson density is now measured by the $\ell$ flux piercing the
associated plaquette on the dual lattice:
\begin{equation}
\hat{S}_{j,z} \sim \epsilon_{\tau\mu\nu} \Delta_{\mu}
\ell_{\bar{\jmath}\nu} \label{u17}
\end{equation}
(generalizing (\ref{o15})). The BC order in (\ref{o16}) needs some
notational modification:
\begin{equation}
Q_{j, j+\hat{x}} \sim
\varepsilon_{\bar{\jmath},\bar{\jmath}+\hat{y}}
\sigma_{\bar{\jmath},\bar{\jmath}+\hat{y}}, \label{u18}
\end{equation}
where the direct lattice link on the left-hand-side intersects
(after projecting the $\tau$ co-ordinate by half a lattice
spacing) the dual lattice link on the right-hand-side. A similar
relation holds with $x \leftrightarrow y$.

Next, we turning describing the physics of the limiting parameter
values in the phase diagram of $Z_{\rm U(1)}$. We will see that
there is much more structure than that in the corresponding
discussion in Section~\ref{sec:obs}.

As before, we first consider the limit $K_d \rightarrow \infty$.
Now the results are essentially identical to those in one
dimension: the values of $a_{\bar{\jmath} \mu}$ are pinned to
$a_{\bar{\jmath} \mu}^0$, and so $\epsilon_{\mu\nu\lambda}
\Delta_{\nu} a_{\bar{\jmath} \lambda} = \eta_j \delta_{\mu \tau}$.
The partition function (\ref{u12}) is exactly the boson
current-loop model studied by Otterlo {\em et al.} \cite{otterlo}.
With the purely on-site coupling $K_{jj'}^{\mu\nu}$ in (\ref{o9}),
they found that the boson-loop model was always a superfluid,
which corresponds to a spin system with long-range XY order. With
a nearest-neighbor coupling in the $K_{jj'}^{\mu\nu}$, Otterlo
{\em et al.} found states with a boson site density wave, which
corresponds here to spin states with I order.

In the limit $g \rightarrow \infty$, $Z_{\rm U(1)}$ reduces to a
non-trivial model, unlike the case in Section~\ref{sec:obs}. From
(\ref{u14}) we deduce that the integer vector field
$\ell_{\bar{\jmath} \mu}$ must be curl-free, implying it can be
expressed as the gradient of an integer-valued field on the sites
of the dual lattice. The relationship (\ref{u15}) then implies
that Ising field $\sigma_{\bar{\jmath},\bar{\jmath}+\hat{\mu}}$
is also `pure gauge' {\em i.e.\/}
\begin{equation}
\sigma_{\bar{\jmath},\bar{\jmath}+\hat{\mu}} =
\varsigma_{\bar{\jmath}} \varsigma_{\bar{\jmath}+\hat{\mu}}
\label{u19}
\end{equation}
where $\varsigma_{\bar{\jmath}} = \pm 1$ is an Ising field on the
sites of the dual lattice. So
\begin{equation}
Z_{\rm U(1)} ( g \rightarrow \infty ) =
\sum_{\{\varsigma_{\bar{\jmath}} = \pm 1\}} \exp \left( K_d
\sum_{\bar{\jmath}, \hat{\mu}}
\varepsilon_{\bar{\jmath},\bar{\jmath}+\hat{\mu}}
\varsigma_{\bar{\jmath}} \varsigma_{\bar{\jmath}+\hat{\mu}}
\right). \label{u110}
\end{equation}
This is the partition function of an ordinary Ising model on a
cubic lattice, with a nearest neighbor exchange of modulus $K_d$,
and the signs of the exchange chosen so that every spatial
plaquette is frustrated. The expression (\ref{u110}) could also
have been obtained by applying the standard Ising duality to the
$Z_2$ gauge theory in (\ref{u11a}) in the limit $g=\infty$; the
Berry phase in (\ref{u11a}) appears as the frustration in
(\ref{u110}). Also note that in the framework of the
parameterization in (\ref{u12}), the steps leading to
(\ref{u110}) correspond precisely to the discussion in
Section~\ref{sec:bo} that the large $g$ limit in the presence of
U(1) symmetry imposed the constraint (\ref{d9}) on (\ref{d7a}).
The Ising model in (\ref{u110}) was studied in
Ref.~\onlinecite{rodolfo2,japan} with the motivation also provided
by $S=1/2$ quantum antiferromagnets. The context of U(1)-symmetric
models is quite different here, and the earlier motivation is
close to the SU(2) symmetric models to be considered in
Section~\ref{planar}. From this earlier work on this frustrated
Ising model, and from methods to be described in detail in
Section~\ref{sec:u12}, we conclude that the ``ferromagnetic''
phase found at large $K_d$ has BC order. Also, the Ising
``paramagnetic'' phase at small $K_d$ is {\em fractionalized} (F)
with no broken lattice symmetry. This F phase had no analog in one
dimension, where the large $g$ limit yielded only the BC phase.

Turning to the limit $K_d\rightarrow 0$, $Z_{\rm U(1)}$ in
(\ref{u14}) becomes a model of interacting {\em half}-integer
currents with short-range interactions. After a rescaling of
currents and coupling constants, this model is formally
equivalently to one with integer currents, which correspond in
turn to boson models at integer filling studied in
Refs.~\onlinecite{sorensen,otterlo}. These have a transition
between a superfluid phase at small $g$ and an insulating phase
at large $g$ which is in the universality class of the O(2)
$\varphi^4$ field theory in 3 dimensions. For the spin model, the
superfluid phase corresponds to the XY phase, while the
insulating phase is F phase.

Finally, as in Section~\ref{sec:obs}, the $g \rightarrow 0$ limit
is trivial and the ground state always has XY order.

\begin{figure}
\epsfxsize=2.6in \centerline{\epsffile{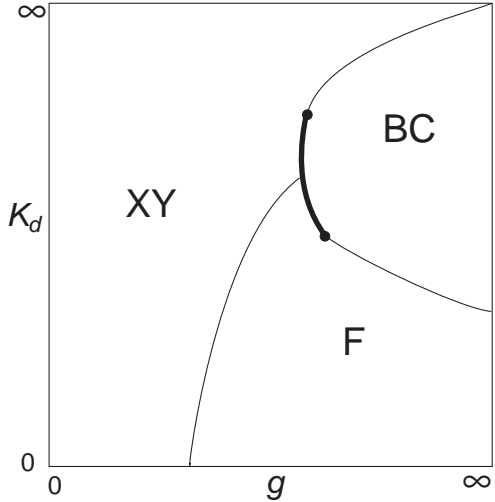}} \vspace{0.2in}
\caption{Phase diagram of two spatial-dimension model $Z_{\rm
U(1)}$ in (\protect\ref{u11a}) or (\protect\ref{u12}) or
(\protect\ref{u14}) for the case of the on-site coupling
(\protect\ref{o9}). The thin lines are second-order phase
boundaries while the thick line is a first-order boundary. The
field theories for the phases and the critical properties are
discussed in Section~\protect\ref{sec:u12}.}\label{phase3}
\end{figure}

Putting all these results together, we can sketch the phase
diagram in Fig~\ref{phase3} for $Z_{\rm U(1)}$.
Actually, determining the structure of the phase boundaries in
this figure required some additional field-theoretic analysis
which appears in Section~\ref{sec:u12}.

\subsection{Field theories}
\label{sec:u12}

We now proceed to a field-theoretic analysis of the model $Z_{\rm
U(1)}$ in (\ref{u14}) following the same method used to analyze
$Z^{(1)}_{\rm U(1)}$ in Section~\ref{sec:kt}. The technical
details of the derivation are similar to those used in the one
dimensional case in Appendix~\ref{e}, and their generalization to
two dimensions is described in Appendix~\ref{g}.

The field theory obtained in this manner turns out to be very
closely related to that studied recently by Lannert, Fisher and
Senthil (LFS) \cite{lannert}, although in a somewhat different
physical context. They were studying charge fluctuations in a
correlated superconductor using models of interacting boson
Cooper pairs. As we noted in Section~\ref{intro:frac}, and again
in Section~\ref{sec:obs}, our antiferromagnetic spin models can
also be interpreted as interacting boson Hamiltonians, but the
boson is now a spin-flip operator, as in (\ref{o14}). Casting
aside the different physical interpretation of the bosons, the
results of LFS can be directly applied to our models: we simply
have to map their superfluid state to a XY phase of
antiferromagnets, their boson site charge-density-wave to an I
phase, while the interpretation of BC order remains the same in
both cases. We will follow the same general approach as LFS, but
will extend their results in several directions to address the
questions of interest here.

In one dimension, as presented in Section~\ref{sec:kt}, the field
theory describing the phases of the antiferromagnet was the
sine-Gordon model in (\ref{o17}). In two dimensions, the
analogous procedure in Appendix~\ref{g} shows that the central
actors are point vortices in the wavefunction of the spin-flip
bosons in (\ref{o14}). We will denote the complex field which
creates or annihilates {\em single\/} vortices by $\phi (r)$,
where $r$ is a spacetime co-ordinate (the boson wavefunction
winds by a phase of $2 \pi$ around such a vortex). Also essential
for a description of the phases, and emerging naturally in the
derivation in Appendix~\ref{g}, are the {\em double\/} vortices
$\Phi (r)$ (the boson wavefunction winds now by $4 \pi$). As
discussed by LFS, and in Appendix~\ref{g}, the Berry phase terms
induce frustration in the hopping Hamiltonian for the $\phi$
vortices: consequently, the low energy vortices have momenta near
both $(0,0)$ and $(\pi, 0)$ and it is necessary to include both
sets in the field theory. We denote a convenient linear
combination of fields representing vortices near these minima by
$\phi_1 (r)$ and $\phi_2 (r)$ (see Appendix~\ref{g}). The motion
of the double vortex, $\Phi$, is unfrustrated, and it only has a
single minimum at momentum $(0,0)$. We list these fields, and
their physical properties, in Table~\ref{t2a}; this table also
contains additional fields to be introduced later in
Section~\ref{sec:su2}.
\begin{table}
\begin{tabular}{c|c}
Field & Description \cr \hline $\phi_1$, $\phi_2$ &
\begin{minipage}{2.8in} \vspace{0.07in} Complex field operators
for $\pm 2 \pi$ vortices in the spin-flip boson $b^{\dagger} =
\hat{S}_{+}$ (see (\protect\ref{o14})). The Berry phase
frustrates the hopping of such vortices, and  their dispersion
(in a convenient gauge) has minima near momenta $(0,0)$ and
$(\pi,0)$. $\phi_1$ and $\phi_2$ are convenient linear
combination of continuum fields representing fluctuations near
these minima (see Appendix~\ref{g}). By the boson-vortex duality,
$\pm 2 \pi$ vortices in $\phi_{1,2}$ change $b$ boson number by
$\pm 1$, and therefore carry $S_z = \pm 1$: such vortices must be
present in both $\phi_1$ and $\phi_2$ together to ensure a finite
energy cost. \vspace{0.07in}
\end{minipage}  \cr
\hline $\Phi$ & \begin{minipage}{2.8in} \vspace{0.07in} Complex
field operator for $\pm 4 \pi$ vortices in the spin-flip boson
$b^{\dagger} = \hat{S}_{+}$ (see (\protect\ref{o14})). Its
hopping is not frustrated and the dispersion has a minimum near
$(0,0)$. By the boson-vortex duality, $\pm 2 \pi$ vortices in
$\Phi$ change $b$ boson number by $\pm 1/2$, and therefore carry
$S_z = \pm 1/2$ {\em i.e.\/} they are spinons.  The relationship
to $\phi_{1,2}$ is $\Phi \sim \phi_1 \phi_2$ ($\sim$ indicates
that the two sides have correlators with identical physical
properties). \vspace{0.07in}
\end{minipage}  \cr \hline
$\Psi_I$ & \begin{minipage}{2.8in} \vspace{0.07in} Real order
parameter for I order, $\Psi_I \sim |\phi_1 |^2 - |\phi_2|^2$
\vspace{0.07in}
\end{minipage} \cr \hline
$\Psi_{BC}$ & \begin{minipage}{2.8in} \vspace{0.07in} Complex
order parameter for BC order, $\Psi_{BC} \sim \phi_1^{\ast}
\phi_2 $ \vspace{0.07in}
\end{minipage}
\end{tabular}
\vspace{0.2in} \caption{ Summary of all the continuum fields
introduced in Section~\protect\ref{sec:u1} and their physical
properties. } \label{t2a}
\end{table}

An explicit derivation of the effective action for $\phi_1$,
$\phi_2$, $\Phi$ is discussed in Appendix~\ref{g}. Here we will
begin by noting the transformations of these fields under the
symmetries of the lattice Hamiltonian, and then use these to
eventually write down the most general effective action
consistent with these symmetries. First, as is always the case in
the boson-vortex duality mapping, the fields $\phi_1$, $\phi_2$,
$\Phi$ all carry charges with respect to an internal, {\em
non-compact\/}, U(1) gauge field $b_{\mu}$ (this U(1) gauge field
is entirely distinct from the {\em compact\/} U(1) gauge fields
${\cal A}_{\mu}$ (Section~\ref{intro}) and $A_{\mu}$
(Section~\ref{planar} and Appendix~\ref{a})). The fields
$\phi_{1,2}$ carry $b$-charge 1, while $\Phi$ carries $b$-charge
2. Lattice symmetries lead to non-trivial mappings among these
fields, as discussed by LFS:
\begin{eqnarray}
&& \mbox{$\pi/2$ rotation about a direct lattice site:} \nonumber
\\ &&~~~~~~~~~~~~~~~~~~~~~~~~~~~~~ \mbox{$\phi_1 \rightarrow e^{i \pi/4} \phi_1$, $\phi_2 \rightarrow
e^{-i \pi/4}
\phi_2$} \nonumber \\
&& \mbox{$x$ translation by a lattice
spacing: $\phi_1 \rightarrow
\phi_2$, $\phi_2 \rightarrow \phi_1$} \nonumber \\
&& \mbox{$y$ translation by a lattice spacing: $\phi_1 \rightarrow
i\phi_2$, $\phi_2 \rightarrow -i\phi_1$} ;\nonumber \\
\label{u111}
\end{eqnarray}
$\Phi$ remains invariant under all of the above lattice
transformations.

Before writing down the effective action, we note the connection
of these fields to some of the orders characterizing the
antiferromagnet as presented in Table~\ref{t1}. The results
follow from simple symmetry considerations and are summarized in
Table~\ref{t2}.
\begin{table}
\begin{tabular}{c|c}
Label & Characterization \\ \hline BC & $\langle \phi_1^{\ast}
\phi_2 \rangle \neq 0$ \\ I & $\langle |\phi_1|^2 - |\phi_2|^2
\rangle \neq 0$  \\ XY & $\langle \phi_1 \rangle = \langle \phi_2
\rangle = \langle \Phi \rangle = 0$ \\ F & $\langle \Phi \rangle
\neq 0$, $\langle \phi_1 \rangle = \langle \phi_2 \rangle =0$
\end{tabular}
\vspace{0.2in} \caption{Characterization of some of the state
properties in Table~\protect\ref{t1} in terms of expectation
values of the fields $\phi_1$, $\phi_2$, and $\Phi$. States which
satisfy more than one property are also possible, and obey the
union of the corresponding expectation values {\em e.g.\/} the
BC+F state has $\langle \phi_1^{\ast} \phi_2 \rangle \neq 0$,
$\langle \Phi \rangle \neq 0$, and $\langle \phi_1 \rangle =
\langle \phi_2 \rangle =0$.} \label{t2}
\end{table}

The BC order breaks only a lattice symmetry, and must clearly be
neutral under the $b$-charge. Examination of (\ref{u111}) and the
symmetries of the states in Fig~\ref{bcorder} then easily shows
that the BC order parameter is the composite field $\phi_1^{\ast}
\phi_2$. The same connection can also be established directly by
applying (\ref{u18}) to the analysis of Appendix~\ref{g}. The
phase of the mean value of this order parameter distinguishes
between the various states in Fig~\ref{bcorder}:
$\arg(\phi_1^{\ast} \phi_2)=0,\pi/2,\pi,3\pi/2$ yields the four
states like Fig~\protect\ref{bcorder}a, $\arg(\phi_1^{\ast}
\phi_2)=\pi/4,3\pi/4,5\pi/4,7\pi/4$ yields the four states like
Fig~\protect\ref{bcorder}b, and all other values correspond to an
8-fold degenerate state with co-existing dimer and plaquette
order.

For the I order, we note from (\ref{u17}) that Ising ordering is
associated with a staggered pattern of boson site density; in the
boson-vortex duality this corresponds to $b$-charge neutral,
staggered, vortex current loops. The structure of the vortex
eigenmodes in Appendix~\ref{g} then shows that the I order is
$|\phi_1|^2 - |\phi_2|^2$.

The XY order corresponds to boson superfluidity. This is only
possible if none of the vortex fields have condensed, and hence
the vanishing of all expectation values in Table~\ref{t2}.

Finally, as discussed by LFS, the F states appear when $\Phi$ is
preferentially condensed over $\phi_{1,2}$. In the dual boson
language this means that excitations with boson charge 1/2 are
liberated (these latter bosons are themselves $\pm 2 \pi$
vortices in the double-vortex field $\Phi$); from (\ref{o14}) we
see that these are simply the $S=1/2$ spinons of a F state. Also,
it is evident from Table~\ref{t2} that co-existence between the F
and XY orders is not possible, while BC+F and I+F states can, in
principle, exist. In the approach of Senthil and Fisher
\cite{senthil} XY+F states did emerge in a larger space of models
and degrees of freedom than those considered here. Here, we will
show that XY+F states do not appear in the models considered so
far, but are present in the models with non-collinear spin
correlations to be considered in Section~\ref{planar}.

The derivation in Appendix~\ref{g}, or application of the
symmetries in (\ref{u111}) shows that action describing the
phases of $Z_{\rm U(1)}$ is the spacetime integral of the
Lagrangian
\begin{eqnarray}
&& {\cal L} = |(\partial_{\mu} - i b_{\mu}) \phi_{1}|^2 +
|(\partial_{\mu} - i b_{\mu}) \phi_{2}|^2 + |(\partial_{\mu} - 2 i
b_{\mu}) \Phi|^2 \nonumber \\ &&~~+ \frac{\tilde{g}}{2}
(\epsilon_{\mu\nu\lambda} \partial_{\nu} b_{\lambda})^2 + s_1
(|\phi_{1}|^2 + |\phi_{2}|^2 ) + s_2 |\Phi |^2 \nonumber \\
&&~~- w \Phi \phi_{1}^{\ast} \phi_{2}^{\ast} + \mbox{c.c.} +
\frac{u_1}{2} (|\phi_{1}|^2 + |\phi_{2}|^2)^2 + \frac{u_2}{2}
|\Phi|^4 \nonumber \\&&~~+ v_1 |\phi_{1}|^2 |\phi_{2}|^2 + v_2
|\Phi|^2 ( |\phi_{1}|^2 + |\phi_{2}|^2) + v_8 (\phi_{1}^{\ast}
\phi_{2} )^4 + \mbox{c.c.} \nonumber \\ \label{u112}
\end{eqnarray}
We have scaled all the field variables to make the co-efficients
of the gradient terms equal to unity; we have also set all
velocities equal to unity, and, for simplicity of notation,
ignored their possibly different values. The parameters $s_1$ and
$s_2$ tune the system across the various phase transitions, and
we will shortly discuss the phase diagram as a function of $s_1$
and $s_2$. The cubic coupling $w$ (not considered by LFS) is
allowed by all the symmetries in (\ref{u111}), and by charge
neutrality: it will play a central role in determining important
features of the phase diagram. The various quartic terms couple
invariant scalar densities and usually only play a peripheral
role in stabilizing the action. Finally, the $v_8$ term serves
only to choose between the different types of BC order.

\begin{figure}
\epsfxsize=3in \centerline{\epsffile{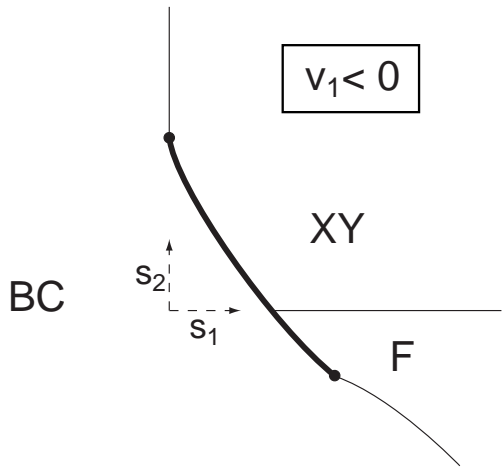}} \vspace{0.2in}
\caption{Mean field phase diagram of ${\cal L}$ in
(\protect\ref{u112}) as a function of $s_1$ and $s_2$ for typical
parameter values and $v_1 < 0$. The non-compact U(1) gauge field
$b_{\mu}$ is ignored in this analysis, and the action is simply
minimized w.r.t. to mean values of the fields $\phi_{1,2}$,
$\Phi$. As in Fig~\protect\ref{phase3}, first-order boundaries
are denoted by thick lines, while second-order boundaries are
thin lines. The BC phase has $\langle \phi_{1,2} \rangle \neq 0$
and $|\langle \phi_{1} \rangle| = |\langle \phi_{2} \rangle|$,
and these can be deduced from the characterizations in
Table~\protect\ref{t2}. See also
Table~\protect\ref{t3}.}\label{phase1}
\end{figure}

\begin{figure}
\epsfxsize=3in \centerline{\epsffile{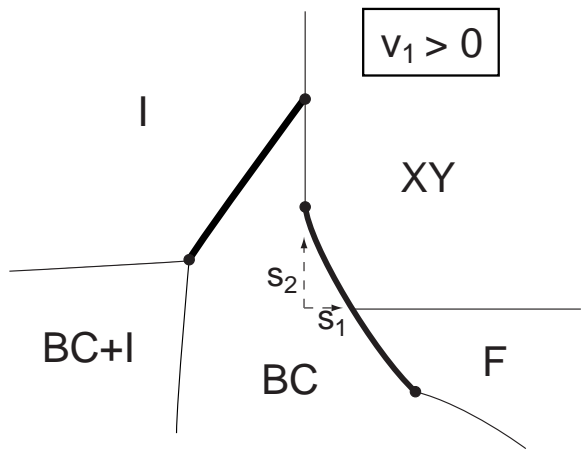}} \vspace{0.2in}
\caption{As in Fig~\protect\ref{phase1} but for $v_2 > 0$. The I
phase has $\langle \phi_{1} \rangle \neq 0$, and $\langle \phi_2
\rangle =0$ or vice-versa, as can be deduced from
Table~\protect\ref{t2}. The BC+I phase has $\langle \phi_{1,2}
\rangle \neq 0$ and $|\langle \phi_{1} \rangle| \neq |\langle
\phi_{2} \rangle|$, as indicated in
Table~\protect\ref{t3}.}\label{phase2}
\end{figure}

The mean-field phase diagram of ${\cal L}$ has a somewhat
different structure depending upon the sign of $v_1$. We sketch
typical phase diagrams as functions of $s_{1,2}$ for these two
cases in Figs~\ref{phase1} ($v_1 < 0$) and~\ref{phase2} ($v_1 >
0$).
The simpler case in Fig~\ref{phase1} has only three phases: XY,
BC, and F, and the saddle point always has $|\phi_1| = |\phi_2|$
(so there is no I order). Any two of the phases can be separated
by second-order phase boundary (whose field theory and order
parameter can be easily determined from Table~\ref{t2} and ${\cal
L}$). However the phase boundaries necessarily become first order
near the point where the three phases meet: this is a consequence
of the cubic $w$ term in ${\cal L}$, as shown recently in
Appendix~A of Ref.~\onlinecite{kwon} in a different context but
with a field theory with a very similar structure. The results of
Fig~\ref{phase1} were used to deduce the phase diagram of $Z_{\rm
U(1)}$ in Fig~\ref{phase3}. Note that the two diagrams have an
identical topology. The absence of an I phase shows that the
purely on-site coupling (\ref{o9}) in $Z_{\rm U(1)}$ corresponds
to a field theory ${\cal L}$ with $v_1 < 0$.

The phase diagram in Fig~\ref{phase2}, with $v_1 > 0$, does allow
for an I phase. So it can only correspond to $Z_{\rm U(1)}$ with
non-on-site couplings $K_{jj'}^{\mu\nu}$. A crucial feature of
this phase diagram is that all phases with I order are well
separated from the F phase: there is no second order transition
between the F phase and the I phase (although a strong first order
transition can always be induced by choosing couplings
judiciously). This feature will play a central role in our
discussion of the restoration of SU(2) symmetry in
Section~\ref{sec:su2}. For now, we present a simple argument that
the separation of phases with F and I order also holds beyond the
mean-field analysis of ${\cal L}$. Let us begin in the
fractionalized $F$ phase. From Table~\ref{t2}, we see that the
$\Phi$ field has condensed. The $b$-gauge fluctuations are
quenched in such a region (by the usual Higgs mechanism), and can
therefore be neglected. The onset of the I phase involves a
confinement transition and so one of the fields $\phi_1$,
$\phi_2$ must condense. Let us examine the structure of the
effective action for $\phi_{1,2}$ in a region where we can safely
set $b_{\mu} = 0$ and $\langle \Phi \rangle \neq 0$; from ${\cal
L}$ the quadratic form controlling $\phi_{1,2}$ fluctuations is
\begin{eqnarray}
|\partial_{\mu} \phi_1 |^2 + |\partial_{\mu} \phi_2 |^2 &+& (s_1 +
v_2 |\langle \Phi \rangle|^2)(|\phi_1|^2 + |\phi_2|^2) \nonumber
\\
&-& w \langle \Phi \rangle \phi_1^{\ast} \phi_2^{\ast} - w \langle
\Phi^{\ast} \rangle \phi_1 \phi_2 \label{u113}
\end{eqnarray}
Near a second order transition where $\phi_{1,2}$ condense, we
must diagonalize the quadratic form (\ref{u113}) and condense the
field combinations in the eigenmode with the lowest eigenvector.
A simple analysis of (\ref{u113}) shows that, for arbitrary
$\langle \Phi \rangle$, this eigenmode has $|\phi_1 | = |\phi_2|
$. From Table~\ref{t2}, it follows immediately that the condensed
phase cannot have I order. Therefore the F phase is surrounded
entirely by with the BC phase (corresponding to condensation of
$\phi_{1,2}$) or by the XY phase (corresponding to removal of the
$\Phi$ condensate, as is presented in Fig~\ref{phase2}.

\subsection{Generalized phase diagrams and restoration of SU(2)
symmetry} \label{sec:su2}

The ultimate objective of our analysis of antiferromagnets with
U(1) symmetry is to find parameter regimes where full SU(2)
symmetry is restored, thus enabling a description of the phases
and phase transitions of SU(2)-invariant antiferromagnets. At the
corresponding stage in our analysis of U(1) antiferromagnets in
one dimension in Section~\ref{sec:chain} we had achieved this
objective. We found there that boundary between the TL and I
phases was a critical line, possessing SU(2) invariant
correlations at long distances with a power-law decay (as in
(\ref{o2}) characterized by exponents $\eta_{XY} = \eta_I = 1$.
This is, of course, the well-known property of the $S=1/2$
antiferromagnetic Heisenberg chain. Furthermore, as indicated in
Fig~\ref{figsg}, the SU(2)-invariant line extends into the gapped
BC phase, and we obtained the correct and complete theory of the
quantum phase transition from the TL to BC phase for
SU(2)-invariant antiferromagnets.

Let us now apply the same strategy in two dimensions. As a first
step, we have to obtain a parameter regime where the ground state
has HN order. Unlike the SU(2) invariant case in one dimension,
the HN state is {\em not} critical: it has true long-range
magnetic order, and an excitation spectrum of two, gapless,
linearly dispersing spin waves. By analogy with one dimension, a
natural candidate for HN order is the boundary between the XY and
I phases in Fig~\ref{phase2}. However, further consideration shows
that this line clearly does not satisfy the necessary criteria.
This is a {\em critical} line describing a second order
transition: the critical degrees of freedom are the fields
$\phi_{1,2}$ coupled to the $b_{\mu}$ non-compact U(1) gauge
field, and the critical field theory is ${\cal L}$ in (\ref{u112})
after we set the massive field $\Phi = 0$. It is expected that all
correlators will have an anomalous power-law decay along this
line, and these characteristics are far removed from those of the
HN phase.

We have to enlarge the parameter space of our field theory to
find the HN phase. To this end, note that the I (and BC) order in
Table~\ref{t2}, were associated with {\em composite} fields built
out of $\phi_{1,2}$. It would clearly pay to elevate the I order
parameter to an elementary degree of freedom, as that would then
permit co-existence of {\em non-critical} XY and I order over a
wide parameter regime, allowing for possible subspace with HN
order. We therefore introduce the elementary Ising order
parameter field $\Psi_I$, and, for completeness, the elementary
BC order parameter field, $\Psi_{BC}$. We have now introduced all
the continuum fields listed in Table~\ref{t2a}, which contains a
summary of their physical interpretation.

The relationships of $\Psi_I$ and $\Psi_{BC}$ to the fields
$\phi_{1,2}$ noted in Table~\ref{t2a} follow immediately from the
state characterizations in Table~\ref{t2}. We can now make a
catalog of possible phases by considering the physical properties
of all the different combinations of condensation among the
fields $\phi_{1,2}$, $\Phi$, $\Psi_{BC}$ and $\Psi_I$. This leads
to the exhaustive listing provided in Table~\ref{t3}.
\begin{table}
\begin{tabular}{c|c}
State & Condensed fields \cr \hline BC & $\phi_1$, $\phi_2$,
$\Psi_{BC}$, $\Phi$ \cr I & $\phi_1$ or $\phi_2$, $\Psi_{I}$ \cr
BC+I & $\phi_1$, $\phi_2$, $\Psi_{BC}$, $\Psi_I$, $\Phi$ \cr XY &
none \cr XY+BC & $\Psi_{BC}$ \cr XY+I & $\Psi_I$ \cr XY+BC+I &
$\Psi_{BC}$, $\Psi_I$ \cr F & $\Phi$ \cr F+BC & $\Phi$,
$\Psi_{BC}$ \cr F+I & $\Phi$, $\Psi_I$ \cr F+BC+I & $\Phi$,
$\Psi_{BC}$, $\Psi_I$
\end{tabular}
\vspace{0.2in} \caption{Exhaustive specification of the states
expected in the generalized parameter space of ${\cal
L}^{\prime}$. Notice that the left column contains state labels,
and not the label of a property that may be obeyed by a state,
which was the case in Table~\protect\ref{t2}. The right-column
contains all the fields that have a non-zero expectation value in
the corresponding state; if a field is not mentioned, it has
vanishing expectation value in that state. The table above can be
deduced directly from the characterizations in
Table~\protect\ref{t2}, and from the field correspondences in
Table~\ref{t2a}.} \label{t3}
\end{table}
Note the appearance of the AF+I phase--this will be our candidate
for possessing HN order along a subspace in its interior.

An understanding of the topology of the phase diagram and the
nature of the quantum phase transitions requires an effective
Lagrangian for all the fields. We can use the same symmetry
considerations discussed near (\ref{u111}) to extend the
Lagrangian ${\cal L}$ in (\ref{u112}) to include $\Psi_{BC}$ and
$\Psi_{I}$. Such a procedure yields
\begin{eqnarray}
&& {\cal L}' = {\cal L} + \frac{1}{2} (\partial_{\mu} \Psi_{I})^2
+ |\partial_{\mu} \Psi_{BC}|^2 + \frac{s_I}{2} \Psi_{I}^2 + s_{BC}
|\Psi_{BC} |^2 \nonumber \\
&&~~- w_I \Psi_I \left(|\phi_{1}|^2 - |\phi_{2}|^{2} \right) -
w_{BC} \Psi_{BC} \phi_1 \phi_2^{\ast} + \mbox{c.c.} \nonumber \\
&&~~+ v_8^{\prime} \Psi_{BC}^4 + \mbox{c.c.} + \ldots .\label{su2}
\end{eqnarray}
The ellipses represents numerous quartic terms involving
couplings between the modulus squared of all the fields like
$|\Psi_{BC}|^2 |\Phi|^2$, $|\Psi_I |^2 (|\phi_1|^2 + |\phi_2|^2)$,
$|\Psi_{BC}|^4 \dots$; these serve mainly to stabilize the action
and we will refrain from writing them out explicitly. We now have
two additional parameters, $s_{BC}$ and $s_I$, which can tune the
system between its phases by controlling the condensation of
$\Psi_{BC}$ and $\Psi_I$ respectively. The cubic couplings $w_I$
and $w_{BC}$ follow directly from Table~\ref{t2a}, and play a
central role in determining the phase diagram.

We can now envisage a four-dimensional phase diagram as a
function of the parameters $s_1$, $s_2$, $s_I$ and $s_{BC}$. This
is a space of immense complexity and we will explore its
structure by describing a few judiciously chosen two-dimensional
cross-sections.

\begin{figure}
\epsfxsize=3in \centerline{\epsffile{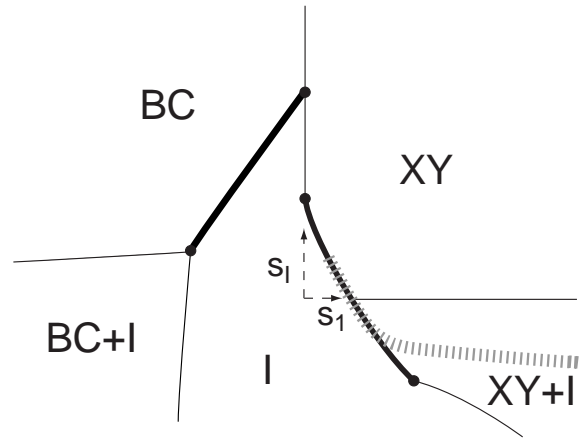}} \vspace{0.2in}
\caption{Schematic mean field phase diagram of ${\cal L}'$ in
(\protect\ref{su2}) as a function of $s_1$ and $s_I$ with $s_2 >
0$, $s_{BC} > 0$, and $v_1 < 0$. The conventions for the phase
boundaries are as in Fig~\protect\ref{phase3}. The grey hatched
line represents a possible line along which SU(2) symmetry is
restored: this line begins within the XY+I phase, and then
continues along the first-order boundary between the XY and I
phases; as argued in the text it cannot reach the end-point of
this first-order boundary and veers off into another direction
not included within this cross-section of the phase diagram.
}\label{phase4}
\end{figure}

Our primary interest is in studying the vicinity of the XY+I
phase. So we describe the cross-section in the $s_1$-$s_I$ plane
with $s_2 > 0$ and $s_{BC} > 0$. The mean-field phase diagram for
this case is shown in Fig~\ref{phase4}.
An XY+I phase appears, as expected, and we will now study its
properties in greater detail.

\subsubsection{HN order in the XY+I phase}
\label{sec:hn}

The XY+I phase only has $\langle \Psi_I \rangle$ non-zero, as was
indicated in Table~\ref{t3}. A simple stability analysis about
such a saddle-point shows that fluctuations of $\Psi_I^{\prime} =
\Psi_I - \langle \Psi_I \rangle$, $\Psi_{BC}$, $\phi_{1,2}$, and
$\Phi$ are all gapped. However, the transverse component of the
non-compact U(1) gauge field $b_{\mu}$ (which is not Higgsed by
the neutral $\Psi_I$) yields a {\em single} gapless,
linearly-dispersing mode: this corresponds to the familiar
spin-wave mode of the XY order. To obtain a state with HN order,
we need {\em two} spin-wave modes, and we can reasonably hope
this may happen for special values of the parameters.

As a prelude to demonstrating this more explicitly, let us write
down the form of the effective Lagrangian that will describe the
HN phase. In its most symmetric form, this is simply the O(3)
$\varphi^4$ field theory:
\begin{equation}
{\cal L}_H = \frac{1}{2} \left( (\partial_{\mu} \varphi_\alpha)^2
+ s_H \varphi_a^2 \right) + \frac{u_H}{4!} (\varphi_a^2)^2
,\label{su3}
\end{equation}
where $a=x,y,z$ and $s_H < 0$. In the ordered phase, we
parameterize the fluctuations of $\varphi_a$ by
\begin{equation}
\varphi_a = (\sqrt{2}\mbox{Re}(\Psi_{XY}), \sqrt{2}\mbox{Im}
(\Psi_{XY}), \langle \Psi_I \rangle + \Psi_I^{\prime} ),
\label{su4}
\end{equation}
where here $\langle \Psi_I \rangle = \sqrt{-6 s_H/u_H}$, and
$\Psi_{XY}$ and $\Psi_I^{\prime}$ are fluctuating complex and
real fields respectively. Inserting (\ref{su4}) into (\ref{su3})
we obtain the Lagrangian controlling the fluctuations about the
ordered state:
\begin{eqnarray}
&& {\cal L}_H = |\partial_{\mu} \Psi_{XY} |^2 +
\frac{1}{2}(\partial_{\mu} \Psi_{I}^{\prime})^2  + |s_H|
(\Psi_{I}^{\prime})^2 \nonumber \\&&~~~~~ + \sqrt{\frac{|s_H|
u_H}{6}} \Psi_I^{\prime} \left((\Psi_{I}^{\prime})^2 + 2
|\Psi_{XY}|^2\right) \nonumber \\
&&~~~~~+ \frac{u_H}{4!} \left((\Psi_{I}^{\prime})^2 + 2
|\Psi_{XY}|^2\right)^2 \label{su5}
\end{eqnarray}
Two properties of this form of ${\cal L}_H$ are especially
noteworthy. First, there is no mass term for the complex field
$\Psi_{XY}$, and so there are two gapless spin-wave modes at
lowest order. Second, a number of cubic and quartic couplings are
present, but their co-efficients have specific algebraic
relations between them, tying them to the gap in the
$\Psi_I^{\prime}$ mode ($\sqrt{2 |s_H|}$) and the single quartic
coupling $u_H$: these relations ensure that the gapless spin wave
modes survive at all orders in perturbation theory.

We now search for an effective Lagrangian like (\ref{su5}) in
${\cal L}^{\prime}$ defined in (\ref{su2}). Without loss of
generality, we take $\langle \Psi_I \rangle >0$ and $w_I > 0$.
Then, notice from (\ref{su2}) that the combination of the $w_I$
cubic term and the presence of the $\Psi_I$ condensate causes the
mass of the $\phi_2$ field to be significantly larger than the
$\phi_1$ field. So we proceed to integrate out $\phi_2$, along
with the double-vortex field $\Phi$ which remains strongly gapped
everywhere in Fig~\ref{phase4}. The resulting effective Lagrangian
for $\phi_1$ and $\Psi_I^{\prime}$ has the $\phi_1$ controlled by
a model of scalar electrodynamics:
\begin{eqnarray}
{\cal L}^{\prime\prime} &=& |(\partial_{\mu} - i b_{\mu})
\phi_{1}|^2 +
 + \frac{\tilde{g}^{\prime}}{2}
(\epsilon_{\mu\nu\lambda} \partial_{\nu} b_{\lambda})^2 +
s_1^{\prime} |\phi_{1}|^2  + \frac{u_1^{\prime}}{2} |\phi_{1}|^4
\nonumber \\
&+& \frac{1}{2} (\partial_{\mu} \Psi_{I}^{\prime})^2 +
\frac{s_I^{\prime}}{2}
 (\Psi_{I}^{\prime})^2 + \ldots, \label{su6}
\end{eqnarray}
where, at lowest order in the non-linearities,
\begin{equation}
s_1^{\prime} = s_1 - 2 w_1 \langle \Psi_I \rangle - 2 w_1
\Psi_I^{\prime}. \label{su6a}
\end{equation}
We are now ready to make our key non-trivial observation. We use
the well-established result \cite{dh} that the scalar
electrodynamics model is dual, in the continuum, to an O(2)
$\varphi^4$ field theory in three spacetime dimensions.
Performing this duality on the $\phi_1$, $b_{\mu}$ degrees of
freedom in (\ref{su6}), we conclude that ${\cal
L}^{\prime\prime}$ is equivalent to the following dual continuum
Lagrangian
\begin{eqnarray}
{\cal L}_{D} &=&  | \partial_{\mu} \Psi_{XY} |^2 + s_{XY}
|\Psi_{XY}|^2  + \frac{u_{XY}}{4!} |\Psi_{XY}|^4
\nonumber \\
&+& \frac{1}{2} (\partial_{\mu} \Psi_{I}^{\prime})^2 +
\frac{s_I^{\prime}}{2}
 (\Psi_{I}^{\prime})^2 + \ldots, \label{su7}
\end{eqnarray}
where the two-component field of the O(2) $\varphi^4$ field
theory has been written as the complex field $\Psi_{XY}$; this
field has a complicated non-local relationship to $\phi_1$ and
$b_{\mu}$. Further, the parameters $s_{XY}$ and $u_{XY}$ are
determined by the couplings $s_1^{\prime}$, $u_1^{\prime}$, and
$\tilde{g}^{\prime}$ in (\ref{su6}). The linear dependence of
$s_1^{\prime}$ on $\Psi_I^{\prime}$ in (\ref{su6a}) implies that
the coupling in ${\cal L}_D$ also have linear dependence on
$\Psi_I^{\prime}$. With this dependence, the similarity between
the (\ref{su7}) and O(3) symmetric model ${\cal L}_H$ in
(\ref{su5}) becomes evident: the $\Phi_I^{\prime}$ dependence of
$s_I^{\prime}$ will lead to cubic terms like those in
(\ref{su5}). A suitable choice of the couplings in ${\cal
L}^{\prime}$ can clearly ensure that its dualized version obeys
the same constraints on its couplings as those required by O(3)
symmetry in ${\cal L}_H$. In particular, the mass $s_{XY}$ in
${\cal L}_D$ remains pinned to 0 along this O(3)-symmetric
subspace, as is needed to obtain two linearly dispersing
spin-wave modes in the HN state. This establishes our claim that
HN order appear in a subspace with the XY+I phase.

We have sketched a possible line with SU(2) symmetry restored in
Fig~\ref{phase4}. A portion of it is in a phase with co-existing
XY+I order, but it also continues along the first-order boundary
between the XY and I phases; all our above arguments also apply
along such a boundary. However, within the phases found in
Fig~\ref{phase4}, there is no simple and natural way to extend
this SU(2)-symmetric line into a non-magnetic phase. If we move
above along the vertical line $s_1 =0$ in Fig~\ref{phase4}, we
eventually reach the second-order boundary between the XY and I
phases: we argued earlier that this line cannot be SU(2)
symmetric. It is possible to choose parameters such the
end-points of the two first-order lines in Fig~\ref{phase4}
co-incide: then the SU(2) symmetric line can be extended into the
non-magnetic BC phase, and we have the possibility of a
second-order transition between the HN and BC phases in a system
with SU(2) symmetry. But this case is  non-generic as there does
not appear to be any fundamental reason for the two end-points to
merge. This conclusion is also entirely consistent with our
earlier considerations in Section~\ref{sec:hnbo}.

To find a more natural scheme for a transition to a non-magnetic
state with SU(2) symmetry we have to search the phase diagram in
a regime beyond the cross-section in Fig~\ref{phase4}. This we
will do in the following subsections.

\subsubsection{SU(2) symmetry with bond order}
\label{sec:subo}

\begin{figure}
\epsfxsize=3in \centerline{\epsffile{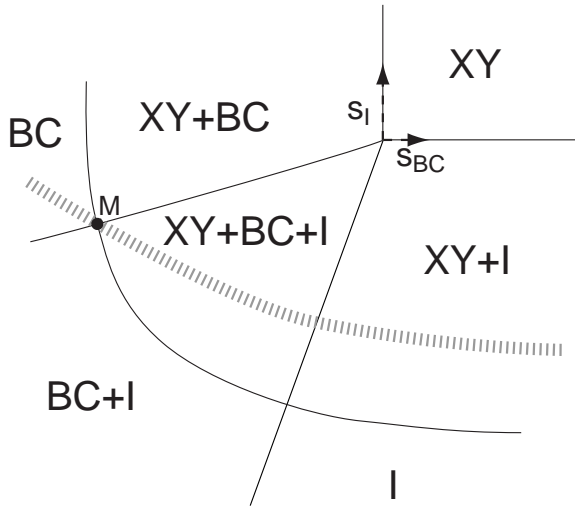}} \vspace{0.2in}
\caption{Schematic mean field phase diagram of ${\cal L}'$ in
(\protect\ref{su2}) as a function of $s_{BC}$ and $s_I$ with $s_1
> 0$ and  $s_2 > 0$. The conventions for the phase
boundaries are as in Fig~\protect\ref{phase3}. The grey hatched
line represents a possible line along which SU(2) symmetry is
restored: this line begins within the XY+I phase, continues into
the XY+BC+I phase, and finally into the non-magnetic BC phase;
note that sequence of transitions along this line (moving right
to left) is identical to that in Fig~\ref{phaseZ} (moving left to
right), and also to that discussed in Section~\ref{sec:hnbo}.
Upon going beyond mean-field theory, it is expected that
tetracritical point M will turn into two bi-critical points
separated by a short first order boundary between the XY+BC and
BC+I phases; the line of SU(2) symmetry will overlap this
first-order line.}\label{phase5}
\end{figure}

A very useful cross-section of the phase diagram is that
controlled by $s_I$ and $s_{BC}$. This is sketched in
Fig~\ref{phase5} in the regime $s_1 > 0$ and $s_2 > 0$.
As in Section~\ref{sec:su2}, the XY+I phase is a reasonable
candidate for SU(2) symmetry and HN order. Here we discuss the
question of whether it is possible to continue the line of SU(2)
symmetry into the XY+BC+I phase, and then further into the
non-magnetic BC phase.

We begin our discussion by the spectrum of excitations in the BC
phase. This is a confining phase with the fields $\phi_{1,2}$
condensed (see Table~\ref{t3}). As noted in Table~\ref{t2a}, the
excitations with non-zero total $S_z$ are gapped $\pm 2 \pi$
vortices in $\phi_{1,2}$ (the relative phases of $\phi_1$ and
$\phi_2$ are pinned to each other): these vortices have total
$S_z = \pm 1$. In a BC phase with SU(2) symmetry this doublet of
excitations must be part of a $S=1$ triplet---this is the
familiar $S=1$ gapped spin exciton expected above a dimerized
insulator. We will now identify the missing ${S}_z=0$ excitation
which completes this degenerate triplet along a special line
within the BC phase. We will also show that condensation of this
triplet leads to the BC+HN phase, and the latter phase naturally
lies along a line within the XY+BC+I phase. Thus we will be able
to justify the entire trajectory of the SU(2) symmetric line in
Fig~\ref{phase5}.

Within the BC or BC+XY+I phases, the field $\Psi_{BC}$ is
condensed. So we can safely replace $\Psi_{BC}$ by $\langle
\Psi_{BC} \rangle$; fluctuations of $\Psi_{BC}$ about this value
are massive and do not play an important role in our
considerations here. Let us examine the effective action for
$\phi_{1,2}$ and $\Psi_I$ fluctuations in phases with $\langle
\Psi_{BC} \rangle \neq 0$. Consider the quadratic form
controlling $\phi_{1,2}$; from (\ref{u112}) and (\ref{su2}), this
has the form:
\begin{eqnarray}
|\partial_{\mu} \phi_1 |^2 &+& |\partial_{\mu} \phi_2 |^2 + s_1
(|\phi_1|^2 + |\phi_2|^2) \nonumber
\\
&-& w_{BC} \left( \langle \Psi_{BC} \rangle \phi_1 \phi_2^{\ast} -
 \langle \Psi_{BC}^{\ast} \rangle \phi_1^{\ast} \phi_2 \right) .\label{su8}
\end{eqnarray}
Without loss of generality, we can assume that $w_{BC} > 0$ and
$\langle \Psi_{BC} \rangle$ is real. With these values, the
quadratic form (\ref{su8}) is diagonalized by the linear
combinations $\phi = (\phi_1 + \phi_2)/\sqrt{2}$ and
$\phi^{\prime} = (\phi_1 - \phi_2)/\sqrt{2}$. The eigenvalue of
$\phi^{\prime}$ is larger than that of $\phi$, and so we
integrate out $\phi^{\prime}$ and express the effective action
${\cal L}^{\prime}$ in terms of $\phi$ and $\Psi_I$. The result
has a very simple structure. In particular, there is no cubic
residual cubic coupling between $\Psi_I$ and $\phi$ which arises
from the $w_I$ term in (\ref{su2}); in terms of $\phi$,
$\phi^{\prime}$, the term proportional to $w_I$ is $\Psi_I ( \phi
\phi^{\ast \prime} + \phi^{\ast} \phi^{\prime})$, and upon
integrating out $\phi^{\prime}$ this leads only to the quartic
coupling $|\phi|^2 |\Psi_I|^2 $ between the modulus-squared of
the two fields. With these considerations, the new form of ${\cal
L}^{\prime}$ in (\ref{su2}), in the phases with $\Psi_{BC}$
condensed is
\begin{eqnarray}
&& {\cal L}_{BC} = |(\partial_{\mu} - i b_{\mu}) \phi|^2 +
 + \frac{\tilde{g}^{\prime}}{2}
(\epsilon_{\mu\nu\lambda} \partial_{\nu} b_{\lambda})^2 + s
|\phi|^2  + \frac{u}{2} |\phi|^4
\nonumber \\
&&+ \frac{1}{2} (\partial_{\mu} \Psi_{I}^{\prime})^2 +
\frac{s_I}{2}
 (\Psi_{I}^{\prime})^2 + \frac{u_I}{2} |\Psi_I|^4 +
 v_I |\phi|^2 |\Psi_I|^2 . \label{su9}
\end{eqnarray}
The $S_z = \pm 1$ excitations are the vortices in $\phi$, and so
we should naturally proceed to a duality transformation on the
$\phi$, $b_{\mu}$ degrees of freedom. These fields are controlled
by scalar electrodynamics and so the transformation is similar to
that discussed below (\ref{su6a}). Scalar electrodynamics is dual
to the O(2) $\varphi^4$ field theory \cite{dh}, and, as before, we
express this in terms of the complex field $\Psi_{XY}$ (which is
nothing but the field operator for $\pm 2 \pi$ vortices in
$\phi$). So ${\cal L}_{BC}$ turns out to be dual to
\begin{eqnarray}
&& {\cal L}_{BC}^{D} = |\partial_{\mu} \Psi_{XY}|^2 +
 + s_{XY} |\Psi_{XY} |^2 + \frac{u_{XY}}{2} |\Psi_{XY}|^4
\nonumber \\
&&+ \frac{1}{2} (\partial_{\mu} \Psi_{I}^{\prime})^2 +
\frac{s_I}{2}
 (\Psi_{I}^{\prime})^2 + \frac{u_I}{2} \Psi_I^4 +
 v_I^{\prime} |\Psi_{XY}|^2 \Psi_I^2 . \label{su10}
\end{eqnarray}
Now the connection between the ${\cal L}_{BC}^{D}$ and the O(3)
$\varphi^4$ fields theory is plainly visible: for $s_{XY} = s_I$
and $u_{XY} = v_I^{\prime} = 4 u_I$ the theory in (\ref{su10}) is
precisely ${\cal L}_H$ in (\ref{su3}) after we identify
$\varphi_a = (\sqrt{2} \mbox{Re} ( \Psi_{XY} ), \sqrt{2}
\mbox{Im} ( \Psi_{XY} ), \Psi_I)$. So we have shown that SU(2)
symmetry is restored with these restrictions on the couplings.
 For $s_H = s_I = s_{XY} > 0$,
the theory ${\cal L}_H$ is in the BC phase and $\varphi_a$ is the
field operator for the required gapped $S=1$ spin exciton; the
gapped, $S_z = 0$ particle associated with $\Psi_I$ has combined
with the $S_z = \pm 1$ vortices in $\phi$ to yield the required
triply degenerate particle. For $s_H < 0$, we have long-range HN
order induced by the condensation of the $S=1$ exciton; this
corresponds to moving along the hatched line in Fig~\ref{phase5}
from the BC phase to the XY+BC+I phase. The critical properties
of the transition from the BC phase to the BC+HN phase are
described by the familiar O(3) $\varphi^4$ field theory in
(\ref{su3}).

The above considerations clearly do not apply to the transition
from the BC+XY+I phase to the XY+I phase: long-range BC order was
assumed to be present above, and this becomes critical here. The
line with SU(2) symmetry plays no special role at this
transition, and it described simply by a field theory for
$\Psi_{BC}$ alone, which is obtained by setting all other fields
in ${\cal L}^{\prime}$ to zero. The coupling to the gapless
spin-wave modes of the XY (or HN) order is irrelevant
\cite{sushkov1}.

All the theoretical justification for the trajectory of the SU(2)
symmetric line in Fig~\ref{phase5} has now been provided. It is
satisfying that the sequence of phases along this line (moving
right to left) is identical to that presented in Fig~\ref{phaseZ}
(moving left to right). Entirely independent theoretical
arguments for the sequence in Fig~\ref{phaseZ} were presented in
Section~\ref{sec:hnbo} using methods that preserved SU(2)
symmetry at all stages, in contrast to those discussed in the
present section.

Further examination of Fig~\ref{phase5} also suggests an
apparently exotic circumstance under which there could be a second
order transition between the HN and BC states. This happens if
parameters are such that the XY+BC+I phase has just been shrunk
to vanishing size {\em i.e.} the three multicritical points in
Fig~\ref{phase5} have just coalesced into one, and the line of
SU(2) symmetry passes through this coalesced point. Such a point
does appear to represent a multicritical point, requiring the
tuning of more than one parameter, but we cannot definitively
rule out that accessing such a point in the subspace of SU(2)
symmetry requires only one tuning parameter. In any event, this
reasoning does lead to a candidate field theory for the critical
point between the HN and BC phases (multicritical or not): it is
the field theory ${\cal L}^{\prime}$ in (\ref{su2}) with the
primary fields $\phi_{1,2}$, $b_{\mu}$, $\Psi_{BC}$ and $\Psi_I$
all critical, while $\Phi$ is massive and can be integrated out.

\subsubsection{Interplay with fractionalization}
\label{sec:sufrac}

The previous section paid particular attention to the role of BC
order in the phase diagram of the field theory ${\cal
L}^{\prime}$ in (\ref{su2}). This section will carry out the
complementary analysis for the case of F order. We will also
address the issue of whether SU(2) symmetry can be found in such
phases. We argued earlier, in Section~\ref{sec:bo} that F phases
are not found in the class of SU(2) symmetric systems that have
been considered so far: our discussion below will further support
this conclusion. However, we remind the reader that in the
following Section~\ref{planar} we will introduce a new class of
models which do have F order and SU(2) symmetry.

\begin{figure}
\epsfxsize=3in \centerline{\epsffile{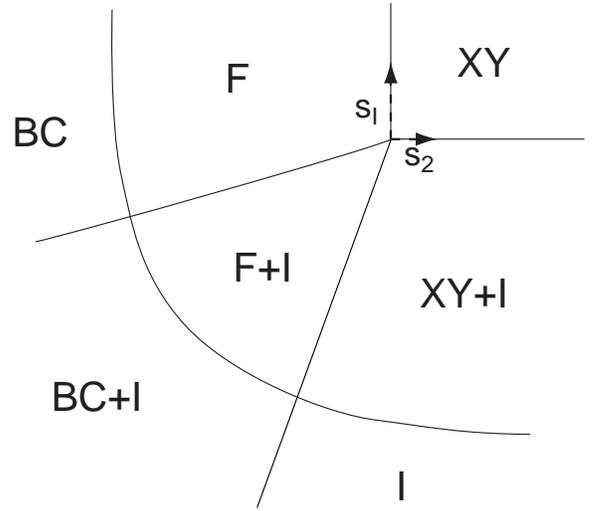}} \vspace{0.2in}
\caption{Schematic mean field phase diagram of ${\cal L}'$ in
(\protect\ref{su2}) as a function of $s_{2}$ and $s_I$ with $s_1
> 0$ and  $s_{BC} > 0$. The conventions for the phase
boundaries are as in Fig~\protect\ref{phase3}. As we discuss in
Section~\protect\ref{sec:sufrac}, there is no reasonable case for
a line of SU(2) symmetry in this phase diagram, apart from within
the XY+I and BC phases.}\label{phase6}
\end{figure}

As a prelude to our discussion, we determine a section of the
mean-field phase diagram of (\ref{su2}) as a function of $s_2$
(which controls condensation of the field $\Phi$ essential to the
appearance of F order) and $s_I$ (which controls the I order
necessary for the presence of phases with SU(2) symmetry). The
other two `masses', $s_1$ and $s_{BC}$ are taken to be positive.
The results are shown in Fig~\ref{phase6}.

Let us focus our attention on the F phase in Fig~\ref{phase6} and
ask if it is possible for it to possess SU(2) symmetry. A simple
argument shows that this is not possible, at least in the
framework of the continuum models described by ${\cal
L}^{\prime}$. From Table~\ref{t3}, only the field $\Phi$ is
condensed in the F phase. As was noted in Table~\ref{t2a},
vortices with phase winding $\pm 2 \pi$ in the field $\Phi$ carry
boson charge 1/2, and are therefore the $S_z = \pm 1/2$ spinon
excitations about the F state. (Note that such vortices were not
possible in the BC states of Fig~\ref{phase5} and~\ref{phase6}
because the field $\phi_{1,2}$ was condensed and this would be
double-valued around a $\pm 2 \pi$ vortex in the $\Phi$ field). A
theory for these spinons follows by setting the massive fields
$\phi_{1,2}=0$ in ${\cal L}$ in (\ref{u112}). The resulting
Lagrangian for $\Phi$ is (again) simply that of scalar
electrodynamics. Just as in Sections~\ref{sec:hn}
and~\ref{sec:subo}, we can dualize this to an $O(2)$ $\varphi^4$
field theory, which we now express in terms of the complex scalar
$\Psi_{\rm sp}$. This $\Psi_{\rm sp}$ is the field operator for
the $S_z = \pm 1/2$ spinons, and is (roughly) the `square-root'
of the field $\Psi_{XY}$ in (\ref{su10}); it is controlled by a
`$|\Psi_{\rm sp}|^4$' field theory identical in form to that for
$\Psi_{XY}$ in (\ref{su10}). Our purpose here was to search for
SU(2) symmetry in the theory of $\Psi_{\rm sp}$ and it is evident
that this is not possible in such a field theory which only has a
U(1) symmetry. The single spinon states with $S_z = \pm 1/2$ are
degenerate and so do form an SU(2) doublet; however, there is no
reason, in the presence of the $|\Psi_{\rm sp}|^4$ interaction,
for the S-matrices of the multi-spinon states to possess SU(2)
symmetry.

To further confirm this conclusion, let us look at states in the
vicinity of the F state and examine whether any of them can have
SU(2) symmetry. While the XY state shares a phase boundary with F
in Fig~\ref{phase6}, note that the I phase does not: this prevents
the XY+I state from being connected to the F state except across a
special multicritical point. The general reasons for the
separation of the I and F states were noted earlier at the end of
Section~\ref{sec:u12} in the discussion around (\ref{u113})---this
conclusion also holds in the presence of BC order: one can show by
a very similar argument that it is not possible for the F+BC and
the F+I states to share a second order phase boundary. As we
discussed in Section~\ref{sec:hn}, the XY+I state does posses a
line of SU(2) symmetry along which there is HN order. We can now
try to connect this to the F state in Fig~\ref{phase6}. We clearly
cannot proceed across either the F+I state or the XY state as
these state have anisotropic magnetic order which explicitly break
SU(2) symmetry. The only remaining possibility is to access the F
state directly across the multicritical point at $s_I=s_2 = 0$.
However this is also not plausible: if the multicritical point was
SU(2) symmetric, we would expect it to respond symmetrically to a
uniaxial anisotropy with respect to I and XY orders, and to be
flanked symmetrically either by I+F and XY+F states, or by I and
XY states. This is clearly not the case in Fig~\ref{phase6}.

So in agreement with the arguments in Section~\ref{sec:bo}, we
have found no plausible line of SU(2) symmetry in
Fig~\ref{phase6} which can extend into the F state. Consequently,
we have also not found a direct transition between the HN and F
states.

\section{Fractionalization with
SU(2) symmetry, and non-collinear spin correlations}
\label{planar}

This section will expand upon the theory of fractionalization in
two-dimensional spin systems originally presented in
Refs~\onlinecite{rsprl2,sr,rodolfo2,triangle,japan}. This
formulation will exhibit the intimate connection between
fractionalized phases and non-collinear spin correlations. Our
treatment here will also incorporate insights gained from the
recent analyses of Refs~\onlinecite{senthil,lannert,five}: in
particular our phase diagrams here contain a new phase (the HN+F
phase) not considered in the original treatment.

The treatment here will apply to frustrated $S=1/2$
antiferromagnets on the square lattice like those in (\ref{H}). In
particular, the model with first, second, and third neighbor
exchange (the so-called $J_1$-$J_2$-$J_3$ model) has magnetically
ground states with spiral spin order, and fractionalized
paramagnetic states could appear in its vicinity. Further, the
same general theory should also apply to magnetically ordered and
paramagnetic phases of other frustrated lattices in two
dimensions, including the triangular lattice. The case of the
triangular lattice was considered in
Refs.~\onlinecite{triangle,ross}: its three sublattice
magnetically ordered state can be viewed as the limiting case of
a spiral state on a distorted square lattice whose wavevector has
been pinned at a special value determined by triangular symmetry.

As in Refs~\onlinecite{rsprl2,sr}, we will present the theory
here using $CP^1$ degrees of freedom. In the body of the paper so
far, we have used the O(3) vector ${\bf n}_j$ to represent the
local orientation of the antiferromagnetic order. The close
relationship of this previous formulation to the $CP^1$ approach
is discussed in Appendix~\ref{a}: a reading of this appendix is a
prerequisite to the discussion in the present section. The
primary actor in $CP^1$ formulation is a complex spinor
$z_{j\alpha}$ ($\alpha = \uparrow, \downarrow$) of unit modulus
($\sum_{\alpha} |z_{j \alpha}|^2 = 1$) which is related to ${\bf
n}_j$ by (\ref{cp2}), which we reproduce here for completeness:
\begin{equation}
n_{ja} = z_{j\alpha}^{\ast} \sigma^a_{\alpha\beta} z_{j \beta}
\label{f1}
\end{equation}
with $a=x,y,z$ a spin index, and $\sigma^a$ the Pauli matrices.
This representation has the disadvantage of introducing redundant
degrees of freedom associated with the local compact U(1) gauge
symmetry $z_{j \alpha} \rightarrow e^{i \phi_j} z_{j \alpha}$.
However, as was also the case in the treatment of Shraiman and
Siggia~\cite{shsi} of doped antiferromagnets, it allows for an
especially simple theory of the appearance of non-collinear,
spiral spin correlations. Furthermore, the $z_{j\alpha}$ quanta
carry spin $S=1/2$, are therefore natural candidates for spinon
excitations in fractionalized paramagnetic phases.

As a prelude to writing down our model, we need to understand the
characterization of non-collinear spin correlations in terms of
the $z_{j \alpha}$. Consider the spin correlations between two
near neighbor sites $j=1,2$. In the context of models like
(\ref{sigma}) and (\ref{cp3}), the optimum configuration of ${\bf
n}_1$ and ${\bf n}_2$ is to have them parallel to each other:
such a state has $|z_{1 \alpha}^{\ast} z_{2 \alpha}|^2 = 1$, and
simultaneously $\epsilon_{\alpha\beta} z_{1 \alpha}^{\ast} z_{2
\alpha}=0$. We would like to an energetic preference for the ${\bf
n}$ vector to rotate by a non-zero angle in moving from site 1 to
site 2: this can be ensured by demanding that
$|\epsilon_{\alpha\beta} z_{1 \alpha}^{\ast} z_{2 \alpha}|^2 \neq
0 $, while $|z_{1 \alpha}^{\ast} z_{2 \alpha}|^2$ remains
non-zero. So, natural order parameters for non-collinear spin
correlations are the two fields
\begin{equation}
Q_{j\varrho} \sim \epsilon_{\alpha\beta} z_{j\alpha}
z_{j+\hat{\varrho},\beta}, \label{f2}
\end{equation}
where the index $\varrho$ will always extend over $\varrho = x,y$.
These fields are not gauge invariant and preserving gauge symmetry
will, of course, be essential in determining the structure of the
effective action. Notice also that in the continuum limit,
$Q_{\varrho} \sim \epsilon_{\alpha\beta} z_{\alpha}
\partial_{\varrho} z_{\beta}$, which is the spiral order parameter
used by Shraiman and Siggia~\cite{shsi} and in
Ref.~\onlinecite{sr}. The form of the order parameter chosen in
(\ref{f2}) corresponds to spin spirals along the $(1,0)$ and
$(0,1)$ directions of the square lattice. Clearly, it is possible
to have antiferromagnets in which the spiral correlations are
oriented along the $(1,\pm 1)$ or other directions: there is a
corresponding modification of the order parameter in these cases,
but we will not present this simple generalization explicitly
here.

To fully explore the phase diagram of our model, we find it
convenient to introduce explicitly the real, N\'{e}el, order
parameter vector ${\bf N}_j$. This is the O(3) symmetric
combination of the orders $\Psi_{XY}$ and $\Psi_{I}$ considered in
Section~\ref{sec:su2}. Clearly
\begin{equation}
N_{j a} \sim z_{j\alpha}^{\ast} \sigma^a_{\alpha\beta} z_{j
\beta}, \label{f3}
\end{equation}
where $a =x,y,z$. Also it is evident that ${\bf N}_{j} \sim {\bf
n}_{j}$, but we prefer to use an independent symbol here as the
physical context is different, and we do not impose a unit length
constraint on ${\bf N}_{j}$.

We can now write down the partition function for quantum
antiferromagnets with SU(2) symmetry. The expression here
generalizes $Z_{cp}$ in (\ref{cp3}) of Appendix~\ref{a} by
allowing for non-collinear spin correlations
\begin{eqnarray}
&& Z_{P} = \sum_{\{q_{\bar{\jmath}\mu}\}} \int_{0}^{2 \pi}
\prod_{j\mu} d A_{j \mu} \int \prod_{j \alpha} d z_{j \alpha}
\delta( |z_{j \alpha}|^2 - 1 ) \nonumber \\
&&\!\!\!\!\!\!\! \int \prod_{j\varrho} d Q_{j\varrho} \int
\prod_{ja} d {\bf N}_{j} \exp\left( - S_z - S_A - \widetilde{S}_B
-S_Q -S_N -S_w\right ) \!\!\!\!\!\!\!\!\!
\nonumber \\
&& S_Q = - \frac{1}{g_Q} \sum_{j \mu \varrho} Q_{j \varrho} e^{2 i
A_{j
\mu}} Q_{j+\hat{\mu},\varrho} + \mbox{c.c.} +\sum_{j\varrho} V_Q(|Q_{j\varrho}|^2) \nonumber \\
&& S_N = - \frac{1}{g_N} \sum_{j\mu} {\bf N}_{j} \cdot {\bf
N}_{j+\hat{\mu}}  + \sum_{j} V_N ({\bf N}_{j}^2) \nonumber \\
&&S_w = -w_Q \sum_{j\varrho} Q_{j\varrho}^{\ast}
\epsilon_{\alpha\beta} z_{j \alpha} z_{j+\hat{\varrho}, \beta}
e^{i A_{j\nu}} + \mbox{c.c.} \nonumber \\
&&~~~~~~~~~~~~~~~~-w_N \sum_j N_{ja} z_{j\alpha}^{\ast}
\sigma^a_{\alpha\beta} z_{j \beta} . \label{f4}
\end{eqnarray}
As stated earlier, all sums over $\varrho$ extend only over
$x,y$, while those over $\mu$ extends over $x,y,\tau$. The
expressions for the first three terms in the action were given
earlier in (\ref{cp3}); these are: $S_z$, the hopping term for the
$z_{\alpha}$, $S_A$, the Maxwell term for the {\em compact} U(1)
gauge field $A_{j \mu}$, and $\widetilde{S}_B$, the Berry phase
of the underlying $S=1/2$ spins. The symmetry of the problem
clearly permits a more anisotropic nearest neighbor coupling in
$S_{Q}$, but we have written down an isotropic form for notational
simplicity. Suitable factors of the exponential of the compact
U(1) gauge field $A_{j\mu}$ have been inserted to ensure that the
action is invariant under the U(1) gauge transformations
\begin{eqnarray}
z_{j \alpha} &\rightarrow& z_{j \alpha} e^{i \phi_j} \nonumber \\
Q_{j \varrho} &\rightarrow& Q_{j \varrho} e^{2 i \phi_j} \nonumber \\
A_{j\mu} &\rightarrow& A_{j \mu} - \Delta_{\mu} \phi_j \label{f5}
\end{eqnarray}
where $\phi_j$ represents an arbitrary phase. The potentials in
$S_Q$ and $S_N$ are given, as usual, by
\begin{equation}
V_{\sharp} (x) = \frac{1}{2} \left( s_{\sharp} x + u_{\sharp} x^2
\right) \label{f6}
\end{equation}
where $ \sharp \equiv Q,N$. The allowed cubic terms between the
fields have been explicitly written out in $S_w$. A number of
quartic terms coupling the modulus squared of all the fields are
also allowed but we have not written them out for simplicity.

The model $Z_P$ in (\ref{f4}) is our global theory for the phases
and phase transitions of quantum antiferromagnets. We believe it
offers a complete theory of the universal properties of $S=1/2$
antiferromagnets in two dimensions with SU(2) symmetry on a
variety of bi-partite and non-bi-partite lattices. Further, all
its phases have the obvious generalization to cases where SU(2)
symmetry is broken weakly by an easy-plane or easy-axis
anisotropy: HN order will transform to the corresponding XY or I
order.

Various symmetries have played a crucial role in determining the
structure of $Z_P$, and will obviously also be important in the
analysis of its phase diagram. The foremost among these is the
compact U(1) gauge invariance associated with (\ref{f5}). Along
with the gauge field $A_{j \mu}$, we have two ``matter'' fields:
the unit charge field $z_{j \alpha}$, and the charge 2 field $Q_{j
\varrho}$. The phases of such a gauge theory were discussed by
Fradkin and Shenker~\cite{fsh}, and we review the essential needed
results. For the pure gauge theory (without matter), the
weak-coupling (small $e^2$) regime is continuously connected to
the strong-coupling (large $e^2$) regime, and both of which
exhibit confinement of unit charges. In the presence of charge 1
matter fields, one might expect a separate ``Higgs'' phase where
the matter field has condensed: however, this charge 1 Higgs
regime is continuously connected to the confining phase, and there
are no intervening phase transitions (at least none associated
with the gauge theory---other symmetries could (and do) lead to
phase transitions). Upon adding the charge 2 fields, their
condensation does lead to a charge 2 Higgs phase, which is
separated from the remainder of the phase diagram by a phase
transition: unit charges are deconfined in this Higgs phase, while
they are confined elsewhere.

The second important symmetry is the global SU(2) invariance:
$z_{j\alpha}$ transforms like a spinor under this symmetry, while
${\bf N}_j$ transforms like a vector; all other fields are SU(2)
singlets. This symmetry is broken in phases with HN or P order,
and their properties will be discussed further in
Section~\ref{sec:mo} below.

Finally, $Z_P$ in invariant under the square lattice point group
$D_4$. These symmetries can also be broken, leading to new phases.
As we have seen in the previous sections of this paper, the Berry
phase term can induce BC order, which breaks $D_4$ symmetry. In a
region where $Q_{j \varrho}$ is suppressed (and so we can neglect
the presence of a charge 2 Higgs fields), the compact U(1) gauge
theory is always in a confining phase, and (as we noted above)
there can be no phase transitions associated with it;
nevertheless, Berry phases can induce a transition between the HN
and HN+BC phases (both of which are confining) associated with
$D_4$ symmetry breaking. In the region where $Q_{j \varrho}$ is
non-negligible, another type of $D_4$ symmetry-breaking becomes
possible. Notice from (\ref{f2}) that $Q_{j \varrho}$ transforms
like a spatial vector (the $E$ representation): it is {\em odd}
under reflections, and its components map into each other under
$90^{\circ}$ rotations. Consequently, the Higgs phase where $Q_{j
\varrho}$ is condensed can also break $D_4$ symmetry: we will see
that the paramagnetic F phase can also have ``bond nematic''
order\cite{sr,ssmodel}, associated with the choice of
condensation between $Q_{j x}$ and $Q_{j y}$. Note also that
although $Q_{j x} \rightarrow - Q_{j x}$ under a reflection along
the $y$ axis, condensation of $Q_{j x}$ does not imply loss of
this reflection symmetry: $Q_{j x}$ also carries a compact U(1)
gauge charge, and this change of sign can be undone by a gauge
transformation.

\begin{figure}
\epsfxsize=3in \centerline{\epsffile{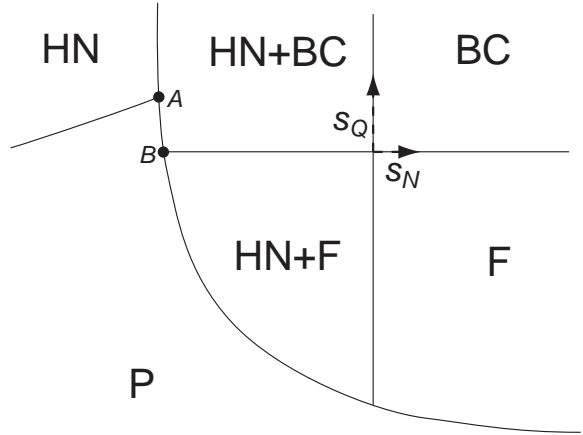}} \vspace{0.2in}
\caption{Section of the proposed global phase diagram of $Z_P$ in
(\protect\ref{f4}) as a function of $s_N$ and $s_Q$ for
moderately large values of $g$, $g_Q$ and $g_N$. The phases with
F order are likely to have ``bond-nematic'' order over a
substantial (and possibly) portion of their domain. Bond-nematic
order corresponds to a preferential polarization of the spin
correlations along the $x$ or $y$ axes, and spontaneously reduces
the lattice symmetry from $D_4$ (tetragonal) to $D_2$
(orthorhombic). The nature of the phase boundaries near the
multi-critical points $A$ and $B$ is uncertain, and a direct
second-order transition between the HN+F and HN states also
appears possible; this is discussed further in Appendix~\ref{h}.
Most of the P phase has coplanar spin order, but portions do have
non-coplanar: the latter order appears in the region contiguous
to the HN+F phase; the phase boundary between coplanar and
non-coplanar spin order is not shown. This phase boundary meets
that between the F and HN+F phases, and so  the point where the
P, HN and HN+F phases meet is tetracritical.}\label{phase7}
\end{figure}

\begin{figure}
\epsfxsize=3in \centerline{\epsffile{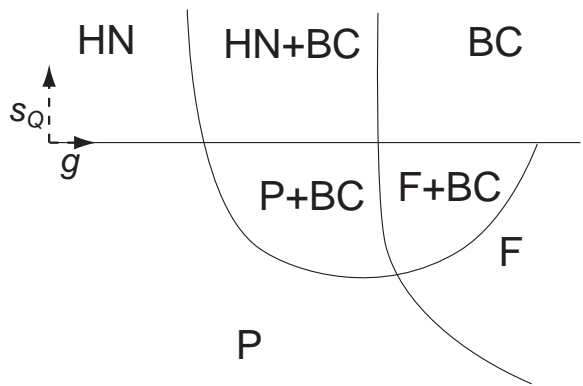}} \vspace{0.2in}
\caption{As in Fig~\protect\ref{phase7}, but as a function of $g$
and $s_Q$ for $s_N$ large and positive, and for moderately large
values of $g_Q$ and $g_N$. }\label{phase7a}
\end{figure}

Careful consideration of the above symmetries, and insights from
the analyses of the previous section, lead to sections of the
proposed phase diagram for $Z_P$ in Fig~\ref{phase7}
and~\ref{phase7a}.
We also present a summary of the characterizations of the various
phases in terms of the fields in $Z_P$ (and in $Z_P^{\prime}$
below) in Table~\ref{tp}.
\begin{table}
\begin{tabular}{c|c}
State & Description \cr \hline HN &
\begin{minipage}{2.8in} \vspace{0.07in} ${\bf N}_j$ is condensed,
$Q_{j \varrho}$ are massive, and
unit charges of the compact $U(1)$ gauge theory $Z_P$ are
confined. As this is continuously condensed to the charge-1-Higgs
phase, we can also view the $z_{\alpha}$ as condensed. This phase
cannot be described by the $Z_2$ gauge theory $Z_P^{\prime}$.
\vspace{0.07in}
\end{minipage}  \cr \hline
BC & \begin{minipage}{2.8in} \vspace{0.07in} ${\bf N}_j$, $Q_{j
\varrho}$ are massive, and unit charges of the compact $U(1)$
gauge theory $Z_P$ (or the $Z_2$ gauge theory $Z_P^{\prime}$) are
confined. The Berry phases induce BC order \vspace{0.07in}
\end{minipage}  \cr \hline
HN+BC & \begin{minipage}{2.8in} \vspace{0.07in} As in the HN
state, but the Berry phases induce BC order, and this phase also
appears in $Z_P^{\prime}$. \vspace{0.07in}
\end{minipage} \cr \hline
F & \begin{minipage}{2.8in} \vspace{0.07in} ${\bf N}_j$ and
$z_{\alpha}$ are massive, while condensation of $Q_{j \varrho}$
puts the compact U(1) gauge theory in its
charge-2-Higgs/deconfined phase; the $Z_2$ gauge theory is also
deconfined and the $z_{\alpha}$ quanta are liberated.
Bond-nematic order is likely to be present over a substantial
portion of the F state. \vspace{0.07in}
\end{minipage} \cr \hline
F+BC & \begin{minipage}{2.8in} \vspace{0.07in} As in the F state,
but BC order is also present. \vspace{0.07in}
\end{minipage} \cr \hline
HN+F & \begin{minipage}{2.8in} \vspace{0.07in} As in the F state,
but ${\bf N}_j$ is condensed. \vspace{0.07in}
\end{minipage} \cr \hline
P & \begin{minipage}{2.8in} \vspace{0.07in} ${\bf N}_j$, $Q_{j
\varrho}$ are condensed, and unit charges of the compact $U(1)$
gauge theory $Z_P$ (or the $Z_2$ gauge theory $Z_P^{\prime}$) are
confined; consequently, as in the HN state, we can also view the
$z_{\alpha}$ as condensed. \vspace{0.07in}
\end{minipage} \cr \hline
P+BC & \begin{minipage}{2.8in} \vspace{0.07in} As in the P state,
but BC order is also present. \vspace{0.07in}
\end{minipage}
\end{tabular}
\vspace{0.2in} \caption{Summary of all the phases in
Figs~\protect\ref{phase7} and~\protect\ref{phase7a} and their
physical properties.} \label{tp}
\end{table}
We will describe various portions of this global phase diagram in
the following subsections.

\subsection{Connection with $Z_2$ gauge theory}
\label{sec:z2}

The phases of the compact U(1) gauge theory noted above provide
the most complete description of the phase diagram of $Z_P$.
However, in some portions of this phase diagram a simpler
description may be provided by a $Z_2$ gauge
theory~\cite{rsprl2,sr,rodolfo2,japan}. This reduction will also
allow us to note the connection to the work of Senthil and
Fisher~\cite{senthil}. However, it is important to note that the
$Z_2$ theory below does {\em not} provide a complete description
of the phases and phase transitions in Figs~\ref{phase7}
and~\ref{phase7a}: we will argue below that it does not contain
the familiar N\'{e}el (HN) state, or its phase boundaries.

The reduction to the $Z_2$ gauge theory is most naturally
observed in the limit $r_Q \ll 0$. Here $Q_{j \varrho}$ is
expected to be large (at least on short scales), and by the Higgs
phenomenon, this will also suppress fluctuations in the flux of
$A_{j \mu}$. However, because $Q_{j \varrho}$ carries charge 2, a
gauge flux of $\pi$ is invisible to it, while the unit charge
fields $z_{j \alpha}$ do observe such a flux. To allow only such
fluxes, we can safely assume that $e^{i A_{j \mu}}$ is pinned to
$\pm 1$ by the large values of $Q_{j \varrho}$. So we write
\begin{equation}
e^{i A_{j \mu}} = s_{j,j+\hat{\mu}} \label{f7}
\end{equation}
where $s_{j,j+\hat{\mu}}=\pm 1$ is an Ising gauge field on the
links of the direct lattice; it is similar to the $Z_2$ gauge
field in (\ref{u11a}), and hence to that in
Ref.~\onlinecite{senthil}. Clearly the field $z_{j \alpha}$
carries unit $Z_2$ gauge charge, while ${\bf N}_j$ and $Q_{j
\varrho}$ are neutral.

Despite the quenching of the gauge charges associated with $Q_{j
\varrho}$, we cannot entirely neglect their fluctuations. This is
because we still have to account for their transformations under
the $D_4$ lattice symmetry and the associated symmetry breaking.
So we introduce additional Ising variables, $\pi_{jx}$ and
$\pi_{jy}$ by the correspondence
\begin{equation}
Q_{j \varrho} \sim \pi_{j \varrho}; \label{f8}
\end{equation}
$\pi_{j \varrho}$ also transforms under the $E$ representation of
$D_4$. Ordering transitions in the $\pi_{j \varrho}$ will signal
$D_4$ symmetry breaking with the appearance of bond-nematic
order. It is important to note that despite being Ising spin
variables, the $\pi_{j \varrho}$ are {\em neutral} under the
$Z_2$ gauge field $s_{j,j+\hat{\mu}}$.

With the above parameterizations, we can reduce $Z_P$ in
(\ref{f4}) to the following $Z_2$ gauge theory
\begin{eqnarray}
&& Z_{P}^{\prime} = \sum_{\{s_{j,j+\hat{\mu}}=\pm 1 \}}
\sum_{\{\pi_{j \varrho} = \pm 1\}} \int \prod_{j \alpha} d z_{j
\alpha} \delta( |z_{j \alpha}|^2 - 1 ) \nonumber \\ && ~~~~~\int
\prod_{ja} d {\bf N}_{j}
\exp\left( - S_z^{\prime}  -S_N -S_w^{\prime}- S_s - S_{\pi}\right ) \nonumber \\
&& S_z^{\prime} = - \frac{1}{g} \sum_{j\mu} z^{\ast}_{j \alpha}
s_{j,j+\hat{\mu}} z_{j+\hat{\mu},\alpha} + \mbox{c.c.} \nonumber \\
&& S_w^{\prime} = -w_Q \sum_{j\varrho} \pi_{j \varrho}
s_{j,j+\varrho} \epsilon_{\alpha\beta} z_{j \alpha}
z_{j+\hat{\varrho}, \beta} + \mbox{c.c.} \nonumber
\\&&~~~~~~~~~~~-w_N \sum_j N_{ja} z_{j\alpha}^{\ast}
\sigma^a_{\alpha\beta} z_{j
\beta} \nonumber \\
&& S_s = - K \sum_{\Box} \prod_{\Box} s_{j,j+\hat{\mu}} + i
\frac{\pi}{2} \sum_j (1 - s_{j,j+\hat{\tau}}) \nonumber \\
&& S_{\pi} =  -\frac{1}{g_{Q}^{\prime}} \sum_{j\mu\varrho}
\pi_{j\varrho} \pi_{j+\hat{\mu},\varrho} + v_{\pi} \sum_{j\mu}
\left(\pi_{jx} \pi_{j+\hat{\mu},x}\right)\left(\pi_{jy}
\pi_{j+\hat{\mu},y}\right) ,\nonumber \\
\label{f9}
\end{eqnarray}
where $S_N$ is as defined in (\ref{f4}), the coupling
$g_{Q}^{\prime}$ is a consequence of $g_{Q}$ in $Z_P$, and $K$
derives from the Maxwell term (proportional to $1/e^2$) in
(\ref{cp3}). We have also added a term proportional to $v_{\pi}$,
a repulsive interaction between the energies of the Ising
variables $\pi_{jx}$ and $\pi_{jy}$: this ensures that when these
fields order, the system will choose between orderings of one or
the other, and the bond-nematic or spiral order will be polarized
along the (1,0) or (0,1) directions. As we noted earlier, a
different choice of variables should be made for polarization of
the spiral order in other directions.

The simplest application of the $Z_2$ gauge theory,
$Z_P^{\prime}$, is in the phases in which the SU(2) symmetry is
preserved. Here we may integrate out the $z_{j\alpha}$ and ${\bf
N}_j$ fields and the action simply becomes $S_s + S_{\pi}$, along
with additional couplings between $s_{j,j+\hat{\mu}}$ and
$\pi_{j\varrho}$ which preserve gauge invariance and $D_4$
symmetry. The model $S_s$ is, of course, identical to the $Z_2$
gauge theory obtained in the $g \rightarrow \infty$ limit of
(\ref{u11a}) and was studied in Section~\ref{sec:u1}; it is dual
to the fully frustrated Ising model in (\ref{u110}). It has a
confining phase at small $K$ with BC order, and a deconfined
phase at large $K$ in which integer $Z_2$ charges (like the
$z_{\alpha}$) are liberated--this is the F phase. So the present
model describes the F and BC phases appearing for large $r_N$ in
Fig~\ref{phase7} and for large $g$ in Fig~\ref{phase7a}. An
extension of this model with more complicated couplings between
the Ising gauge spins could also lead to the F+BC phase in
Fig~\ref{phase7a}. The ordering of the $\pi_{j \varrho}$
variables leads to bond-nematic order: such order is already
present in the BC phase and is therefore not significant there.
On the other hand, in F phase the bond-nematic yields a
preferential polarization of spin correlations along the $x$ or
$y$ axes, and breaks the $D_4$ symmetry down to $D_2$. In
principle, the bond-nematic ordering and the deconfinement
transitions are independent and can occur at distinct positions.
However, in the large $N$ theory of Refs.~\onlinecite{rsprl2,sr}
these two transitions co-incided and the F state had bond-nematic
order over its entire domain; we expect here that bond-nematic
order will be present over a substantial, if not complete,
portion of the phases with F order.

A similar description of these two non-magnetic phases can also
be obtained starting from the compact $U(1)$ gauge theory in
(\ref{f4}). Again, we integrate out the $z_{j\alpha}$ and ${\bf
N}_j$ and end up with the theory of a compact U(1) gauge field
coupled to a charge 2 scalar. The latter theory was described in
Ref.~\onlinecite{japan} and possesses a transition between BC and
F phases identical to that described by the $Z_2$ gauge theory
$S_s$.

We now attempt to extend the above study of the $Z_2$ gauge
theory, $Z_P^{\prime}$, to the remaining phases in
Figs~\ref{phase7} and~\ref{phase7a} with broken SU(2) symmetry,
and will quickly run into a problem; we will have to return to
the original compact U(1) gauge theory, $Z_P$, to rectify the
situation. As we will see in Section~\ref{sec:mo}, all the
magnetically ordered phases have $\langle {\bf N}_j \rangle \neq
0$. So let us assume that $\langle {\bf N} \rangle = N_0
(0,0,1)$. Gauge-invariant fluctuations of ${\bf N}$ about this
direction will lead to the two required spin-wave modes. We now
examine the gauge-dependent portions to check for the presence of
other orderings. In the presence of ${\bf N} = (0,0,1) N_0$ and
$w_N
>0$, we see from (\ref{f9}) that among the $CP^1$ fields, the
$z_{j\uparrow}$ fluctuations will be preferred from the
$z_{j\downarrow}$ fluctuations; this corresponds to an easy-axis
anisotropy to the $CP^1$ model, in contrast to the easy-plane
anisotropy considered in Appendix~\ref{f}. So we can safely
neglect $z_{j\downarrow}$, and model the other component by a
phase variable $z_{j\uparrow} = e^{i \theta_{j\uparrow}}$. The
resulting effective action for the angle $\theta_{j\uparrow}$ and
the Ising gauge field $s_{j,j+\hat{\mu}}$ in (\ref{f9}) is
essentially identical to the model (\ref{u11a}) which was studied
in great detail in Section~\ref{sec:u1} (the $\pi_{j\varrho}$
Ising fields are innocuous spectators controlling the
bond-nematic order). So we may directly apply the earlier
results, especially the phase diagram in Fig~\ref{phase3} to the
present situation. Because there is background N\'{e}el order
with $\langle {\bf N} \rangle \neq 0$, the physical
interpretation of the phases is now different. In particular, it
is obvious that the BC phase in Fig~\ref{phase3} corresponds to
the HN+BC phase in Figs~\ref{phase7} and~\ref{phase7a}, and the F
phase of Fig~\ref{phase3} maps to the HN+F phase of
Fig~\ref{phase7}. The interpretation of the XY phase in
Fig~\ref{phase3} is slightly more subtle: this state has a
gapless Goldstone mode associated with the XY order, and in the
present situation this must be combined with the two spin-wave
modes of the ${\bf N}$ order to yield a total of 3 gapless modes.
This can only correspond to the P phase in Figs~\ref{phase7}
and~\ref{phase7a}. As we will see below in Section~\ref{sec:hnf}
the ${\bf N}$ order also gets spatially modulated in the P phase,
and this is an effect that the present simple analysis misses.
Similarly the XY+BC phase found in Fig~\ref{phase5} maps here
onto the P+BC phase of Fig~\ref{phase7a}.

The most striking feature of the above analysis of phases with
broken SU(2) symmetry of the $Z_2$ gauge theory, $Z_P^{\prime}$ in
(\ref{f9}), is that we achieved a description of all the phases
and phase transitions in Figs~\ref{phase7} and~\ref{phase7a}
except those associated with the HN phase. In the above mapping
of its properties to $Z_{\rm U(1)}$ in (\ref{u11a}), the HN phase
requires a phase of $Z_{\rm U(1)}$ with no broken symmetries and
no fractionalization or topological order. No such phase was found
in entire analysis of Section~\ref{sec:u1}, and so we conclude
that $Z_P^{\prime}$ in (\ref{f9}) does not possess a HN phase. So
we are in the slightly embarrassing position of having a
sophisticated theory for the exotic states in Figs~\ref{phase7}
and~\ref{phase7a}, but are not able to capture the simple and
familiar N\'{e}el state. This situation is rectified by returning
to the original compact U(1) gauge theory $Z_P$ in (\ref{f4}). As
we will show in Section~\ref{sec:mp} and Appendix~\ref{h}, the
compact U(1) gauge boson (unlike the $Z_2$ gauge field) is able
to quench the $\theta_{\uparrow}$ fluctuations without breaking
any symmetries or inducing additional gapless modes, as is
required in the HN state.

\subsection{Phases with magnetic order}
\label{sec:mo}

Now we describe more completely the phases in Figs~\ref{phase7}
and~\ref{phase7a} that break the spin rotation symmetry by a
spontaneous magnetic polarization: such states will possess
gapless spin-wave excitations. The phases in
Sections~\ref{sec:phn} and~\ref{sec:hnf} below can also co-exist
with BC order, and this will (obviously) not change the structure
of the gapless modes.

\subsubsection{HN phase}
\label{sec:phn}

This familiar N\'{e}el state is found in the regime with $r_Q >
0$ and $r_N < 0$. So the $Q_{\varrho}$ fields are massive and can
be neglected, while the ${\bf N}$ field is condensed. The
remaining degrees of freedom, $A_{\mu}$ and $z_{\alpha}$, obey
the standard action of 2+1 dimensional compact U(1) gauge field
coupled to a complex scalar of charge 1, along with an additional
Berry phase term. This latter theory is expected \cite{fsh} to
always be in a confining phase: the Higgs phase where the unit
charge complex scalar $z_{\alpha}$ is condensed is continuously
connected to the confining phase of the pure compact U(1) gauge
theory. So the present phase with $Q_{\varrho}$ massive and ${\bf
N}$ condensed, also has $z_{\alpha}$ condensed. We will present a
more complete analysis of this important phenomenon in
Section~\ref{sec:mp} and Appendix~\ref{h}: we will find that the
Berry phases can also induce a region with co-existing BC order.

This starting point of condensation of ${\bf N}$ and $z_{\alpha}$
provides a straightforward description of the fluctuations in the
HN phase. We perform a standard Gaussian fluctuations analysis
about a mean-field saddle point at which ${\bf N}$ and
$z_{\alpha}$ are non-zero, and all other fields are zero. The
fluctuations of the gauge field $A_{\mu}$ are `Higgsed' by the
$z_{\alpha}$ condensate, and are consequently massive. However,
there are two gapless spin-wave eigenmodes which arise from
eigenvectors made up the transverse components of ${\bf N}$ and
$z_{\alpha}$. These are just the properties expected in the
familiar collinear N\'{e}el state.

\subsubsection{P phase}
\label{sec:p}

\begin{figure}
\epsfxsize=3.5in \centerline{\epsffile{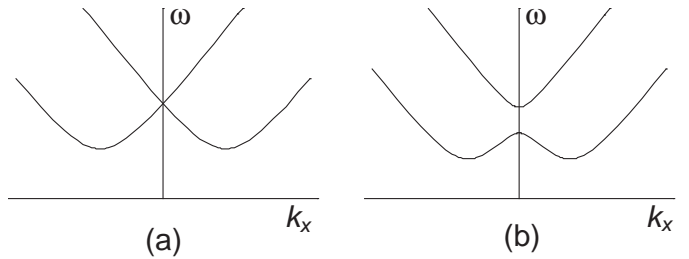}}
\vspace{0.2in} \caption{Plot of the spinon dispersion in (a) the
F phase (Eqn (\protect\ref{f11})), and (b) the HN+F phase (Eqn
(\protect\ref{f13})). For larger $N_0$, the latter has minima
only at $k=0$. There is one complex (or two real) bosonic mode at
each frequency.}\label{dispersion}
\end{figure}

Let us approach the P phase from the F phase, where as we noted
in Table~\ref{tp}, the $Q_{j\varrho}$ are condensed. For
simplicity, let us assume that $Q_{j x}$ is condensed while $Q_{j
y}$ is zero. By the Higgs phenomenon, the $A_{\mu}$ fluctuations
are then quenched and can be neglected. From $Z_P$ we see then
that long-wavelength action controlling the spectrum of the
$z_{\alpha}$ spinons is then controlled by the action
\begin{eqnarray}
&& S_z = \int d^2 x d \tau \Big[ |\partial_{\mu} z_{\alpha}|^2 +
\Delta^2 |z_{\alpha}|^2 \nonumber \\
&&~~~~~~~~~~~~~~~~~~~- w_Q Q_x \epsilon_{\alpha\beta}z_{\alpha}
\partial_x z_{\beta} + \mbox{c.c.} \Bigr]; \label{f10}
\end{eqnarray}
we have relaxed the length-constraint on the $z_{\alpha}$ and
replaced it by a quadratic mass term which induces a spin gap
$\Delta$. Note that the action for spinons in (\ref{f10}) is
explicitly SU(2) invariant, in contrast to the discussion of the
F phase in Section~\ref{sec:sufrac}. It is a simple matter to
diagonalize the quadratic form in (\ref{f10}): we obtain two
complex modes with dispersion
\begin{equation}
\omega = \left( k^2 + \Delta^2 \pm w_Q |k_x Q_x| \right)^{1/2} ;
\label{f11}
\end{equation}
there are two degenerate real modes at each frequency. This
dispersion relation is sketched in Fig~\ref{dispersion}a; note
that it always has minima at non-zero wavevectors $k = \pm k_{\rm
min} \equiv \pm ( w_Q |Q_x|/2, 0)$ and the minimum energy is
$\left( \Delta^2 - w_Q^2 |Q_x|^2 /4 \right)^{1/2}$.
For large enough $Q_x$ this minimum energy will vanish, and the
$z_{\alpha}$ will condense at the wavevectors $\pm k_{\rm min}$.
By a determination of the eigenvectors we see that the most
general condensate has the spatial dependence (assuming, without
loss of generality, that $Q_x$ is real)
\begin{equation}
z_{\alpha} = \left( \begin{array}{cc} d_1 & i d_2 \\
i d_2^{\ast} & d_1^{\ast} \end{array} \right)
\left(\begin{array}{c} e^{i k_{\rm min} x} \\ -i e^{-ik_{\rm min}
x} \end{array}\right), \label{f11a}
\end{equation}
where $d_{1,2}$ are complex numbers, with only $|d_1|^2 + |d_2|^2$
determined by higher order terms in the action. We should now
compute the staggered spin polarization by inserting (\ref{f11a})
into (\ref{f3}). This is simplified by noting that the first
factor on the right hand side is simply proportional to a general
SU(2) matrix, and will only lead to an overall rotation in ${\bf
N}$. So up to an arbitrary O(3) rotation, we can set $d_1 = 1$,
$d_2=0$, and then the spin-polarization in the condensed phase is
seen to have the form
\begin{equation}
{\bf N} \propto (\sin(2 k_{\rm min} x), -\cos(2 k_{\rm min} x),
0); \label{f11b}
\end{equation}
this is an incommensurate coplanar spiral in which ${\bf N}$
precesses at the wavevector $k = 2 k_{\rm min}$, as is expected in
the P state (recall that for the underlying spins, this wavevector
is measured relative to $(\pi,\pi)$). The critical properties of
the transition at which the $z_{\alpha}$ condense (which is
expected to have O(4) symmetry), and that of fluctuations in the
P states can be addressed as in Ref.~\onlinecite{book} and will
not be discussed further here. We merely note that the $A_{\mu}$
gauge fluctuations are quenched across the transition, and so the
theory can be expressed in terms of the $z_{\alpha}$ alone; the
$Q_{\varrho}$ are also not critical, while ${\bf N}$ is pinned to
the $z_{\alpha}$ via (\ref{f3}).

\subsubsection{HN+F phase}
\label{sec:hnf}

The spectrum of spin excitations in the HN+F phase can be analyzed
as in Section~\ref{sec:p} for the F phase, but we now have to
condense ${\bf N}$. Let us assume, as in Section~\ref{sec:z2},
that $\langle {\bf N} \rangle = N_0 (0,0,1)$. Fluctuations of
${\bf N}$ about this state will lead, as usual, to the two
familiar gapless spin-wave modes. For the gapped spinon states we
use the same strategy as that leading to (\ref{f10}); the
presence of the ${\bf N}$ condensate now modifies this to
\begin{eqnarray}
S_z &=& \int d^2 x d \tau \left[ |\partial_{\mu} z_{\alpha}|^2 +
\Delta^2 |z_{\alpha}|^2 - w_Q Q_x
\epsilon_{\alpha\beta}z_{\alpha} \partial_x z_{\beta} +
\mbox{c.c.}\right.\nonumber\\
&~&~~~~~~~~~~~~~~~~\left. - w_N N_0 (|z_{\uparrow}|^2 -
|z_{\downarrow}|^2) \right]. \label{f12}
\end{eqnarray}
This can now be diagonalized, and we again find two complex modes
with dispersion
\begin{equation}
\omega = \left( k^2 + \Delta^2 \pm \sqrt{ (w_Q k_x Q_x)^2 + (w_N
N_0)^2} \right)^{1/2} . \label{f13}
\end{equation}
These dispersions are sketched in Fig~\ref{dispersion}b. Unlike
(\ref{f11}), the expression (\ref{f13}) can either have minima at
non-zero wavevectors or only at zero wavevector; the latter
applies for $w_N N_0 > w_Q^2 |Q_x|^2 /2$. Although each frequency
has doubly degenerate real modes, there is now no SU(2) symmetry
in the S-matrices that describes the scattering of these modes.
Condensation of the spinons at the non-zero wavevector minima of
the dispersion (in the case where they exist) will lead again to
the P state. Analysis corresponding to that in (\ref{f11a}) and
(\ref{f11b}) now shows that the ${\bf N}$ spin polarization is
{\em non-coplanar} in the P phase. Further, the ${\bf N}$
condensate has components at wavevectors at $k = \pm 2k_{\rm
min}$ and at $k=0$ (note again that for the underlying spins,
these wavevectors are measured relative to $(\pi,\pi)$).
Condensation of the spinons at the zero wavevector minimum (if it
exists) leads to the HN state which has a ${\bf N}$ condensate
only at $k=0$.

\subsection{Multicritical points}
\label{sec:mp}

The discussion in the previous subsections, combined with the
analyses in Section~\ref{sec:u1} has given a complete description
of the phases in Figs~\ref{phase7} and~\ref{phase7a}, and of most
the boundaries between them. It remains to address the nature of
some of the multicritical points.

The tetracritical points in Figs~\ref{phase7} and ~\ref{phase7a}
all appear to be fairly conventional. As an example, consider the
point where the four phases, BC, HN+BC, HN+F, and F meet in
Fig~\ref{phase7}. The transitions in the vertical direction are
described by the BC order parameter associated with the frustrated
Ising model in (\ref{u110}), while that in the vertical direction
is controlled by the O(3) $\varphi^4$ field theory in (\ref{su3}).
The two order parameters have only an quartic energy-energy
coupling involving quadratic terms which are respectively
invariant under $D_4$ and O(3) symmetries.

It remains to sort out the physics near the multicritical points
A and B in Fig~\ref{phase7}. This appears to be a problem of
great complexity. However, some progress can be made by using the
observation that all four phases in the vicinity of A and B have
$\langle {\bf N} \rangle \neq 0$. This permits a description in
terms of a simplified compact U(1) gauge theory model which is
amenable to a complete duality transformation, and which does not
suffer from the problem of $Z_2$ gauge theory $Z_P^{\prime}$ of
omitting the HN phase. Details of the analysis along these lines
are provided in Appendix~\ref{h}.

\section{Conclusions}
 \label{conc}

This paper has presented an extensive study of the paramagnetic
states of two-dimensional, $S=1/2$, quantum antiferromagnets with
both U(1) and SU(2) spin rotation symmetries. We have focused on
two central properties that characterize such states:
bond-centered charge (BC) order ({\em e.g.} spin-Peierls order)
and fractionalization (F).

Our point of departure was the `minimal model' of $S=1/2$ spin
systems with local antiferromagnetic spin correlations: the O(3)
non-linear sigma model with Berry phases. We were strongly
influenced in this choice by the remarkable success of such an
approach in producing a theory of one-dimensional antiferromagnets
(reviewed and extended in Section~\ref{sec:1d}), with all results
in complete agreement with those obtained by bosonization methods
and also with the solution of a variety of integrable models. We
wrote a lattice version of the minimal model in two dimensions,
the partition function $Z$ in (\ref{z}), and
Sections~\ref{sec:bo}-\ref{sec:u1} were devoted, directly or
indirectly, to addressing its properties. We argued in
Section~\ref{sec:bo} that the strong-coupling limit of the
SU(2)-invariant $Z$ had BC order, while that of U(1)-invariant
systems was shown in Sections~\ref{sec:bo} and~\ref{sec:u1} to
also allow F states. We also discussed, in
Sections~\ref{sec:hnbo} and~\ref{sec:u1}, how the familiar
magnetically ordered N\'{e}el (HN) state was connected to the BC
state in systems with SU(2) symmetry: we presented a number of
independent arguments which supported an intermediate
co-existence (HN+BC) state.

We also presented an in-depth discussion of a number of routes to
fractionalization that have appeared in the literature
\cite{rsprl2,sr,wenold,senthil,five}. These can be divided, on
physical grounds, into at least three distinct classes which we
review below. Near the transition to magnetically ordered states
(or to superconducting states in class (B) below), these routes
produce fractionalized states with clearly distinct low energy
spectra. We do not address the more subtle question of whether
ground states in distinct classes can be connected to each other
via some path in parameter space without an intervening phase
transition---there can be considerable rearrangement between the
excited states so that one can connect states with very different
low energy structures without a singular change in the ground
state. The routes are:
\newline\newline
(A) In systems with a $U(1)$ spin symmetry, we approach the F
state from a magnetically ordered state with XY order, which we
view as `superfluidity' in the $S_z=1$ spin-flip boson
$b^{\dagger} = \hat{S}_{+}$ (see (\ref{o14})). This superfluidity
can be destroyed by condensation of vortices in the phase of
$b^{\dagger}$, and the F phase is produced~\cite{senthil,five} by
a preferential condensation of {\em double vortices}, $\Phi$;
near its transition to the XY phase, the F phase has a massive
$S_z=1/2$ spinon excitation described by the complex, bosonic
field operator $\Psi_{\rm sp}$, as was discussed in
Section~\ref{sec:sufrac}. We have argued in Section~\ref{sec:u1}
that this route to fractionalization does not survive the
restoration of SU(2) spin symmetry.
\newline\newline
(B) The second route~\cite{bfn,senthil} is technically similar to
(A), but has a very different physical interpretation, and applies
to systems with SU(2) symmetry. The boson, $b^{\dagger}$ is
re-interpreted as the creation operator for Cooper pairs, and the
XY phase is now a true superconductor. The condensation of double
vortices (with flux $hc/e$) then leads to a fractionalized
insulator with $S=1/2$ fermionic spinons (which are possibly
gapless) . At half-filling, this route requires proximity to
superconducting states, and so is only likely in systems with
strong charge fluctuations which can allow the Mott charge gap to
collapse.
\newline\newline
(C) The third route~\cite{rsprl2} was discussed in
Section~\ref{planar}. Non-collinear spin correlations were shown
to liberate bosonic $S=1/2$ spinons described by the actions
(\ref{f10}) and (\ref{f12}). Note that (\ref{f10}) has an
explicit SU(2) symmetry, and the number of spinon excitations is
doubled with respect to (A).

We now use the perspective of the above `classification' to
present a critical review other recent studies of $S=1/2$
antiferromagnets in two dimensions.

The Paris group~\cite{claire1,claire2,claire3,claire4,claire5} has
performed large scale numerical studies on a variety of lattices.
They have found examples of BC states, in proximity to
magnetically ordered states with collinear order, in keeping with
the discussion in Sections~\ref{sec:bo} and~\ref{sec:hnbo}. On the
analog of the pyrochlore lattice in two dimensions, they have
recently found~\cite{claire5} plaquette BC order, in accordance
with the predictions of the sine-Gordon model of
Appendix~\ref{sgm}~\cite{ssmodel} and the quantum dimer
model~\cite{sondhi2}. They also find good candidates for
fractionalized states with a well-developed gap to all
excitations: these are in regimes where the classical ground state
is non-collinear, and so we propose that these states are in class
(C). The case of kagome lattice remains mysterious as it has a
large density of very low energy singlet excitations.

We have already mentioned recent numerical
evidence~\cite{sushkov1,leiden} for BC order in frustrated square
lattice antiferromagnets with first and second neighbor exchange
interactions (the $J_1$-$J_2$ model). Capriotti and Sorella
originally claimed~\cite{sorella1} plaquette BC order in the
$J_1$-$J_2$ model, but have recently argued~\cite{sorella2} in
favor of a state with no broken symmetries near $J_2 /J_1 = 0.5$.
Our studies here and the earlier work~\cite{rsprl2,sr} do not, in
principle, rule out a F state in the $J_1$-$J_2$ model, but do
require a BC state to intervene (or, more exotically a HN+F
state) as one moves towards the HN state at smaller $J_2$. The
variational approach in Ref.~\onlinecite{sorella2} does not work
well in the region with HN order; as in our theory the BC order is
intimately connected to fluctuating HN order, we suspect that the
approach of Ref.~\onlinecite{sorella2} is also insensitive to the
development of BC correlations.

Moessner, Sondhi, and collaborators~\cite{sondhi1,sondhi2} have
examined generalizations of the quantum dimer model~\cite{qd}
(discussed in Appendix~\ref{qdm}) to a variety of lattices. They
present very convincing evidence for BC phases on bipartite
lattices, and for an F state on the triangular lattice; these
results are in general accord with the discussion of the present
paper and the route (C) to fractionalization. The dimer model
yields an appealing picture of the F state as Anderson's RVB
phase~\cite{pwa}, and so this agreement suggests that there is a
natural evolution from deconfined phases (C) with appreciable
non-collinear spin correlations to the short-range spin singlet
correlations of the RVB state.

Kashima and Imada~\cite{imada} have examined a Hubbard model with
first and second neighbor hopping near the metal-insulator
transition. At large enough Hubbard repulsion, $U$, this model
reduces to the $J_1$-$J_2$ antiferromagnet, and so the
corresponding region could have BC order. However the Hubbard
model also induces additional ring-exchange terms at smaller $U$,
and this should increase the tendency towards F states of class
(C). Nearer the metal-insulator transition, the strong charge
fluctuations could induce F states in class (B).

The striking experimental evidence~\cite{radu} for a F state in
Cs$_2$CuCl$_4$ appears to place it very naturally in class (C),
as has been discussed by Chung, Marston, and
McKenzie~\cite{ross}: the neutron scattering measurements show
clear evidence for incommensurate, spiral spin correlations.

It has recently become possible~\cite{troyer} to perform
large-scale numerical studies of a class of frustrated $S=1/2$
antiferromagnets with a U(1) spin symmetry. It would be very
useful to study these in great detail, and to compare the phase
diagrams with those in Section~\ref{sec:u1}.

Starykh {\em et al.}~\cite{starykh} and Moessner {\em et
al.}~\cite{sondhi2} have recently proposed a most interesting
two-dimensional paramagnetic phase which does not fall into any
of the classes discussed here: they have found an example of a
``sliding Luttinger liquid'' (``sliding ice~\cite{sondhi2}''), in
which the low energy excitations are well described by an
infinite set of essentially decoupled one-dimensional
antiferromagnets in the TL phase of Figs~\ref{figh}
and~\ref{figsg}. However, it remains to be seen if such a phase
is stable in a parameter regime with full SU(2) spin symmetry.

Zaanen and collaborators~\cite{jan} have introduced a $Z_2$ gauge
theory of ``sublattice parity order'' in the paramagnetic states
of a class of doped antiferromagnets. Their work is in a spirit
similar to (C) but differs in important aspects: the paramagnetic
state has incommensurate spin correlations, but the spins remain
collinear; further, there do not appear to be any neutral $S=1/2$
excitations.

Rantner and Wen~\cite{rantner} and Wen~\cite{wennew} have recently
provided a comprehensive classification of ``spin-liquid'' ground
states in two dimensions. Many of these are fractionalized states
which fall into our classes (B) and (C). However, they also
propose an ``algebraic spin liquid'' (related to the earlier
``flux phase''~\cite{affmar}) in which there is a gapless U(1)
gauge boson strongly interacting with gapless fermionic spinons.
We are skeptical about the stability of such a state. The
original analysis~\cite{rantner} dealt with a field theory of a
non-compact U(1) gauge field, whereas a microscopic derivation
always yields a compact U(1) field. The latter possesses
instantons which have been the center of attention in our
analysis here, and which have been the driving force behind the
appearance of BC order. Rantner and Wen~\cite{rantner} neglected
the instantons, whereas Wen~\cite{wennew} argued for the
irrelevance of the operator creating a {\em single} instanton
under suitable conditions. However, this irrelevance is not a
sufficient criterion for neglecting the effects of a {\em finite
density} of instantons. The latter effects can be addressed by
the Kosterlitz recursion relations~\cite{kost}, which examine the
renormalization of the instanton action by integrating out a a
tightly bound instanton/anti-instanton dipole. In spacetime
dimensions $D>2$ such renormalizations are disruptive, and cannot
be simply absorbed into an innocuous renormalization of a
``stiffness'' (as is the case in $D=2$ in the superfluid phase of
the Kosterlitz-Thouless transition). A number of models of the
statistical mechanics of ``charges'' in $D=3$ have been examined
in the literature (with interactions between opposite charges
separated by spacetime distance $R$  behaving like $\sim -1/|R|$
(Ref.~\onlinecite{kost}), $\sim \ln (|R|)$ (Ref.~\onlinecite{gm}),
and $\sim |R|$ (Ref.~\onlinecite{drn})), and all results are the
same: there is a strong renormalization induced by the dipoles,
and this always liberates the charges at large scales; this
liberation includes the regime where a naive analysis  of the
single vortex fugacity suggests that unbound charges are
suppressed. The instanton gas is then in the Debye-H\"{u}ckel
plasma screening phase, and there is no phase of confined charges
for $D>2$. It is this screened plasma of instantons which is
responsible for the BC order and confined spinons we have studied
here, and we believe it will also disrupt the algebraic spin
liquid.

\acknowledgments We thank E.~Demler, M.~P.~A. Fisher, S.~Girvin,
B.~I.~Halperin, P.~Lecheminant, S.~L.~Sondhi, O.~Starykh, and
O.~Sushkov for helpful discussions. We are especially grateful to
T.~Senthil for numerous insightful remarks. This research was
supported by US NSF Grant DMR 0098226.


\appendix

\section{$CP^1$ formulation}
\label{a}

We discuss here an alternative lattice model for the ground
states of quantum antiferromagnets in two dimensions. It is
closely related to the model $Z$ in (\ref{z}), which was in turn
derived from the coherent state path integral of quantum spin
systems. The approach here is closer to the model of
Ref.~\onlinecite{rodolfo}: it has the advantage of allowing a
simple generalization which accounts for incommensurate spin
correlations and fractionalized phases in systems with SU(2) spin
symmetry (discussed in Section~\ref{planar}), and of having a
simpler large $g$ limit. Its main disadvantage is that it
enlarges the number of degrees of freedom on the lattice scale,
and the connection to the underlying quantum spin Hamiltonian is
not as direct.

The central actors in this formulation are complex spinors $z_{j
\alpha}$ ($\alpha=\uparrow, \downarrow$) on the sites of the
direct spacetime lattice. These obey the constraint
\begin{equation}
\sum_{\alpha} |z_{j \alpha}|^2 = 1 \label{cp1}
\end{equation}
on every site, and are related to the ${\bf n}_j$ by
\begin{equation}
n_{ja} = z_{j\alpha}^{\ast} \sigma^a_{\alpha\beta} z_{j \beta}
\label{cp2}
\end{equation}
where $a=x,y,z$ is a spin index, and the $\sigma^a$ are the Pauli
matrices. Trading in the ${\bf n}_j$ for the $z_{j \alpha}$
introduces a compact U(1) gauge redundancy under which $z_{j
\alpha} \rightarrow z_{j \alpha} e^{i \phi_j}$, with $\phi_j$
arbitrary. This is closely related to (\ref{gauge}), but will now
be an explicitly realized in every term in the Hamiltonian.

A companion player in the present formulation is a compact $U(1)$
gauge field $A_{j \mu}$. This transforms under the above gauge
transformation in the obvious manner $A_{j \mu} \rightarrow A_{j
\mu} - \Delta_{\mu} \phi_j$, and the action is invariant under
$A_{j \mu} \rightarrow A_{j \mu} + 2 \pi$. As we will see below,
$A_{j \mu}$ is closely related to ${\cal A}_{j \mu}$.

We can now write down the new formulation of the partition
function:
\begin{eqnarray}
Z_{cp} &=& \sum_{\{q_{\bar{\jmath}\mu}\}} \int_{0}^{2 \pi}
\prod_{j, \mu} d A_{j \mu} \int \prod_{j \alpha} d z_{j \alpha}
\delta( |z_{j \alpha}|^2 - 1 )\nonumber \\
&~&~~~~\times
\exp\left( - S_z - S_A - \widetilde{S}_B \right ) \nonumber \\
S_z &=& - \frac{1}{g} \sum_{j, \mu} z^{\ast}_{j \alpha} e^{i A_{j
\mu}} z_{j+\hat{\mu},\alpha} + \mbox{c.c.} \nonumber \\
S_{A} &=& \frac{1}{2e^2} \sum_{\Box} (\epsilon_{\mu\nu\lambda}
\Delta_{\nu} A_{j\lambda}-2 \pi q_{\bar{\jmath}\mu} )^2
\nonumber \\
\widetilde{S}_B &=& \sum_j \eta_j A_{j \tau} \label{cp3}
\end{eqnarray}
The parallel between $Z_{cp}$ and $Z$ should clear: $S_z$
replaces $S_{{\bf n}}$, $S_A$ replaces $S_{\cal A}$, and
$\widetilde{S}_B$ replaces $S_B$. The phase diagram of $Z_{cp}$ as
a function of $g$ and $e^2$ should be similar to that of $Z$, as
we argue below and in Appendix~\ref{f}.

It is tempting to analyze $Z_{cp}$ by taking the continuum limit
of $S_A$ valid for small $e^2$: in this case $S_A$ reduces to the
standard continuum action for a $U(1)$ gauge field, and $Z_{cp}$
reduces to the theory of a complex spinor, $z_{\alpha}$, coupled
to a non-compact U(1) gauge field. However, this approach is not
correct for any non-zero $e^2$: it ignores the compactness of the
U(1) gauge field, and the associated instantons, which confines
charged fields in 2+1 dimensions.

A proper understanding of $Z_{cp}$ is achieved by an analysis
appropriate for large $e^2$. A crucial identity in this limit is
the exact expression
\begin{equation}
\sum_{\alpha} z^{\ast}_{j \alpha} z_{j+\hat{\mu},\alpha} = \left(
\frac{1 + {\bf n}_j \cdot {\bf n}_{j + \hat{\mu}}}{2}
\right)^{1/2} e^{i {\cal A}_{j \mu}}; \label{cp4}
\end{equation}
a suitable gauge choice has to be made on the left-hand-side,
corresponding to the choice of ${\bf n}_0$ in the definition of
${\cal A}_{j \mu}$. Using this, we can explicitly perform the
integral of $A_{j \mu}$ in $Z_{cp}$ and map it into a model which
has a structure similar to $Z$. This is most easily done at
$e^2=\infty$, when the integral over $A_{j \mu}$ can be done
independently on every site, without any preliminary decoupling
procedure (we have already argued that the $e^2 \rightarrow
\infty$ limit should be smooth, and the physics at $e^2=\infty$
should not differ from that at finite $e^2$). A simple evaluation
of the integral shows that
\begin{eqnarray}
Z_{cp} (e^2 = \infty) &=& \int \prod_j d {\bf n}_j \delta({\bf
n}_j^2 - 1) \exp\left( - S_B - \widetilde{S}_{\bf n} \right)
\nonumber \\
e^{- \widetilde{S}_{\bf n}} &=& \prod_{j} I_1 \left(
\frac{\sqrt{2( 1 + {\bf n}_j \cdot {\bf n}_{j + \hat{\tau}})}}{g}
\right) \nonumber \\
&~&\times\prod_{\mu=x,y} I_0 \left( \frac{\sqrt{2( 1 + {\bf n}_j
\cdot {\bf n}_{j + \hat{\mu}})}}{g} \right), \label{cp5}
\end{eqnarray}
where $I_{0,1}$ are modified Bessel functions which are
monotonically increasing functions of the their arguments, and
increase exponentially fast at large arguments. In the evaluation
of the integral it is useful to shift $A_{j \mu}$ by ${\cal A}_{j
\mu}$ and this leads to the Berry phase $S_B$ expressed in terms
of the ${\cal A}_{j \tau}$ rather than the $\widetilde{S}_B$
expressed in terms of the $A_{j \tau}$ in (\ref{cp3}). For small
$g$ it is also interesting to note that the integral over $A_{j
\mu}$ is dominated by values for which
\begin{equation}
A_{j \mu} \approx {\cal A}_{j \mu}, \label{cp5a}
\end{equation}
although this approximation was not used in the evaluation of
$\widetilde{S}_{\bf n}$--- thus the $A$ flux is pinned to values
close to the flux associated with ${\bf n}$ fluctuations. The
explicit form of differs $\widetilde{S}_{\bf n}$ differs from
that of $S_{\bf n}$, but the two actions should have essentially
identical physical properties for any finite $g$. Both actions
are minimized when the ${\bf n}_j$ are all parallel (a state with
HN order), and both increase monotonically as the angle between
any two neighboring ${\bf n}_j$ is increased. The action
$\widetilde{S}_{\bf n}$ does not have precisely the same form
along the spatial and temporal links, but this should lead only
to an innocuous and uninteresting renormalization of a spin-wave
velocity. This establishes the close connection between the
properties of $Z$ and $Z_{cp}$.

Another connection can be seen in the $g\rightarrow \infty$ limit
of $Z_{cp}$. In this limit, a direct application of the duality
methods developed in Section~\ref{sec:bo} establishes the exact
result
\begin{equation}
Z_{cp} (g=\infty) = Z_h, \label{cp6}
\end{equation}
where $Z_h$  is the height model partition function defined in
(\ref{d11}). So all of the results for $Z_h$ discussed in
Appendix~\ref{c} apply directly to $Z_{cp}$.

\section{Large $\lowercase{g}$ limit with SU(2) symmetry}
\label{b}

This Appendix will describe the evaluation of $W(a_{\mu})$ in
(\ref{d7}) for the SU(2) symmetric action $S_{\bf n}$ in
(\ref{sn}) for large $g$.

One way to proceed is to accept the equivalence between $Z$ and
$Z_{cp}$ just discussed, and perform the average not over $S_{\bf
n}$, but over $S_z$ with ${\cal A}_{j \mu}$ replaced by $A_{j
\mu}$: this will lead to essentially the same results as those
appearing below.

Alternatively, we can follow a procedure that is roughly the
inverse of that discussed towards the end of Appendix~\ref{a}. We
first, formally, write down an effective action for ${\cal A}_{j
\mu}$ fluctuations by
\begin{equation}
e^{-S'_A (A_{j\mu}) } \equiv \int \prod_j d {\bf n}_j \delta({\bf
n}^2 - 1) \delta(A_{j \mu} - {\cal A}_{j \mu}) e^{-S_{\bf n}}.
\label{g1}
\end{equation}
The resulting action will be invariant under gauge transformation
of the $A_{j \mu}$ and also periodic under shifts of $A_{j \mu}$
by $ 2 \pi$. However, because of the complicated form of
(\ref{area}), this action is difficult to write down explicitly,
even at $g=\infty$. Nevertheless, it is evident from a ``high
temperature'' expansion of $S_{\bf n}$ that correlations of the
flux decay exponentially fast {\em i.e.}
\begin{equation}
\left\langle \exp \left( i E_{\mu} (r_{\bar\jmath}) - i E_{\mu}
(0) \right) \right\rangle_{S'_A} \sim e^{|r_{\bar\jmath}|/\xi},
\label{g2}
\end{equation}
where $E_{\mu} (r_{\bar\jmath}) =
(\epsilon_{\mu\nu\lambda}\Delta_{\nu}
A_{\lambda})_{\bar{\jmath}}$, and the correlation length $\xi =
1/(2 \ln g)$. So it is reasonable to model $S'_A$ by the simplest
lattice action which is consistent with all the symmetries and
which reproduces (\ref{g2}) at long scales:
\begin{equation}
S'_A = \frac{1}{2g^2} \sum_{\Box} (\epsilon_{\mu\nu\lambda}
\Delta_{\nu} A_{j\lambda}-2 \pi q_{\bar{\jmath}\mu} )^2
.\label{g3}
\end{equation}
Now we can easily evaluate $W(a_{\mu})$ by the same duality
transformations developed in Section~\ref{sec:bo} and obtain
\begin{eqnarray}
W(a_{\mu}) &=& \left\langle \exp \left(- i \sum_{\Box}
\epsilon_{\mu\nu\lambda} a_{\bar{\jmath}\mu} \Delta_{\nu} { A}_{j
\lambda} \right) \right\rangle_{S'_A} \nonumber \\
&=& \sum_{\{h_{\bar{\jmath}}\}} \exp \left ( - \frac{g^2}{2}
\sum_{\bar{\jmath}} \left( a_{\bar{\jmath}\mu} - \Delta_{\mu}
h_{\bar{\jmath}} \right)^2 \right) \label{g4}
\end{eqnarray}
This last expression is our main result for $W(a_{\mu})$, and it
clearly shows the $a$-Meissner nature of the large $g$ regime;
notice its similarity to the London-limit action for the
electromagnetic field in a superconductor---after allowing the
discrete variables $a$ and $h$ to be continuous, we see that $a$
plays the role of the electromagnetic vector potential, and $h$
that of the phase of the superconducting order. So (\ref{g4})
indicates that $a$-flux will be expelled, and the phase is well
characterized by (\ref{d8}). More precisely, we can easily
evaluate the expression for $Z$ in (\ref{d7a}) and obtain
\begin{equation}
Z = \sum_{\{h_{\bar{\jmath}}\}} \exp\left( - f(\Delta_{\mu}
h_{\bar{\jmath}}-a_{\bar{\jmath}\mu}^0)\right) \label{g5}
\end{equation}
where the function $f(n)$ is given by
\begin{equation}
e^{-f(n)} \equiv \sum_{m=-\infty}^{\infty} \exp \left( -
\frac{e^2}{2} m^2 - \frac{g^2}{2} (m-n)^2 \right). \label{g6}
\end{equation}
The expression (\ref{g5}) for $Z$ is very closely related to that
for the height model partition function, $Z_h$, in (\ref{d11});
the latter corresponds to the choice $f(n) = e^2 n^2/2$. More
generally, $f(n)$ is an even, monotonically function of $n$, and
we expect that the properties of the height model should be quite
insensitive to the detailed form of $f(n)$. The expression
(\ref{g6}) can be viewed as a renormalization of the coupling
$e^2$ due to ${\bf n}$ fluctuations.

\section{Properties and mappings of the height model}
\label{c}

This appendix will review the properties of the height model,
$Z_h$, in (\ref{d11}), and also present its equivalence to a
number of other models of paramagnetic states of quantum
antiferromagnets.

\begin{figure}
\epsfxsize=2.4in \centerline{\epsffile{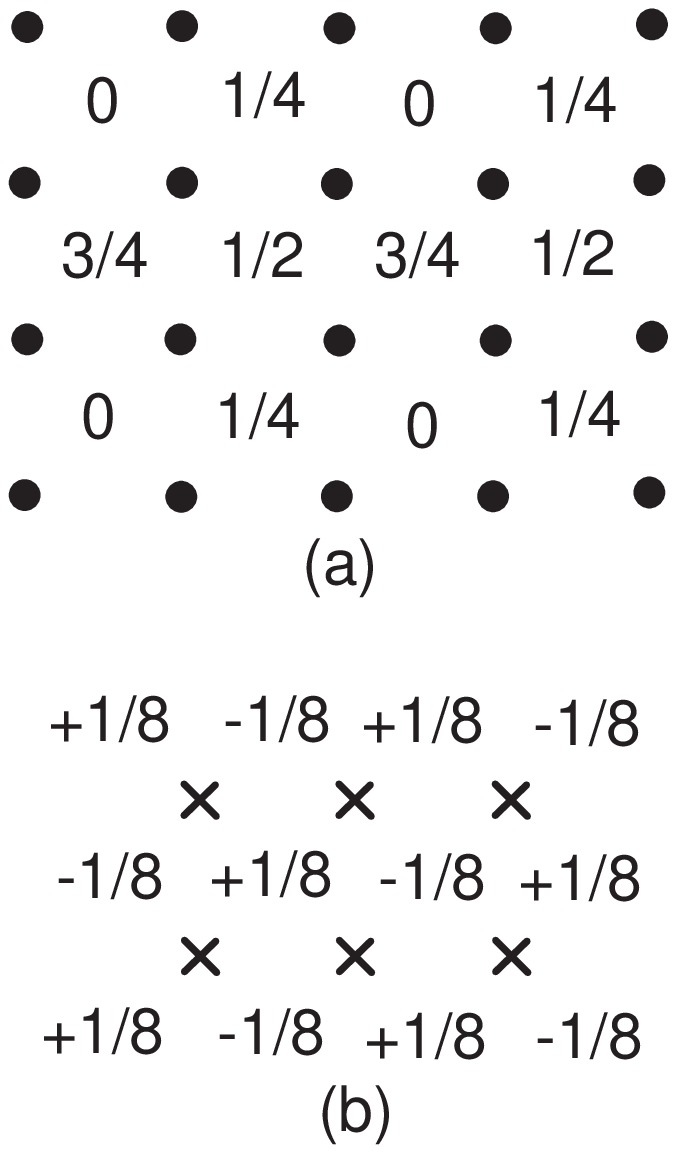}} \vspace{0.2in}
\caption{Specification of the non-zero values of the fixed fields
(a) ${\cal X}_{\bar{\jmath}}$ and (b) ${\cal Y}_{j \mu}$
introduced in (\protect\ref{XY}). The notational conventions are
as in Fig~\protect\ref{figa0}. Only the $\mu=\tau$ components of
${\cal Y}_{j \mu}$ are non-zero, and these are shown in (b).
}\label{figXY}
\end{figure}

It is useful to first rewrite $Z_h$ is a form closer to that used
in earlier work. We decompose the fixed field
$a_{\bar{\jmath}\mu}^{0}$ into curl and divergence free parts by
writing it in terms of new fixed fields, ${\cal X}_{\bar{\jmath}}$
and ${\cal Y}_{j \mu}$ as follows:
\begin{equation}
a_{\bar{\jmath}\mu}^{0} = \Delta_{\mu} {\cal X}_{\bar{\jmath}} +
\epsilon_{\mu\nu\lambda} \Delta_{\nu} {\cal Y}_{j  \lambda}.
\label{XY}
\end{equation}
The values of these new fields are shown in Fig~\ref{figXY}.
Inserting (\ref{XY}) into (\ref{d11}), we can now write the
height model in its simplest form
\begin{equation}
Z_h = \sum_{\{H_{\bar{\jmath}}\}} \exp \left ( - \frac{e^2}{2}
\sum_{\bar{\jmath}} \left( \Delta_{\mu} H_{\bar{\jmath}} \right)^2
\right), \label{he1}
\end{equation}
where
\begin{equation}
H_{\bar{\jmath}} = h_{\bar{\jmath}} - {\cal X}_{\bar{\jmath}}
\label{he2}
\end{equation}
is the new height variable we shall work with. Notice that the
${\cal Y}_{j \mu}$ have dropped out, while the ${\cal
X}_{\bar{\jmath}}$ act only as fractional offsets to the integer
heights. From (\ref{he2}) we see that the height is restricted to
be an integer on of the four sublattices, an integer plus 1/4 on
the second, an integer plus 1/2 on the third, and an integer plus
3/4 on the fourth; the fractional parts of these heights are as
shown in Fig~\ref{figXY}a. The steps between neighboring heights
are always an integer plus 1/4, or an integer plus 3/4.

\subsection{Quantum dimer model}
\label{qdm}

\begin{figure}
\epsfxsize=2.4in \centerline{\epsffile{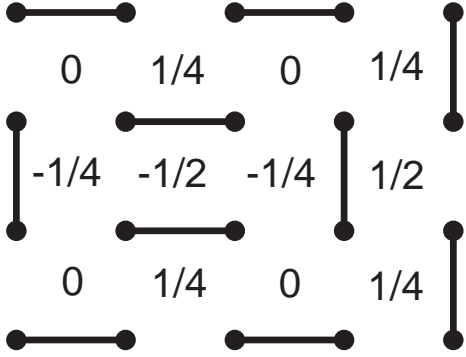}} \vspace{0.2in}
\caption{Mapping between the quantum dimer model and $Z_h$. Each
dimer on the direct lattice is associated with a step in height
of 3/4 on the link of the dual lattice that crosses it. All other
height steps are 1/4.}\label{dimer}
\end{figure}

The connection\cite{zheng} with the quantum dimer
model\cite{qd,sondhi1,sondhi2} emerges upon examining the large
$e^2$ limit of $Z_h$. In this limit, the action is minimized by
choosing the smallest possible modulus of steps between
neighboring heights: these are either 1/4 or 3/4. It is not
possible to reduce the number of 3/4 steps to zero, and height
configurations with the minimal number of 3/4 steps have an exact
one-to-one correspondence with the set of dimer coverings of the
direct lattice (heights which differ merely by a uniform shift of
all heights by the same integer are treated as equivalent). This
correspondence is illustrated in Fig~\ref{dimer}.
Temporal fluctuations of the heights between two such minimal
action configurations correspond to quantum tunnelling between
different dimer states, and hence the connection with the quantum
dimer model.

\subsection{Coulomb plasma of instantons}

Elevate the height variable to a continuous real field by the
Poisson summation formula; this allows us to rewrite (\ref{he1})
as
\begin{eqnarray}
Z_h &=& \sum_{\{m_{\bar{\jmath}}\}} \int_{-\infty}^{\infty}
\prod_{\bar{\jmath}} d H_{\bar{\jmath}}\exp \Biggl( -
\frac{e^2}{2} \sum_{\bar{\jmath}} \left( \Delta_{\mu}
H_{\bar{\jmath}}
\right)^2 \nonumber \\
&~&~~~~~~~~~~ - 2\pi i \sum_{\bar{\jmath}} m_{\bar{\jmath}} (
H_{\bar{\jmath}} + {\cal X}_{\bar{\jmath}} )    \Biggr)
\label{he3}
\end{eqnarray}
Now perform the Gaussian integral over the $H_{\bar{\jmath}}$ and
obtain the height model partition function written as a sum over
the plasma of ``charges'' $m_{\bar{\jmath}}$.
\begin{eqnarray}
Z_h &=& \sum_{\{m_{\bar{\jmath}}\}}  \exp \Biggl( - \frac{2
\pi^2}{e^2} \sum_{\bar{\jmath} \bar{\jmath}'} m_{\bar{\jmath}}
G(\bar{\jmath} - \bar{\jmath}' ) m_{\bar{\jmath}'} \nonumber \\
&~&~~~~~~~~~~- 2\pi i \sum_{\bar{\jmath}} m_{\bar{\jmath}} {\cal
X}_{\bar{\jmath}} \Biggr), \label{he4}
\end{eqnarray}
where $G$ is the lattice Green's function in (\ref{m1}). For
large $|r|$, $G(r) \sim 1/|r|$, and hence this is a Coulomb
plasma of charges in 2+1 dimensions. Working backwards through
the series of duality transformations, we can deduce that these
charges are the `hedgehogs' in the ${\bf n}$ field. Each charge
also carries a Berry phase of 1, $i$, -1, $-i$, on the four dual
sublattices, represented by the last term in (\ref{he4})---these
phases are identical to the hedgehog Berry phases computed by
Haldane\cite{berryh} by entirely different methods. The plasma is
always in its ``Debye-screened'' phase for all values of $e^2$,
and so isolated hedgehog are ``free''. As we review in the
following subsection, this free plasma of charges always has BC
order, induced by the Berry phases.

\subsection{Sine-Gordon model}
\label{sgm}

The most practical formulation of the height model is in terms of
a frustrated sine-Gordon model. To obtain this, we insert a
phenomenological ``core'' action $e^{-E_c m_{\bar{\jmath}}^2}$ in
(\ref{he3}), and in the limit of large $E_c$, sum over just
$m_{\bar{\jmath}} = 0, \pm 1$ on every site. This expresses $Z_h$
as the functional integral over what is now the sine-Gordon field,
$H_{\bar{\jmath} \mu}$, with the action
\begin{equation}
S_{sG} = \frac{e^2}{2} \sum_{\bar{\jmath}} \left[ (\Delta_{\mu}
H_{\bar{\jmath}})^2 - y \cos(2\pi ( H_{\bar{\jmath}} + {\cal
X}_{\bar{\jmath}})) \right] \label{he5}
\end{equation}
The properties of the model $S_{sG}$ have been discussed at
length in Ref.~\onlinecite{rsprb}, and so we need not reproduce
the analysis here. The main result is that the height
$H_{{\bar{\jmath}}}$ is always in its ``smooth'' phase (and the
plasma above is therefore in its Debye screening phase),
characterized by a definite value of the average height
$\overline{H} \langle H_{\bar{\jmath}}$ (where the average is over
quantum fluctuations and over all sites), and such a phase
necessarily has BC order. The fractional value of $\overline{H}$
specifies the nature of the BC order. As in Section~\ref{sec:su2},
we introduce the BC order parameter
\begin{equation}
\Psi_{BC} = \sum_{\bar{\jmath}} e^{2 \pi i H_{\bar{\jmath}}}.
\label{he6}
\end{equation}
Then, from the results of Ref.~\onlinecite{rsprb,srprl} we see
that the values $\arg \langle \Psi_{BC} \rangle = \pi/4, 3\pi/4,
5\pi/4, 7 \pi/4$ correspond to the four BC states in
Fig~\ref{bcorder}a, the values $\arg \langle \Psi_{BC} \rangle =
0, \pi/2, \pi, 3\pi/2$ correspond to the four BC states in
Fig~\ref{bcorder}b, while any other value of $\arg \langle
\Psi_{BC} \rangle$ yields an eight-fold degenerate state
involving superposition of the orders in Fig~\ref{bcorder}a and b
(lattice symmetries show that the states with $\Psi_{BC}
\rightarrow \Psi_{BC}^{\ast}$, and $\Psi_{BC} \rightarrow e^{i n
\pi/2} \Psi_{BC}$ ($n$ integer) are all equivalent).

\section{Correlations of defects in the $\varphi^4$ field theory}
\label{d}

This appendix shows how defect correlations may be computed in a
continuum $\varphi^4$ field theory. The methods used here are a
generalization of an approach due to Halperin~\cite{bert}. We
will outline the technical steps in the calculation, as no
details were given in Ref.~\onlinecite{bert}.

We are interested in a field theory with action
\begin{equation}
S_{\varphi} = \frac{1}{2} \int \frac{d^D k}{(2 \pi)^D}
\frac{|\varphi_a (k) |^2}{Q(k)} + \frac{u}{4!} \int d^D r
(\varphi_a^2 (r))^2 \label{h1}
\end{equation}
where $D=3$ is the dimension of spacetime, and $a=1 \ldots N$ for
a theory with O($N$) symmetry. The propagator $Q^{-1} (k) = m_0^2
+ k^2$ at small $k$, while at $k$ larger than some cutoff,
$\Lambda$, $Q(k)$ decreases sufficiently fast with increasing $k$
to make the interacting theory ultraviolet finite. The `mass',
$m_0^2$ is tuned by varying the coupling $g$ in the underlying
model $Z$.

We consider first the case $N=D=3$. Then we have point hedgehog
defects. In the present continuum formulation, Halperin showed
that the topological charge density can be written as
\begin{equation}
\rho_h (r) = \delta^D \left[ \varphi_a (r) \right] \det \left[
\partial_{\mu} \varphi_a (r) \right] \label{h2}
\end{equation}
We are interested in evaluating correlators of $\rho_h (r)$ under
the action $S_{\varphi}$ in a perturbation theory in $u$. The main
technical difficulty in such a calculation arises from the delta
function in (\ref{h2}). However this may be easily treated by a
standard Lagrange multiplier technique.

We illustrate the method by computing the two-point correlator of
$\rho_h (r)$ at $u=0$. In the standard field theoretic method,
correlators of $\varphi$, and its derivatives, may be easily
obtained by first evaluating the generating function
\begin{equation}
\left\langle \exp \left( - \int d^D r J(r) \varphi (r) \right)
\right\rangle_{S_{\varphi}}. \label{h3}
\end{equation}
For simplicity of notation, we will work in this paragraph with a
single component field $\varphi$; the generalization to multiple
values of $\alpha$ is immediate. At $g=0$, (\ref{h3}) evaluates to
\begin{equation}
\exp \left( \frac{1}{2} \int d^D r_1 d^D r_2 J(r_1) J(r_2) Q(r_1
- r_2) \right), \label{h4}
\end{equation}
where $Q(r)$ is the Fourier transform of $Q(k)$. Now generalize
(\ref{h3}) by including two delta functions of $\varphi$
\begin{equation}
\left\langle \delta(\varphi(r_3)) \delta(\varphi(r_4)) \exp \left(
- \int d^D r J(r) \varphi (r) \right) \right\rangle_{S_{\varphi}}
\label{h5}
\end{equation}
with $r_3 \neq r_4$. We write the first delta function as
\begin{equation}
\delta(\varphi(r_3)) = \int_{-\infty}^{\infty} \frac{d \lambda}{2
\pi} e^{- i \lambda \varphi(r_3)}, \label{h6}
\end{equation}
and similarly for the second delta function. We can now evaluate
the $\varphi$ averages in (\ref{h5}), and then explicitly perform
the integrals over the two $\lambda$'s. In this manner we obtain
our key result for the value of (\ref{h5}) at $g=0$:
\begin{eqnarray}
(\ref{h5}) && = \frac{1}{2 \pi \sqrt{\Delta(r_3 - r_4)}}
\nonumber \\
&& \times \exp \left( \frac{1}{2} \int d^D r_1 d^D r_2 J(r_1)
J(r_2) \widetilde{Q}(r_1 , r_2) \right) \label{h7}
\end{eqnarray}
where $\Delta (r) \equiv Q^2 (r=0) - Q^2 (r)$ and
\begin{eqnarray}
\widetilde{Q}(r_1 , r_2 ) &=& Q(r_1 - r_2) - \frac{1}{\Delta(r_3 -
r_4)} \Bigl[
\nonumber \\
&~& Q(r_1 - r_3) Q(r_2 - r_3) Q(0) \nonumber \\
&+& Q(r_1 - r_4) Q(r_2 - r_4) Q(0) \nonumber \\
&-& Q(r_1 - r_3) Q(r_2 - r_4) Q(r_3-r_4) \nonumber \\
&-& Q(r_1 - r_4) Q(r_2 - r_3) Q(r_3-r_4)\Bigr] \label{h8}
\end{eqnarray}
Observe that (\ref{h8}) is a Gaussian in $J$, and so $\varphi$
correlations remain Gaussian even in the presence of the delta
function. In comparing (\ref{h8}) with (\ref{h4}), we notice only
two differences. First, the overall amplitude of the correlators
has been modified by the prefactor $1/(2 \pi \sqrt{\Delta})$.
Second, $Q$ has been replaced by $\widetilde{Q}$; notice that the
additional terms in (\ref{h8}) become subdominant as $r_1$ and
$r_2$ move far apart provided $Q(r)$ is a smoothly decreasing
function of $r$ (which is the case even at the critical point).
This latter fact indicates that at large distances we may as well
neglect the delta functions entirely: they merely renormalize the
correlators with an overall prefactor, which is not of interest
in the universal properties anyway. It is now a straightforward
matter to compute the two-point correlator of $\rho_h$ by repeated
applications of (\ref{h7}) and (\ref{h8}). First by sending $r_3
\rightarrow r_1$ and $r_4 \rightarrow r_2$ we can obtain the
correlator of the gradients of $\varphi$:
\begin{eqnarray}
&& \left\langle \delta(\varphi(r)) \partial_{\mu} \varphi(r)
\delta(\varphi(0)) \partial_{\nu} \varphi (0)
\right\rangle_{S_{\varphi}} = \nonumber \\
&&~~\frac{1}{2 \pi \sqrt{\Delta(r)}} \left[ - \partial_{\mu}
\partial_{\nu} Q(r) - \frac{r_{\mu} r_{\nu}}{r^2} \frac{ Q(r)
(Q'(r))^2}{\Delta(r)} \right], \label{h9}
\end{eqnarray}
where $r \neq 0$, we have assumed that $Q(r)$ is a function only
of $|r|$, and the prime indicates a derivative with respect to the
argument of $Q(|r|)$. Now we can compute $C_h (r)$, defined in
(\ref{m2}), by a straightforward application of Wick's theorem,
and we obtain Halperin's result\cite{bert}:
\begin{equation}
C_h (r \neq 0 ) = 6 P_1(r) P_2^2 (r) \label{h10}
\end{equation}
where
\begin{eqnarray}
P_1 (r) &=& \frac{- Q''(r) \Delta(r) - (Q'(r))^2 Q(r)}{2 \pi
\Delta^{3/2} (r) } \nonumber \\
P_2 (r) &=& \frac{- Q'(r)}{2 \pi r \Delta^{1/2}(r) }. \label{h11}
\end{eqnarray}
There is additional singular contribution to $C_h (r)$ at $r=0$,
proportional to $\delta (r)$, which has been computed by
Halperin: this contribution is precisely that needed for $C_h
(r)$ to satisfy the neutrality condition (\ref{m3}).

Our approach above can be easily generalized to develop a
perturbation theory for $C_h (r \neq 0)$ in $u$. We expect that
the neutrality condition (\ref{m3}) will be satisfied at each
order in $u$, as it is a topological constraint imposed by the
existence of the smooth field $\varphi (r)$.

We now have the machinery to determine the long-distance decay of
$C_h (r)$ at the critical point of $S_{\varphi}$. The key
observation, noted above, is that the delta functions may be
neglected in determining the leading functional form of this
decay (note however, that the delta-function related contributions
were essential in satisfying the neutrality condition (\ref{m3}),
as it depended also on the short distance behavior of $C_h (r)$).
The subdominance of the delta-function related contributions at
long distances holds also at each order in $u$, and so we
conclude that the scaling dimension of $\rho_h$ is given simply by
\begin{eqnarray}
\mbox{dim}[\rho_h (r)] &=& \mbox{dim} \left[ \det[\partial_{\mu}
\varphi_a ] \right] \nonumber \\
&\equiv& (9 + \eta_h)/2, \label{h12}
\end{eqnarray}
where the second expression defines the anomalous dimension
$\eta_h$. The expression on the right-hand-side of the first
equation is a local composite operator, and its scaling dimension
can be determined in an expansion in $\epsilon=4-D$ by standard
field theoretic methods. It is not difficult to show that this
composite operator has no renormalizations to two loops (although
they do appear at higher loop order); hence we immediately obtain
(\ref{m4},\ref{m5}). We also discussed in Section~\ref{sec:hnbo}
that we do not expect the conservation law (\ref{m3}), which is
special to $D=3$, to disrupt the validity of the $\epsilon$
expansion at $\epsilon=1$. The scaling dimension in (\ref{h12})
also specifies the singularity in the density of hedgehogs at the
critical point of $S_{\varphi}$; for the original model $Z$ this
critical point is at $g=g_{2c}$ (see Fig~\ref{phaseZ}) and
\begin{equation}
\bar\rho_h = \langle |\rho_h| \rangle = C_1 (g) \Lambda^{-3} +
C_{\pm} |g-g_{2c}|^{(9 + \eta_h)/2} \label{h12a}
\end{equation}
where $C_1 (g)$ is a smooth function of $g$, finite at
$g=g_{2c}$. The non-universal constants $C_{\pm}$ apply on the
two sides of the transition. Notice that the singularity is very
weak and essentially unobservable.

We conclude this Appendix by stating the generalization of these
results to the O(2) case, with $N=2$, $D=3$; this applies to the
spin systems with U(1) symmetry. In this case, the generalization
of the expression (\ref{h2}) for the defect density is the
following expression\cite{bert} for the vortex current ${\cal
J}_{\mu} (r)$:
\begin{equation}
{\cal J}_{\mu} (r) = \delta^2 \left[\varphi_a (r) \right]
\epsilon_{\mu\nu\lambda} \partial_{\nu} \varphi_1 (r)
\partial_{\lambda} \varphi_2 (r).
\label{h12b}
\end{equation}
The above methods show that at $u=0$, the correlator of ${\cal
J}_{\mu} (r)$ is given by
\begin{eqnarray}
&& C_{v,\mu\nu}(r) = \langle {\cal J}_{\mu} (r) {\cal J}_{\nu} (r)
\rangle \nonumber \\
&& = 2 P_2^2 (r) \frac{r_{\mu} r_{\nu}}{r^2} + 2 P_1 (r) P_2 (r)
\left( \delta_{\mu\nu} - \frac{r_{\mu} r_{\nu}}{r^2} \right),
\label{h13}
\end{eqnarray}
where $P_{1,2}$ were defined in (\ref{h11}). This result corrects
an error in Halperin's result, quoted in (6.33) of
Ref.~\onlinecite{bert}. It can be checked that the result in
(\ref{h13}) satisfies
\begin{equation}
\int d^3 r C_{v,\mu\nu}(r) = 0~~~~;~~~~\partial_{\mu}
C_{v,\mu\nu}(r) =0. \label{h14}
\end{equation}
The first result states that the net vorticity vanishes, while
the second expresses the conservation of the vortex current. Both
these properties will hold at each order in $u$, and this allowed
us to obtain (\ref{m14}). The long-distance behavior of ${\cal
J}_{\mu}$ correlations at the critical point is controlled by the
scaling dimension
\begin{eqnarray}
\mbox{dim}[{\cal J}_{\mu} (r) ] &\equiv& (6 + \eta_v)/2 \nonumber
\\ &=& \mbox{dim} \left[ \epsilon_{\mu\nu\lambda}
\partial_{\nu} \varphi_1
\partial_{\lambda} \varphi_2 \right] \nonumber \\
&=& 4- \epsilon + \eta + {\cal O} (\epsilon^3). \label{h15}
\end{eqnarray}
In the present situation, as was discussed in
Section~\ref{sec:hnbo}, we do expect the conservation law
(\ref{h14}) to play a crucial role in $D=3$, and to impose the
exact result $\eta_v = -2$ for $\epsilon =1$.

\section{Field theory in one dimension}
\label{e}

This appendix outlines the steps which relate the lattice model
$Z^{(1)}_{\rm U(1)}$ in (\ref{o11}) to the field theory in
(\ref{o17}).

We first elevate the height $\ell_{\bar{\jmath}}$ to a continuous
real field by the Poisson summation identity. The presence of the
parity variables, $\sigma_{\bar{\jmath}}$, requires a slight
generalization of this identity to
\begin{eqnarray}
&& \sum_{\ell=-\infty}^{\infty} f (\ell, \sigma) = \frac{1}{2}
\lim_{E_c \rightarrow 0} \int_{-\infty}^{\infty} d \phi
\sum_{m=-\infty}^{\infty} e^{i \pi m \phi} \nonumber \\
&&~~~~~~~~~\times \Bigl[ f(\phi, 1) + (-1)^m f (\phi, -1) \Bigr]
e^{-E_c m^2} \label{appe1}
\end{eqnarray}
where $\sigma \equiv 1 - 2 (\ell (\mbox{mod 2}))$, and $f$ is an
arbitrary function. As is conventional, it is safe to analyze
these interface models in the limit of large ``core'' energy
$E_c$, which accounts for short-distance renormalizations. Then,
application of (\ref{o17}) to (\ref{o11}) shows that
\begin{eqnarray}
&& Z^{(1)}_{\rm U(1)} = \int \prod_{\bar{\jmath}}
d\phi_{\bar{\jmath}} \exp \Biggl[ - \sum_{\bar{\jmath}} \Bigl(
\frac{g}{8} \left( \epsilon_{\mu\nu} \Delta_{\nu}
\phi_{\bar{\jmath}} \right)^2 \nonumber \\
&&- 2 e^{-4E_c}  \cos (2 \pi \phi_{\bar{\jmath}}) - 2 e^{-E_c}
\tanh (K_d)  \varepsilon_{\bar{\jmath}} \cos(\pi
\phi_{\bar{\jmath}}) \Bigr)\Biggr].\label{appe2}
\end{eqnarray}
In terms of these variables, the BC order parameter is
 $Q_{j,j+\hat{x}} \sim \varepsilon_{\bar{\jmath}} \cos(\pi \phi_{\bar{\jmath}})$
and I order is determined by $\hat{S}_{jz} \sim \Delta_{x}
\phi_{\bar{\jmath}}$.

It is useful to now integrate out some short-distance, high-energy
fluctuations in (\ref{appe2}). We write
\begin{equation}
\phi_{\bar{\jmath}} = \phi_0 (r_{\bar{\jmath}}) +
\varepsilon_{\bar{\jmath}} \phi_1 (r_{\bar{\jmath}}) \label{appe3}
\end{equation}
and minimize the action with respect to the massive field
$\phi_1$. This yields
\begin{equation}
\phi_1 (r_{\bar{\jmath}}) = - \frac{2\pi}{g} \tanh(K_d)  e^{-E_c}
\sin ( \pi \phi_0 (r_{\bar{\jmath}})) \label{appe4}
\end{equation}
Inserting this back into (\ref{appe2}), and performing the
gradient expansion for the field $\phi_0$ we get the action of the
sine-Gordon theory in (\ref{o17}) with
\begin{equation}
Y = 2 e^{-4 E_c} - \frac{\tanh^2 (K_d) \pi^2 e^{-2 E_c}}{g}
\label{appe5}.
\end{equation}
The expression (\ref{appe4}) now shows that the I order parameter
(measured by a staggered $\hat{S}_{j,z} $) is $\sim \sin(\pi
\phi_0)$, while the BC order (measured by a staggered
$Q_{j,j+\hat{x}}$) is $\sim \cos(\pi \phi_0)$.

\section{$CP^1$ models with easy-plane anisotropy}
\label{f}

In the body of the paper, we applied the easy-plane anisotropy
implied by (\ref{o3}) to the central partition function $Z$ in
(\ref{z}), and so obtained $Z_{\rm U(1)}$. This latter model was
studied in great detail in Section~\ref{sec:u1}. We also claimed
that the physics of the phases of the SU(2) invariant $Z$ was
essentially identical to the $CP^1$ model $Z_{cp}$ in
Appendix~\ref{a}. As a consistency check, we will apply the
easy-plane anisotropy to $Z_{cp}$ here: it is then reassuring to
find below properties that are very similar to those in
Section~\ref{sec:u1}.

Comparing (\ref{o3}) and (\ref{cp2}), we see that the easy-plane
limit corresponds to $|z_{j \uparrow}| = |z_{j \downarrow}|$. So
we parameterize $z_{j\alpha} = e^{i \theta_{\alpha}}$, and then
the gauge-invariant angle $\theta = \theta_{\uparrow}-
\theta_{\downarrow}$. Inserting this in (\ref{cp3}), and writing
the periodic couplings in Villain form, we obtain the form of
$Z_{cp}$ with U(1) symmetry and easy-plane anisotropy:
\begin{eqnarray}
Z_{cp,{\rm U(1)}} &=&
\sum_{\{q_{\bar{\jmath}\mu},m_{j\uparrow\mu},m_{j\downarrow\mu}\}}
\int_{0}^{2 \pi} \prod_{j\mu} d A_{j \mu} \int \prod_{j } d
\theta_{j\uparrow} d \theta_{j\downarrow}  \nonumber \\
&~&~~~~\times
\exp\left( - S_A - \widetilde{S}_B -S_{\uparrow} -S_{\downarrow} \right ) \nonumber \\
S_{\uparrow} &=& \frac{1}{2g} \left( \Delta_{\mu}
\theta_{j\uparrow} - 2 \pi m_{j\uparrow\mu} - A_{j \mu} \right)^2
\nonumber \\
S_{\downarrow} &=& \frac{1}{2g} \left( \Delta_{\mu}
\theta_{j\downarrow} - 2 \pi m_{j\downarrow\mu} - A_{j \mu}
\right)^2 , \label{appf1}
\end{eqnarray}
where $S_A$ and $\widetilde{S}_B$ are defined in (\ref{cp3}). We
can easily apply the duality methods of Sections~\ref{sec:bo}
and~\ref{sec:u1} to (\ref{appf1}) and obtain a partition function
for integer valued vectors fields
$a_{\bar{\jmath}\mu},p_{\bar{\jmath}\mu}$ residing on the links of
the dual lattice:
\begin{eqnarray}
Z_{cp,{\rm U(1)}} &=&
\sum_{\{a_{\bar{\jmath}\mu},p_{\bar{\jmath}\mu}\}} \exp \Biggl( -
\frac{e^2}{2} (a_{\bar{\jmath}\mu} - a_{\bar{\jmath}\mu}^0 )^2 -
\frac{g}{2} \sum_{\Box} (\epsilon_{\mu\nu\lambda} \Delta_{\nu}
p_{\bar{\jmath}\lambda})^2 \nonumber \\
&~&~~~~~~~~~- \frac{g}{2} \sum_{\Box}
\left(\epsilon_{\mu\nu\lambda} \Delta_{\nu}
p_{\bar{\jmath}\lambda}- \epsilon_{\mu\nu\lambda} \Delta_{\nu}
a_{\bar{\jmath}\lambda}\right)^2 \Biggr). \label{appf2}
\end{eqnarray}
This model is closely related to (\ref{u12}) for the case of the
on-site coupling (\ref{o9}). The connection becomes evident when
we rewrite the last two terms in (\ref{appf2}) as
\begin{equation}
 g \sum_{\Box}
\left(\epsilon_{\mu\nu\lambda} \Delta_{\nu}
p_{\bar{\jmath}\lambda}- \frac{1}{2} \epsilon_{\mu\nu\lambda}
\Delta_{\nu} a_{\bar{\jmath}\lambda}\right)^2 +
\frac{g^{\prime}}{4} \sum_{\Box} (\epsilon_{\mu\nu\lambda}
\Delta_{\nu} a_{\bar{\jmath}\lambda})^2  \label{appf3}
\end{equation}
with $g^{\prime} = g$. For $g^{\prime} =0$, the model
(\ref{appf2}) is identical to the model (\ref{u12}). Adding such a
`Maxwell' term with $g^{\prime}=g$ does appear to be an innocuous
change, and it is clear that the basic physics of the class of
models described by (\ref{appf2}) and (\ref{u12}) should be the
same.

\begin{figure}
\epsfxsize=3.6in \centerline{\epsffile{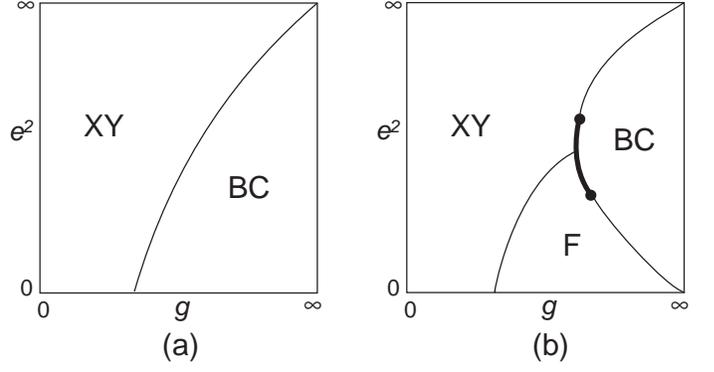}}
\vspace{0.2in} \caption{As in Fig~\protect\ref{phase3}, but for
the model $Z_{cp,{\rm U(1)}}$ in (\protect\ref{appf2}). Two
possible phase diagrams are shown. In (a), the BC order vanishes
as $e^2 \rightarrow 0$, and the $e^2=0$ line of the BC state is
fractionalized.} \label{phase3a}
\end{figure}

For the specific value $g'=g$, the phase diagram of (\ref{appf2})
does have some differences from that of (\ref{u12}) in
Fig~\ref{phase3} (however, the topological relationship of the
phases remains the same), and two possible phase diagrams are
shown in Fig~\ref{phase3a}.
Some insight is gained by looking at limiting values of the
parameters in (\ref{appf2}). For $e^2=\infty$, (\ref{appf2}) is
identical to the corresponding limit of (\ref{u12}): both models
reduce to the boson-current loop model of Otterlo {\em et
al.}~\cite{otterlo} which has XY order for all values of $g$; this
XY order corresponds to `superfluidity' in the gauge-invariant
angle $\theta$. Similarly, the $g=0$ limit is also identical in
the two models with XY order always present. A difference appears
in the limit $g=\infty$: now (\ref{appf2}) becomes equivalent the
height model in (\ref{d11}) whose ground state always has BC
order, while (\ref{u12}) reduced to the frustrated Ising model in
(\ref{u110}) which had both F and BC ground states. Finally, the
limit $e^2=0$ becomes clearest in the form (\ref{appf1}): now the
gauge field $A_{\mu}$ is entirely quenched and we have two
independent XY models associated with the angles
$\theta_{\uparrow,\downarrow}$. These have an XY ordering
transition, and $z_{\uparrow,\downarrow}$ quanta become free
massive spinons in the disordered phase: this is evidently an F
phase.

An important unanswered question is the fate of the F phase found
above at $e^2=0$ for non-zero values of $e^2$. In the case of
(\ref{u12}) we also found the F phase along the line $g=\infty$
and so it was clear that F order survived over a finite domain for
$e^2 > 0$. This is not so clear here for
(\ref{appf1},\ref{appf2}). The compact U(1) gauge theory is
confining, and the liberation of a finite density of instantons
for $e^2 > 0$ could immediately confine the spinons and lead to BC
order. This possibility is reflected in Fig~\ref{phase3a}a.
However, in the absence of numerical studies of (\ref{appf2}) we
leave open the possibility in Fig~\ref{phase3a}b that the F phase
survives for a range of value of $e^2$.

Similar considerations apply to the easy-plane $CP^1$ model in
one dimension, and its properties are essentially identical to
those discussed in Section~\ref{sec:1d} for $Z_{\rm U(1)}^{(1)}$.

\section{Field theory in two dimensions}
\label{g}

This appendix will provide the missing steps between the model
(\ref{u14}) and the field theory (\ref{u112}). The method is a
combination of that applied in one dimension in Appendix~\ref{e},
and the analysis of LFS~\cite{lannert}.

First, we follow the two-dimensional analog of the steps leading
to (\ref{appe2}). The main change is that the real scalar field
$\phi_{\bar{\jmath}}$ is replaced by a real vector field which we
denote $b_{\bar{\jmath}\mu}/\pi$. By these methods (\ref{u14})
maps to
\begin{eqnarray}
&& Z_{\rm U(1)} = \int \prod_{\bar{\jmath}} db_{\bar{\jmath}\mu}
\exp \Biggl[ -  \frac{g}{8\pi^2} \sum_{\Box} \left(
\epsilon_{\mu\nu\lambda} \Delta_{\nu}
b_{\bar{\jmath}\lambda} \right)^2 \nonumber \\
&&+ \sum_{\bar{\jmath}\mu} \Bigl(2 e^{-4E_c}  \cos (2
b_{\bar{\jmath}\mu}) + 2 e^{-E_c} \tanh (K_d)
\varepsilon_{\bar{\jmath},\bar{\jmath}+\hat{\mu}} \cos(
b_{\bar{\jmath}\mu}) \Bigr)\Biggr],\nonumber \\
\label{appg1}
\end{eqnarray}
where $\varepsilon_{\bar{\jmath},\bar{\jmath}+\hat{\mu}}$ is
defined in (\ref{u16}). We can now make all the terms in
(\ref{appg1}) invariant under a non-compact U(1) symmetry by
introducing an angular variable $\vartheta_j$ and mapping
\begin{equation}
b_{j \mu} \rightarrow -\Delta_{\mu} \vartheta_j + b_{j \mu}
\label{appg2}
\end{equation}
in (\ref{appg1}). The representation (\ref{appg1}) then
corresponds to a particular gauge choice in the non-compact U(1)
gauge theory. The remaining analysis is then identical to that in
LFS. First, we introduce two complex scalars, the field operators
for single vortices, $\phi_{\bar{\jmath}} \sim e^{i
\vartheta_j}$, and the field operator for double vortices
$\Phi_{\bar{\jmath}} \sim e^{2 i  \vartheta_j}$. Then, the second
term in (\ref{appg1}) is proportional to a simple hopping term
for $\Phi_{\bar{\jmath}}$
\begin{equation}
- \sum_{j,\mu}\Phi_{\bar{\jmath}}^{\ast} e^{2 i b_{j \mu}}
\Phi_{\bar{\jmath}+\hat{\mu}}. \label{appg3}
\end{equation}
This $\Phi_{\bar{\jmath}}$ quanta clearly have a minimum near
$k=0$, and so a naive continuum limit may be taken on
(\ref{appg3}). In contrast, the dispersion of
$\phi_{\bar{\jmath}}$ is provided by the last term in
(\ref{appg1}), and is
\begin{equation}
- \sum_{j,\mu} \varepsilon_{\bar{\jmath},\bar{\jmath}+\hat{\mu}}
\phi_{\bar{\jmath}}^{\ast} e^{i b_{j \mu}}
\phi_{\bar{\jmath}+\hat{\mu}}. \label{appg4}
\end{equation}
The $\varepsilon_{\bar{\jmath},\bar{\jmath}+\hat{\mu}} $
frustrates the hopping of the $\phi_{\bar{\jmath}}$ and it is
necessary to properly diagonalize (\ref{appg4}) in momentum space
for $b_{j \mu}=0$. This was performed by LFS who showed that for
a suitable gauge choice in (\ref{u16}), the $\phi_{\bar{\jmath}}$
had equivalent minima near spatial momenta $(0,0)$ and $(\pi,0)$.
We choose the same linear combinations of fields near these
minima as defined by LFS. Now we are prepared to take the
continuum limit of all the terms in (\ref{appg1}), and the result
leads to (\ref{u112}).

\section{Gauge theory with $\langle {\bf N} \rangle \neq 0$}
\label{h}

This appendix will study the vicinity of the points A and B in
Fig~\ref{phase7}. We will use the same strategy as that followed
in Section~\ref{sec:z2} but will apply it to the compact U(1)
gauge theory $Z_P$ in (\ref{f4}) rather than to the $Z_2$ gauge
theory $Z_P^{\prime}$ in (\ref{f9}).

The initial strategy is the same as that followed in
Section~\ref{sec:z2} for the $Z_2$ gauge theory. We quench the
field ${\bf N}_j$ at its condensate $N_0 (0,0,1)$ and neglect its
spin-wave fluctuations (as in the discussion in
Section~\ref{sec:sw}, the latter will lead to weak additional
power-law tails to the interactions considered below and do not
alter the basic physical properties). This allows us to
preferentially focus on $z_{j\uparrow}$ and neglect $z_{j
\downarrow}$; we write the former in terms of an angular variable
as $z_{j\uparrow}=e^{i\theta_{j\uparrow}}$. Note that this is an
easy-axis anisotropy to the $CP^1$ fields, in contrast to the
easy-plane limit considered in Appendix~\ref{f}. For the $Q_{j
\varrho}$ degrees of freedom in (\ref{f4}), we assume, for
simplicity, that the spiral correlations have preferentially
polarized in the $x$ direction: so we write $Q_{j x} = e^{i
\theta_{jQ}}$ and neglect $Q_{j y}$. With this transformations
and (innocuous) approximations we can obtain from $Z_P$ the
following effective action for the angular variables $\theta_{j
\uparrow}$, $\theta_{j Q}$ and the compact U(1) gauge field $A_{j
\mu}$ (all terms have been written in a Villain form to
facilitate duality transforms):
\begin{eqnarray}
Z_{N} &=&
\sum_{\{q_{\bar{\jmath}\mu},m_{j\uparrow\mu},m_{jQ\mu}\}}
\int_{0}^{2 \pi} \prod_{j\mu} d A_{j \mu} \int \prod_{j } d
\theta_{j\uparrow} d \theta_{jQ}  \nonumber \\
&~&~~~~\times
\exp\left( - S_A - \widetilde{S}_B -S_{\uparrow} -\widetilde{S}_{Q}\right ) \nonumber \\
\widetilde{S}_Q &=& \frac{1}{2g_Q} \left( \Delta_{\mu}
\theta_{jQ} - 2 \pi m_{jQ\mu} - 2A_{j \mu} \right)^2  ,
\label{apph1}
\end{eqnarray}
where $S_A$ and $\widetilde{S}_B$ are defined in (\ref{cp3}), and
$S_{\uparrow}$ is defined in (\ref{appf1}). Note that this action
differs in structure from (\ref{appf1}) only in the prefactor of 2
for the gauge field in $\widetilde{S}_Q$; this double charge is,
of course, crucial to the structure explored here. The duality
methods developed in Section~\ref{sec:bo} and~\ref{sec:u1} are
easily applicable to (\ref{apph1}): we obtain the following
action for integer valued vectors fields
$a_{\bar{\jmath}\mu},p_{\bar{\jmath}\mu}$ residing on the links
of the dual lattice
\begin{eqnarray}
Z_{N} &=& \sum_{\{a_{\bar{\jmath}\mu},p_{\bar{\jmath}\mu}\}} \exp
\Biggl( - \frac{e^2}{2} (a_{\bar{\jmath}\mu} -
a_{\bar{\jmath}\mu}^0 )^2 - \frac{g_Q}{2} \sum_{\Box}
(\epsilon_{\mu\nu\lambda} \Delta_{\nu}
p_{\bar{\jmath}\lambda})^2 \nonumber \\
&~&~~~~~~~~~- \frac{g}{2} \sum_{\Box} \left(2
\epsilon_{\mu\nu\lambda} \Delta_{\nu} p_{\bar{\jmath}\lambda}-
\epsilon_{\mu\nu\lambda} \Delta_{\nu}
a_{\bar{\jmath}\lambda}\right)^2 \Biggr). \label{apph2}
\end{eqnarray}
Some physical properties of $Z_N$ will be more transparent in
this dual formulation. Note also that (\ref{apph2}) has only
positive weights and so should be amenable to Monte Carlo
simulations.

\begin{figure}
\epsfxsize=3.6in \hspace{-0.3in} \epsffile{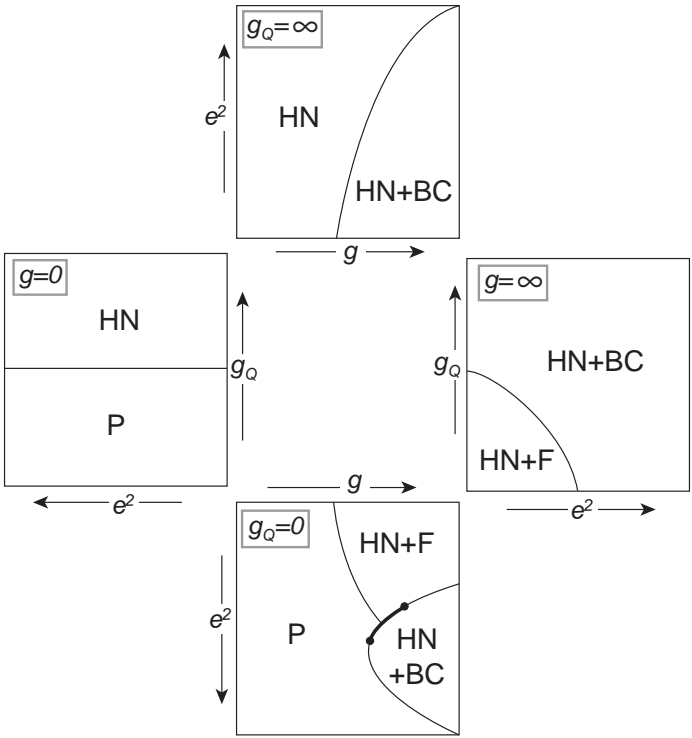}
\vspace{0.2in} \caption{Phase diagram of $Z_N$ in
(\protect\ref{apph1}) and (\protect\ref{apph2}) as a function of
the three couplings $g$, $g_Q$, and $e^2$. The labels of the
states account for the background ${\bf N}$ condensate needed in
the derivation of $Z_P$, and are appropriate to the phases of
$Z_P$ in (\protect\ref{f4}). The figure is to be visualized by
imagining that these couplings run along the axes of a
three-dimensional cube; the sides of the cube have been opened out
flat into the plane of the paper. The arrows indicate that the
associated coupling runs from 0 to $\infty$ along its
direction.}\label{phase8}
\end{figure}

We are interested here in the phase diagram of $Z_P$ as a
function of the three couplings $g$, $g_Q$, and $e^2$. Much
insight can be gained by looking at various limiting cases: the
results of such an analysis of the phase diagram are displayed in
Fig~\ref{phase8}.
Our primary is interest is in the phases at center of the cube
shown in Fig~\ref{phase8} where the states found at the
boundaries will meet and compete. A complete understanding of
this strongly coupled region will probably require Monte Carlo
simulations of (\ref{apph2}), which we have not undertaken.

The identification of the phases in Fig~\ref{phase8} accounts for
the presence of the background ${\bf N}_0$ condensate, and so the
phase labels are appropriate to the theory $Z_P$ in (\ref{f4}).
However, taken by itself, $Z_N$ is a model of a compact U(1)
gauge theory coupled to two angular variables, and its phases can
also be described in the language of the models of
Section~\ref{sec:u1}. $Z_N$ possesses a gauge-invariant angular
variable, $\theta_{\uparrow} - \theta_Q/2$, which can acquire
``superfluid'' or XY order; as in Section~\ref{sec:z2} this XY
order maps to the P phase after accounting for the ${\bf N}_0$
condensate. Also as in Section~\ref{sec:z2}, the gauge degrees of
freedom can induce BC or F phases, which become HN+BC and HN+F
phases in the context of (\ref{f4}) and Fig~\ref{phase8}.

The relationship between $Z_N$ and the models of
Section~\ref{sec:u1} can be made explicit by examining the limit
$g_Q \rightarrow 0$, where the charge 2 scalar $e^{i \theta_Q}$
condenses. Then it is  evident from (\ref{apph2}) that $Z_N$ at
$g_Q=0$ is exactly equivalent to (\ref{u12}) which is in turn
related by duality to the $Z_2$ gauge theory (\ref{u11a}). The
results of Section~\ref{sec:u1} then lead to the phase diagram at
the bottom of Fig~\ref{phase8}. We had argued in
Section~\ref{sec:z2} that $Z_P^{\prime}$ mapped onto $Z_{\rm
U(1)}$ in the region with $\langle {\bf N} \rangle \neq 0$; as
$Z_P^{\prime}$ was also obtained from $Z_P$ in the limit of a
strong charge 2 condensate, the connection between $Z_N$ and the
$Z_2$ gauge theory (\ref{u11a}) for small $g_Q$ is just as
expected.

We also discussed in Section~\ref{sec:z2} that $Z_P^{\prime}$, and
hence the small $g_Q$ limit of $Z_N$, misses a crucial piece of
physics: it is not able to describe the HN phase. The present
model $Z_N$ is able to repair this key deficiency in the opposite
limit of large $g_Q$. The most important property of $Z_N$ in this
limit is that it possesses a phase without any broken symmetries
or fractionalization, and this leads to the HN phase in
Fig~\ref{phase8}. In this respect the compact U(1) gauge theory
$Z_N$ is far removed from the $Z_2$ gauge theory $Z_{\rm U(1)}$ in
(\ref{u11a}) which did not have any such featureless phase; this
phase appears in the region corresponding to that with XY order in
the $Z_2$ gauge theory--unlike the discrete $Z_2$ gauge fields,
the $U(1)$ are able to quench the superfluid order by ``Higgsing''
it into a massive phase. The appearance of HN order is, of course,
satisfactory, and was the reason we stated our preference for
$Z_P$ over $Z_P^{\prime}$. To demonstrate the existence of this
featureless phase, consider the limit of large $g_Q$ in
(\ref{apph2}): there we can set $\epsilon_{\mu\nu\lambda}
\Delta_{\nu} p_{\bar{\jmath}\lambda} = 0$, and $Z_N$ reduces to
precisely the model studied in Ref.~\onlinecite{rodolfo} by Monte
Carlo simulations--this is a model of a compact U(1) gauge field
$A_{\mu}$ coupled to a single charge-1 scalar. In its simplest
form, such a model has only a single featureless phase
continuously connecting the confining and Higgs regions; however,
with the Berry phases present here a BC ordering transition can
occur. The resulting phase diagram is shown at the top of
Fig~\ref{phase8}.

Moving on, we  can also look at the limit of large $g$ exhibited
in Fig~\ref{phase8}. Here, the charge-1 scalar can be ignored, and
we have the theory of a charge-2 scalar coupled to a compact U(1)
gauge field with Berry phases. This was examined in
Ref.~\onlinecite{japan} and the resulting phase diagram appears
on the right of Fig~\ref{phase8}. Its phase transition involving
competition between BC and F order is described by the fully
frustrated Ising model in (\ref{u110}), and this last mapping
becomes exact for small $g_Q$.

Finally, we consider the limit of small $g$ appearing on the left
of Fig~\ref{phase8}. Now the charge-1 scalar completely quenches
the gauge field, and the theory becomes that of an ordinary
three-dimensional XY model; the XY ordering transition is that
between the P and HN phases.


\end{document}